\documentclass[apj,twocolumn]{openjournal}

\usepackage{xcolor}
\usepackage{textgreek}
\usepackage[utf8]{inputenc}
\usepackage[english]{babel}

\usepackage{hyperref}
\hypersetup{
    unicode, 
    colorlinks=true,
    linkcolor=linkcolor,
    citecolor=linkcolor,
    filecolor=linkcolor,
    urlcolor=linkcolor,
}
\usepackage{color,colortbl}
\definecolor{linkcolor}{rgb}{0.0,0.3,0.5}
\usepackage{tensind}
\tensordelimiter{?}
\DeclareGraphicsExtensions{.bmp,.png,.jpg,.pdf}
\usepackage{verbatim}
\usepackage[normalem]{ulem}
\usepackage{orcidlink}
\usepackage{soul}

\graphicspath{{Plots/Timing/}}

\newcommand{\timeddetected}{22}
\newcommand{\timedmonitored}{17}
\newcommand{\timedpreeph}{10}
\newcommand{\timedneweph}{\number\numexpr\timedmonitored-\timedpreeph\relax}
\newcommand{\timedtotalobserved}{33}



\usepackage[separate-uncertainty=true]{siunitx}
\DeclareSIUnit{\parsec}{pc}
\DeclareSIUnit{\deg}{deg}
\DeclareSIUnit{\sky}{sky}
\DeclareSIUnit{\jansky}{Jy}
\DeclareSIUnit{\gauss}{G}
\DeclareSIUnit{\year}{yr}

\usepackage[acronym,style=super,nonumberlist,toc]{glossaries}
\glsdisablehyper
\usepackage{fancyvrb}

\newcommand{\ac}[1]{\gls{#1}}
\newcommand{\acp}[1]{\glspl{#1}}

\newcommand{\hmsangle}{\ang[angle-symbol-over-decimal,angle-symbol-degree=\textsuperscript{h},angle-symbol-minute=\textsuperscript{m},angle-symbol-second=\textsuperscript{s},minimum-integer-digits = 2,]}
\newcommand{\dmsangle}{\ang[angle-symbol-over-decimal,minimum-integer-digits = 2,]}
\newcommand{\galangle}{\ang[angle-symbol-over-decimal]}

\newacronym{ilt}{ILT}{International Low Frequency Array Telescope}
\newacronym{lofar}{LOFAR}{Low Frequency Array}
\newacronym{lta}{LTA}{LOFAR Long Term Archive}
\newacronym{ilofar}{I--LOFAR}{The Irish Low Frequency Array Station}
\newacronym{lotaas}{LOTAAS}{LOFAR Tied-Array All-Sky Survey}
\newacronym{cep}{CEP}{Central Processing}

\newacronym{dm}{DM}{Dispersion Measure}
\newacronym{snr}{S/N}{Signal-to-Noise Ratio}
\newacronym{toa}{TOA}{Time of Arrival}
\newacronym{fwhm}{FWHM}{Full Width at Half Maximum}
\newacronym{rfi}{RFI}{Radio Frequency Interference}

\newcommand{\newacronymS}[1]{
  \newacronym{s#1}{S\textsubscript{#1}}{Source flux density at \unexpanded{\SI{150}{\mega\hertz}}}
}

\newcommand{\newacronymw}[1]{
  \newacronym{w#1}{w\textsubscript{#1}}{Source emission width at \unexpanded{\SI{150}{\mega\hertz}}}
}

\newacronymS{50}
\newacronymS{150}
\newacronymS{1250}

\newacronymw{50}
\newacronymw{150}
\newacronymw{1250}

\begin{document}
\title{Pulsar timing solutions for 17 pulsars at 150~MHz from the Irish LOFAR station}
\author{D.~J.~McKenna\orcidlink{0000-0001-7185-1310}}
\email{mckenna@astron.nl}
\affiliation{ASTRON, The Netherlands Institute for Radio Astronomy, Oude Hoogeveensedijk 4, 7991 PD Dwingeloo, The Netherlands}

\author{E.~F.~Keane\orcidlink{0000-0002-4553-655X}}
\affiliation{School of Physics, Trinity College Dublin, College Green, Dublin 2, D02 PN40, Ireland.}

\author{P.~T.~Gallagher\orcidlink{0000-0001-9745-0400}}
\affiliation{Astronomy \& Astrophysics Section, School of Cosmic Physics, Dublin Institute for Advanced Studies, DIAS Dunsink Observatory, Dublin D15 XR2R, Ireland}

\author{J.~McCauley\orcidlink{0000-0003-4399-2233}}
\affiliation{School of Physics, Trinity College Dublin, College Green, Dublin 2, D02 PN40, Ireland.}

\begin{abstract}
Pulsar timing is a foundational part of pulsar research to triage the most interesting systems and to characterise properties (rotational or otherwise) of the population of these extreme objects. Due to the efficiency of a number of sensitive and/or wide-field surveys in recent years, the number of new pulsars discoveries is growing year-on-year, and most of these lack even basic timing parameter measurements. This work aims to demonstrate the capabilities of international Low Frequency Array (LOFAR) stations operating as single telescopes to follow-up, time and characterise these sources, offering new insight into the emission properties of these neutron stars, and support efforts to build timing models for these sources. Between 2020 and 2023 we used the local-mode allocation of the Irish LOFAR station to follow-up $\timedtotalobserved$ pulsar candidates announced from various surveys at different observing frequencies to determine if an international LOFAR station has sufficient sensitivity to detect and time these sources. From the $\timedtotalobserved$ pulsars selected, $\timeddetected$ pulsars were detected and $\timedmonitored$ were selected for long-term monitoring across \SI{590}{\hour} of observing time. This has resulted in coherent timing solutions for all of these sources at \SI{150}{\mega\hertz} --- $\timedneweph$ of these have never had any reported timing solutions, the remaining $\timedpreeph$ solutions agree well with announcements from others since the beginning of our project. For a fraction of sources announced by surveys each year, the 14 international LOFAR stations are well placed to follow-up survey candidates for long-term pulsar monitoring beyond the standard timing campaigns performed at these telescopes to date, reducing the pressure on observing time availability at these observatories, and enabling the full scientific potential of these pulsars to be realised.
\end{abstract}


\maketitle

\section{Introduction}\label{sec:timingintro}
Pulsar timing is essential for characterising the rotational behaviour of pulsars, and is a foundational step in enabling a plethora of scientific applications~\citep{Lorimer2004,2021hgwa.bookE...4V}. Without timing it is almost impossible to identify the most interesting systems or to understand population properties. However regular monitoring of hundreds to thousands of pulsars, additionally ensuring broad frequency coverage, is intractable for most observatories given the over-subscription rates for the scarcest resource --- observing time.

International LOFAR stations are commonly used to perform pulsar observations during local mode operations. They are well-equipped for the task, with large fractional bandwidths and reasonably good gain at their frequencies~\citep{haarlemLOFARLOwFrequencyARray2013a}. The available observing time is also considerable, especially as there are currently 14 such stations which make up the extended baselines of the \acrfull{ilt}. 
 
Several international \ac{lofar} stations utilise a significant amount of their weekly time allocation to observe and monitor a wide range of pulsars. Recent works demonstrating these efforts range from follow-up of \textit{Fermi} gamma-ray pulsars~\citep{Griessmeier2021}, monitoring dispersion measure variations to detect free-electron variability associated with the interstellar medium and solar wind~\citep{Donner2020,susarla2024}, and the characterisation of the sub-\SI{100}{\mega\hertz} population of pulsars~\citep{Bondonneau2020}.

Over the past decade a large number of pulsars have been detected and have not received sufficient (or any) follow-up observations or timing analysis to determine their underlying rotation models. This means that any ephemerides for these sources can contain large uncertainties in the estimates for sky position, dispersion measure uncertainty and rotation period, often with no higher-order rotation terms quantified. At the beginning of this project (late 2020) the pulsar catalogue\footnote{\texttt{https://www.atnf.csiro.au/research/pulsar/psrcat/}}~\citep[v1.64,][]{manchesterATNFPulsarCatalogue2005} contained $452$ sources that met this description.

Since then the pulsar discovery rates have increased so that this situation continues: v1.70 (v2.6.0) listed $743$ ($957$) pulsars with no period derivative measured. 
These sources may not have been timed for many reasons, most likely from a lack of available observing time. 

The Five-Hundred-Metre Aperture Speherical Telescope (FAST), operating at L band, is responsible for most recent discoveries~\citep{2021RAA....21..107H,2023RAA....23j4001Z,2025RAA....25a4001H} meaning other telescopes, even 100-m class telescopes, cannot feasibly time these pulsars at L band as significantly more observing time per source is needed both for timing \textit{and} to overcome the gain disparity. However as pulsar spectra are often quite steep~\citep{2018MNRAS.473.4436J} a possible avenue for progress is to time these sources at lower frequency.

In this paper we describe our efforts to detect and monitor a fraction of this population of pulsars that were without timing-derived ephemerides at the start of this project; our observations used the Irish LOFAR station. We begin by describing the criteria for our source selection in \S~\ref{sec:timingsources} and the processing methodology in \S~\ref{sec:timingmethod}. We then present an overview of the source properties we measured in this project, and discuss the implications of this work in \S~\ref{sec:timingresults}, before concluding in \S~\ref{sec:timingconclusion}.

\section{Source Selection}\label{sec:timingsources}
Even before the recent influx of pulsar discoveries it was clear that LOFAR had potential for pulsar timing. \ac{lotaas} reported nearly 400 pulsars, detectable with LOFAR~\citep{sanidasLOFARTiedArrayAllSky2019b}, many of which were not observed on a regular basis. 

The sources selected for this work were initially derived from the catalogues used for a rotating radio transient (RRAT, \citealt{mclaughlinTransientRadioBursts2006a}) census at I--LOFAR described in~\citet{2024MNRAS.527.4397M}, but was later expanded to use normal pulsars reported by various catalogues produced during the same surveys, or using the same telescopes, in June 2021. The sources observed consisted of those within the \ac{lotaas} survey without reported parameters, from the Pushchino Radio Astronomy Observatory catalogues~\citep[PRAO, blind searches and PUMPs; \SI{111}{\mega\hertz}; ][]{2016ARep...60..220T,10.1093/mnras/stae070}, Green Bank North Celestial Cap survey~\citep[GBNCC; \SI{350}{\mega\hertz}; ][]{2014ApJ...791...67S}, Arecibo Pulsar Survey using the Arecibo L-band Feed Array~\citep[PALFA; \SI{1420}{\mega\hertz}]{2006ApJ...637..446C} and Arecibo \SI{327}{\mega\hertz} drift scan survey~\citep[AO327; ][]{2013ApJ...775...51D}. We note some of these sources were later re-detected during the Targeted search, Using LoTSS Images, for Polarised Pulsars project~\citep[TULIPP; overlapping bandwidth with this work; ][]{2022A&A...661A..87S} and FAST Galactic Plane Pulsar Snapshot survey~\citep[FAST GPPS; \SI{1250}{\mega\hertz}; ][]{2021RAA....21..107H}.

Reported source pointings, periods, and dispersion measures were cross-referenced between the sampled catalogues to ensure a lack of duplicated sources. Many surveys provide initial estimates for source parameters on their websites, but we found that candidate plots, produced by programs such as \texttt{prepfold}~\citep{ransomPRESTOPulsaRExploration2011}, also provided alongside such tables often provided more significant figures for the source parameters, even after accounting for instrumental uncertainty and source evolution over time. A number of initial positions, rotation periods and dispersion measures were thus manually extracted from these candidate plots. 

This resulted in $\timedtotalobserved$ (see Table~\ref{tab:timingobservedsources}) sources being selected as they had wide uncertainties on their existing basic ephemeris parameters and lacked published timing solutions for observations with I--LOFAR. Initially, this consistent of observations to determine if they could be detected, and if so, to monitor these for characterisation by performing pulsar timing on a long-term basis.

\section{Methodology}\label{sec:timingmethod}

\subsection{Observations}
Observations for sources to monitor as a part of this work that had not been observed as a part of the RRAT census discussed in~\cite{2024MNRAS.527.4397M} began in June 2021 and continued until August 2022. Observations of sources were initially \SI{59}{\minute} centered on the transit time of the source, although in the sources where weak emission was identified (as opposed to a non-detection), follow-up observations were extended to \SI{89}{\minute}, nominally reducing the noise floor by 18\%, to see if the observation criteria were met (see~\S\ref{sec:timingcriteria}), before keeping or removing this source from the potential observing pool for longer term monitoring. The maximum observing window of \SI{89}{\minute} was chosen, as longer observations on a regular basis with the station were deemed to be an excessive use of the \SI{31}{\hour} of weekly observing time typically available for standalone operations at the time. Timing observations proceeded for these sources, however updated sky positions from pulsar timing were not utilised for observations until after this work (July 2023). 

This was to ensure a consistent source pointing for the observations during follow-up observations to simplify later (spectral-) flux density analysis, as opposed to continuously deployed incrementally better sky positions over time. Dispersion measures were updated for coherent dedispersion as soon as deviations between the candidate ephemerides and detected emission were noted.

Observations were pre-processed using the standard I--LOFAR observing system, and RRAT processing pipeline~\citep{2024MNRAS.527.4397M}. Observations were taken with the \ac{lofar} HBA tiles in band `110\_190', with subbands 12 to 499 beam-formed on the best-known position of the sources. This covers \SIrange{102.2}{197.56}{\mega\hertz}, for a total bandwidth of \SI{95.51}{\mega\hertz}. Typically, 5\% of the band edges are unusable due to the high- and low-pass filters at \SI[parse-numbers = false]{100 \text{~and~} 200}{\mega\hertz}, and additional FM-band radio emissions near \SI{100}{\mega\hertz}. The analysed data are further restricted, as described below in~\S\ref{sec:timinganalysis}.

Data are recorded to disk and processed using a modified version of \textsc{cdmt}\footnote{\url{https://github.com/David-McKenna/cdmt}}~\citep{bassaEnablingPulsarFast2017}, using \textsc{udpPacketManager}~\citep{2024JOSS....9.5517M} to prepare voltages instead of reading \ac{cep}-produced \texttt{HDF5} files. \texttt{cdmt\_udp} is used to perform channelisation, coherent dedispersion and filterbanking of raw voltages on the Nvidia Tesla V100s in the local compute cluster~\citep{murphyFirstResultsREALtime2021}. The data were processed in a common manner across all observations, with a channelisation factor of $8$, producing \SI{24.41}{\kilo\hertz} channels, and temporal down-sampling at a factor of $16$, producing \SI{655.36}{\micro\second} samples. Each datum is stored as a 32-bit floating point value by default, before being re-sampled using the \texttt{digifil} tool, which is part of the \textsc{dspsr} software suite~\citep{vanstratenDSPSRDigitalSignal2011}, to re-sample the data as 8-bit samples, using the default \SI{10}{\second} re-scaling interval.

The 32-bit floating point filterbank for each observation is compressed using \textsc{zstandard}~\citep{Collet2018} and archived by default. In the case that a source is not detected after 6 months, the archived filterbank may undergo spectral downsampling by a factor of 8, using \texttt{digifil} without interval re-scaling, to reduce the on-disk size of the archived filterbank, and return it to the original spectral resolution of the data.

\subsection{Detection and Timing Criteria}\label{sec:timingcriteria}
The data were initially processed with \textsc{presto}'s \texttt{prepfold} tool to search for indications of emission after each observation. Two searches were performed per observation, one narrow search focusing on a fixed-\ac{dm} trial width\footnote{Using flags \texttt{-nodmsearch -nopdsearch -fine}}, while further searches were manually performed with numerous period trials, and a narrow \ac{dm} search window\footnote{\texttt{-npfact} was varied from 2--8, while \texttt{-ndmfact -dmstep} were kept at 3 and 1}. A source was deemed to be sufficiently bright to be timed if it could produce a \ac{snr} of $6$ or greater within the detected observing window, after generating a folded archive with \texttt{dspsr} and then preforming \ac{snr} optimisation with \texttt{pdmp}~\citep{hotanPSRCHIVEPSRFITSOpen2004}. An alternative criterion was applied to nulling or intermittent sources in the case that emission \textit{windows} provided sufficient \ac{snr} to be independently selected and timed when off-modes were zapped from the initial observation. Following this, sources were investigated to determine if there were ongoing (or completed) efforts to time the source by investigators associated with the surveys by using public observation databases, such as the \ac{lta}. In the case that the sources were not a part of observing campaigns elsewhere, follow-up observations were scheduled for the weeks post-detection to begin timing the sources at \ac{ilofar}. 

\subsection{Data Processing and Analysis}\label{sec:timinganalysis}
The 8-bit filterbank of observations were initially folded using trial parameters derived from the survey candidate announcement and best-fit \texttt{prepfold} parameters in an ephemeris using \texttt{dspsr}. Archives were formed with 256 bins per frequency channel across 30-second integrations. \texttt{clfd}~\citep{morelloHighTimeResolution2019} was then used to remove \ac{rfi} from each archive. \acp{toa} were generated using \textsc{psrchive}~\citep{hotanPSRCHIVEPSRFITSOpen2004}, whereby the longest observation was used to generate a analytic template model with \texttt{paas} to determine \ac{toa} measurements for each observation with \texttt{pat}, for initial timing efforts. During this early stage, it was found that the parabolic interpolation shift (PIS) method that was typically used for generating single pulse \ac{toa} was not performing as expected for the observed periodic emission. The measurement uncertainties, especially for weaker sources, appeared to be severely underestimated by the algorithm\footnote{A pointing of a non-detected intermittent source had an uncertainty below \SI{20}{\micro\second}, less than 5\% of the width of a single time sample of the input, and below 10\% of the bin width}. As a result, for the data described here, the Fourier-Domain Markov Chain Monte-Carlo (FDM) method was used to produce \ac{toa} measurements, following a comparison of the uncertainties associated with available options. Timing and ephemeris production was performed using \textsc{tempo2}~\citep{hobbsTEMPO2NewPulsartiming2006}.

When timing a pulsar from scratch it is necessary to perform an iterative process to finalise solutions. Once timing solutions were obtained for all sources, in July 2023, each observation was re-folded with 1024 bins per frequency channel with 10-second integrations to improve the archive quality and offer the opportunity to investigate the nulling behaviour of some sources across shorter durations. Additionally, in the case that a source had undergone spectral down-sampling after a non-detection, each full-resolution filterbank was re-processed to achieve a common spectral resolution. The archives were not down-sampled as it was found that the \verb|fscrunch| function of \texttt{psrsh} did not appear to re-normalise the channel data when archives were re-sampled, causing archives generated from spectrally down-sampled filterbanks to have weak contributions in \ac{snr} calculations. After a coherent timing solution was found, final 1-dimensional analytic timing templates were formed using the full set of observations combined with \texttt{psradd} using \texttt{paas}, and sources were re-fit using the final \acp{toa} measurements.

Source spectral flux density calculations were performed on the full corpus of observations for each source, in 10-MHz bandwidth blocks centred on frequencies in decreasing steps from \SIrange{181}{121}{\mega\hertz}. Given the known off-sets of each source inside the LOFAR stations beam, a corrected value is produced considering the offset from a Gaussian beam of known \ac{fwhm} (including the 1.02 correction factor, see Table~B.1 of~\cite{haarlemLOFARLOwFrequencyARray2013a} for details), for the spectral flux densities. Flux densities are derived from the modified radiometer equation in Equation~\ref{eq:pulsartimingrad}, which for \ac{lofar} is typically derived from the work of~\cite{kondratievLOFARCensusMillisecond2016a}. The specific modifications made for this work were previously discussed in \citet{2024MNRAS.527.4397M}, which gives:
\begin{equation}
    S(\nu, l, b) = \text{S/N} \frac{2k_{\rm B} \left(T_\text{sky}(\nu,l,b) + T_\text{ant}\right)}{A_\text{eff}(\nu)m_\text{II}(\nu, l, b, t)\beta\gamma} \Delta\;,
    \label{eq:pulsartimingrad}
\end{equation}
\begin{equation*}
    \Delta = \frac{1}{\sqrt{n_p  \Delta \nu_\text{eff} t_\text{obs}}}\sqrt{\frac{W}{P - W}}\;,
\end{equation*}
where $l$ and $b$ are Galactic coordinates. Several factors like $k_{\rm B}$ the Boltzmann constant, $n_{\rm p}=2$ the number of polarisations, the processing correction factor $\beta$ accounting for the 8-bit digitisation losses (below 1\%) and \ac{snr} losses due to using a boxcar as a reference \citep[about 6\%]{morelloHighTimeResolution2019}, are constant between all calculations. For a frequency, $\nu$, we sample the sky temperature, $T_\text{sky}$, using a primary-beam-lobe-convolved copy of the Low Frequency Sky Model~\citep[LFSM,][]{2017MNRAS.469.4537D}, as provided by \textsc{pygdsm}~\citep{priceGlobalSkyModels2021} and both the antenna temperature, $T_\text{ant}$ and effective collecting area at zenith, $A_\text{eff}$, for the LOFAR system~\citep{wijnholdsSituAntennaPerformance2011}. A Mueller matrix is derived from a Jones matrix calculated by \textsc{dreamBeam}
~\citep{carozzi2baOrNot2baDreamBeam2020} to correct for pointing losses with the factor $m_{\rm II}$ as a function of both frequency and altitude and azimuth, averaged across the galactic coordinates for the observing window of the observation. A source pointing error correction, $\gamma$, is calculated using a 2-dimensional Gaussian to correct for errors between the initial candidate and final timed location.

The post-\ac{rfi} cleaning bandwidth ($\Delta\nu_\text{eff}$), is extracted from the \textsc{clfd} report files, while the pulse width ($W$), dispersion measure and \ac{snr} are calculated using \texttt{pdmp}. \texttt{vap} is used to determine the total integrated time in the final output archive. The spectral behaviour of the data described above was fit using power laws and least-squared modelling in \textsc{Python} using the \textsc{lmfit} package~\citep{newvilleLmfitLmfitpy2021}. The above procedures combined result in a measurement of the source flux density with an absolute scale uncertainty of 50\%~\citep{2016A&A...591A.134B,kondratievLOFARCensusMillisecond2016a}.

\begin{table*}[!tp]
        \centering
        \begin{tabular}{lcclcrcr|rc|rrc}
\hline\hline
& & & & & & & & \multicolumn{2}{c|}{Measured} & \multicolumn{3}{c}{Corrected} \\
Source & Cat. & Year & Period & DM & T\textsubscript{obs} & $w_{50}$ & D.C. & $S_{150}$ &  $\alpha$ & $\theta_\text{off}$ & $S_{150}$ &  $\alpha$ \\
   & & & (\SI{}{\second}) & (\SI{}{\parsec\per\cubic\centi\metre})  & (\SI{}{\hour}) & (\SI{}{\milli\second}) & \% & (\SI{}{\milli\jansky}) &  & (\SI{}{\deg}) & (\SI{}{\milli\jansky}) & \\
\hline
J0104+6438\textsuperscript{$\dagger$} & GL & 2012 & 1.3862 & 42.026(9) & 20.4 & 37(7) & 2.7 & 5.8 & -0.9(2) & 0.10 & 5.8 & -0.9(2) \\
J0146+3055 & P & 2016 & 0.9381 & 25.791(6) & 24.6 & 26(5) & 2.8 & 7.2 & -3.1(3) & 0.34 & 7.6 & -3.0(3) \\
J0220+3626\textsuperscript{$\dagger$} & GL & 2016 & 1.0298 & 45.45(3) & 19.1 & 141(12) & 13.8 & 13.9 & -1.63(12) & 0.00 & 13.9 & -1.63(12) \\
J0226+3356 & C & 2020 & 1.2401 & 27.386(10) & 26.0 & 50(8) & 4.1 & 3.4 & -0.4(3) & 0.05 & 3.4 & -0.4(3) \\
J0317+1328\textsuperscript{$\dagger$} & LP & 2014 & 1.9742 & 12.779(12) & 12.6 & 23(7) & 1.2 & 4.2 & -1.4(3) & 0.28 & 4.4 & -1.4(3) \\
J0355+2838 & GL & 2014 & 0.3649 & 48.734(6) & 28.6 & 30(3) & 8.5 & 6.9 & -1.5(4) & 0.18 & 7.0 & -1.5(4) \\
J0608+1635 & AL & 2013 & 0.9458 & 86.146(6) & 18.3 & 47(7) & 5.0 & 23.7 & -2.9(3) & 0.21 & 24.2 & -2.8(3) \\
J1132+2513 & LP & 2015 & 1.0021 & 23.716(6) & 23.7 & 21(5) & 2.1 & 9.3 & -2.3(3) & 0.30 & 9.7 & -2.2(3) \\
J1243+3946 & LP & 2016 & 1.3103 & 26.462(8) & 19.8 & 35(7) & 2.7 & 7.9 & -1.13(17) & 0.24 & 8.1 & -1.08(18) \\
J1327+3423\textsuperscript{$\dagger$} & GLP & 2014 & 0.0415 & 4.1850(3) & 6.1 &  2.4(7) & 5.7 & 76.2 & -2.9(3) & 0.63 & 90.8 & -2.5(3) \\
J1536+1759\textsuperscript{$\dagger$} & P & 2017 & 0.9331 & 28.594(7) & 47.7 & 19(4) & 2.1 & 1.6 & -2.8(4) & 0.19 & 1.6 & -2.8(4) \\
J1832+2749 & AL & 2014 & 0.6317 & 47.387(5) & 34.2 & 31(4) & 5.0 & 6.6 & -0.80(12) & 0.07 & 6.6 & -0.79(12) \\
J1837+5156 & GL & 2014 & 0.6919 & 43.819(6) & 32.9 & 48(6) & 7.0 & 8.1 & 0.0(4) & 0.22 & 8.3 & 0.0(4) \\
J2000+2920 & AL & 2015 & 3.0738 & 132.151(19) & 38.4 & 102(18) & 3.3 & 3.7 & 0.8(4) & 0.18 & 3.7 & 0.8(4) \\
J2105+1908\textsuperscript{$\dagger$} & P & 2018 & 3.5298 & 34.45(2) & 17.6 & 84(17) & 2.4 & 2.7 & -2.6(5) & 0.03 & 2.7 & -2.6(5) \\
J2202+2134\textsuperscript{$\dagger$} & P & 2020 & 1.3583 & 17.742(8) & 28.5 & 13(4) & 1.0 & 5.6 & -1.89(18) & 0.21 & 5.7 & -1.85(18) \\
J2347+0300 & AL & 2016 & 1.3861 & 16.124(9) & 18.1 & 25(6) & 1.9 & 13.5 & -2.4(6) & 0.15 & 13.6 & -2.4(6) \\

\hline\hline
        \end{tabular}
        \caption[Timed pulsar properties and emission characteristics.]{The observed properties of the sources chosen for monitoring as a part of this work. The table contains columns for the original source catalogs, the year of discovery, a rounded rotation period, their \acrfull{dm}, the total observation time (T\textsubscript{obs}), the \acrfull{fwhm} of emission (w\textsubscript{50}), the duty cycle (D.C.), and both corrected and uncorrected flux density at \SI{150}{\mega\hertz} and spectral power law exponent for the emission, correcting for off-axis observations with respect to the final timing location ($\theta$\textsubscript{off}). Sources with a dagger suffix have shown signs of nulling or highly variable emission. In the catalogues column, we note surveys where parameters were reported for these sources during the observing campaign. ``A'' for work from the Arecibo telescope, ``G'' for the GBNCC survey, ``L'' for the LOTAAS survey, and ``P'' is the PRAO catalogues.}
        \label{tab:timingmonitoredproperties}
    \end{table*}

\section{Results and Discussion}\label{sec:timingresults}
From observations of $\timedtotalobserved$ sources, $\timedmonitored$ were chosen to be monitored out of $\timeddetected$ sources with detectable emission (full list of observations available in Tables~\ref{tab:timingobservedsources} and~\ref{tab:timingobservedsourcesnondetected}). Over \SI{400}{\hour} of telescope time were used to time these sources between December 2020 and July 2023. The overall source properties can be found in Table \ref{tab:timingmonitoredproperties}. These sources were announced as pulsar candidates between 2012 and 2020, cover a rotation period range of \SIrange{0.04}{3.53}{\second} with pointing-corrected flux densities at \SI{150}{\mega\hertz} ranging from \SIrange{1.6}{90.8}{\milli\jansky}. Duty cycles ranged from 1.0\% to 13.8\%. 

While there was insufficient sensitivity to perform a nulling analysis, a number of sources were observed to have variability in their waterfall plots when heavily downsampled, and are noted in Table~\ref{tab:timingmonitoredproperties} with a dagger $\dagger$ suffix. Sources that appeared to have some nulling characteristics include PSRs J0104+6438, J0220+3626, J0317+1328 and J2105+1908. This was expected for PSRs J0317+1328 and J2105+1908 as these were previously reported as RRATs by \cite{tyulbashevDetection25New2018}. PSRs J1327+3423, J1536+1759 and J2202+2134 were observed to have high flux density variability between observations, but were reasonably stable within single observing windows;  PSR J1536+1759 is discussed further in \S~\ref{sec:1536}.

\textsc{DM\_phase}~\citep{Seymour2019} was investigated as an alternative method to calculate dispersion measures, however it was found to provide at least one order of magnitude higher uncertainties for smooth, single component pulse profiles, which make up the majority of the sources in this work (see Fig.~\ref{fig:timingprofiles}). Given it was originally authored with the goal of optimising complex fast radio burst pulses from the CHIME telescope~\citep{2019ApJ...885L..24C}, this is not a completely unexpected result.

Pulsar timing was performed for all sources, and coherent solutions were found in all cases. Ephemerides for each of the sources can be found in Tables \ref{tab:timingephemeridesprereport} and \ref{tab:timingephemerides}. Source positions are reported with uncertainties at arc-second level or better. While estimates have been provided for the source distances based on the \cite{Ocker_2026} NE2025 model, one source (PSR~J1243$+$3946) has only a lower limit distance estimate as it exceeds the maximum expected dispersion measure for its line of sight within the Milky Way. The average summed profiles after dedispersion and \ac{rfi} flagging are provided in Figure~\ref{fig:timingprofiles}.

In this work a population analysis was not conducted given these sources represent a filtered selection of sources from multiple telescopes, observed at multiple frequencies, each of which have their own selection criteria for detecting new pulsars, which were then further filtered by the selection criteria of the \ac{lofar} telescope.

\begin{figure*}[!t]
    \centering
    \includegraphics[width=\textwidth]{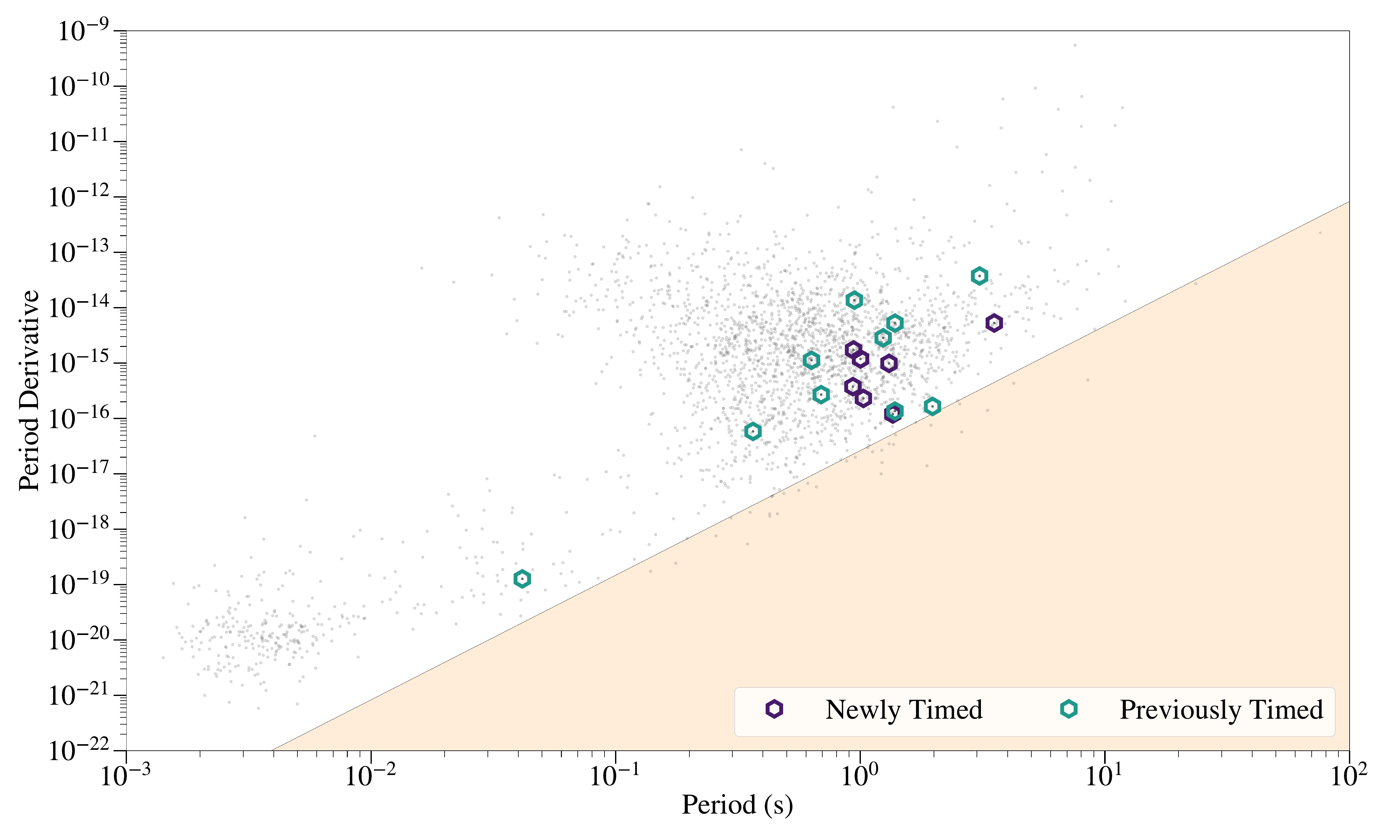}
    \caption[Period-period derivative plot highlighting sources timed with \ac{ilofar}.]{A period-period derivative phase space plot of known pulsars, with the sources timed as a part of this work highlighted. This plot was generated with the aid of \textsc{psrqpy}~\citep{psrqpy}.}
    \label{fig:timingppdot}
\end{figure*}

\begin{table*}[!ht]
        \centering
        \begin{tabular}{lcccc}
\hline\hline
Source  &  J0146+3055  &  J0220+3626  &  J1132+2513  &  J1243+3946 \\
\\
Survey  &  PRAO,  &  GBNCC, PRAO,  &  GBNCC, PRAO,  &  PRAO, \\
& FAST GPPS & LOTAAS, TULIPP & AR327, FAST GPPS & FAST GPPS \\
\hline
Right Ascension (hms)  &  \hmsangle{01;46;41.029}(13)  &  \hmsangle{02;20;42.16}(12)  &  \hmsangle{11;32;29.947}(12)  &  \hmsangle{12;43;3.509}(10) \\
Declination (dms)  &  \dmsangle{+30;55;19.1}(4)  &  \dmsangle{+36;26;55}(2)  &  \dmsangle{+25;13;32.6}(4)  &  \dmsangle{+39;46;9.17}(13) \\
Galactic Longitude ($^\circ$)  &  \galangle{136.67773}(5)  &  \galangle{142.3172}(5)  &  \galangle{214.61428}(5)  &  \galangle{130.24022}(4) \\
Galactic Latitude ($^\circ$)  &  \galangle{-30.470883}(98)  &  \galangle{-23.0404}(6)  &  \galangle{72.20598}(11)  &  \galangle{77.23989}(4) \\
Dispersion Measure  (\SI{}{\parsec\per\centi\metre\cubed})  &  25.791(6)  &  45.45(3)  &  23.716(6)  &  26.462(8) \\
Distance (\SI{}{\parsec})  &  1793  &  3244  &  5216  &  $>$8300 \\
\\
Period (\SI{}{\second})  &  0.938062139694(16)  &  1.029795190912(97)  &  1.002131964901(10)  &  1.310291834197(14) \\
Period Derivative (\SI{e-15}{\second\per\second})  &  1.7250(17)  &  0.23(3)  &  1.18842(97)  &  0.9873(18) \\
Characteristic Age (\SI{}{\mega\year})  &  8.62  &  71.7  &  13.4  &  21.0 \\
Magnetic Field (\SI{e12}{\gauss})  &  1.29  &  0.490  &  1.10  &  1.15 \\
\\
Timing Start (MJD)  &  59528  &  59528  &  59206  &  59521 \\
Timing End (MJD)  &  60123  &  60010  &  60094  &  60107 \\
Reference Epoch (MJD)  &  59857  &  59750  &  59750  &  59750 \\
N\textsubscript{TOAs}  &  20  &  19  &  33  &  21 \\
RMS Residuals (\SI{}{\micro\second})  &  705  &  4659  &  692  &  520 \\
\hline\hline

\\

\hline\hline
 Source  &  J1536+1759  &  J2105+1908  &  J2202+2134 \\
 \\
 Survey  &  PRAO  &  PRAO  &  PRAO, GBNCC \\
 \hline
 Right Ascension (hms)  &  \hmsangle{15;36;34.32}(3)  &  \hmsangle{21;05;30.61}(8)  &  \hmsangle{22;02;16.984}(17) \\
 Declination (dms)  &  \dmsangle{+17;59;24.7}(7)  &  \dmsangle{+19;08;34.8}(1.1)  &  \dmsangle{+21;34;33.6}(4) \\
 Galactic Longitude ($^\circ$)  &  \galangle{28.28365}(12)  &  \galangle{66.8555}(3)  &  \galangle{78.62145}(7) \\
 Galactic Latitude ($^\circ$)  &  \galangle{50.97946}(19)  &  \galangle{-18.3455}(3)  &  \galangle{-26.39566}(12) \\
 Dispersion Measure  (\SI{}{\parsec\per\centi\metre\cubed})  &  28.594(7)  &  34.48(2)  &  17.742(8) \\
 Distance (\SI{}{\parsec})  &  4266  &  2622  &  1407 \\
 \\
 Period (\SI{}{\second})  &  0.93312798225(9)  &  3.5298270014(3)  &  1.35830601223(2) \\
 Period Derivative (\SI{e-15}{\second\per\second})  &  0.376(9)  &  5.29(4)  &  0.1202(15) \\
 Characteristic Age (\SI{}{\mega\year})  &  39.4  &  10.6  &  179 \\
 Magnetic Field (\SI{e12}{\gauss})  &  0.599  &  4.37  &  0.409 \\
 \\
 Timing Start (MJD)  &  59626  &  59569  &  59226 \\
 Timing End (MJD)  &  60101  &  60108  &  60108 \\
 Reference Epoch (MJD)  &  59750  &  59850  &  59819 \\
 N\textsubscript{TOAs}  &  27  &  14  &  16 \\
 RMS Residuals (\SI{}{\micro\second})  &  1915  &  2521  &  518 \\
\hline\hline

        \end{tabular}
        \caption[Novel pulsar timing ephemerides.]{Novel timing ephemerides produced at \ac{ilofar} as a part of this work. The distance measurement of PSR~J1243$+$3946 indicates that it exceeds the maximum dispersion measure expected along the line of sight within the Milky Way from the NE2025 model.}
        \label{tab:timingephemerides}
    \end{table*}

    \begin{table*}[!ht]
        \centering
        \begin{tabular}{lcccc}
\hline\hline
 Source  &  J0104+6438  &  J0226+3356  &  J0317+1328  &  J0355+2838 \\
 \\
 Survey  &  GBNCC  &  CHIME  &  LOTAAS  &  GBNCC \\
 Solution & \href{https://github.com/GBNCC/data}{GBNCC/data} &  \cite{dongCHIME2022}  &  \cite{Wateren2023}  & \href{https://github.com/GBNCC/data}{GBNCC/data} \\
 \hline
 Right Ascension (hms)  &  \hmsangle{01;04;37.21}(3)  &  \hmsangle{02;26;57.14}(3)  &  \hmsangle{03;17;48.71}(8)  &  \hmsangle{03;55;22.89}(4) \\
 Declination (dms)  &  \dmsangle{+64;38;30.0}(2)  &  \dmsangle{+33;56;26.7}(7)  &  \dmsangle{+13;28;32}(4)  &  \dmsangle{+28;38;42}(2) \\
 Galactic Longitude ($^\circ$)  &  \galangle{124.34340}(14)  &  \galangle{144.65706}(13)  &  \galangle{168.7491}(3)  &  \galangle{164.78042}(16) \\
 Galactic Latitude ($^\circ$)  &  \galangle{1.80604}(7)  &  \galangle{-24.8698}(2)  &  \galangle{-36.0598}(10)  &  \galangle{-18.9122}(6) \\
 Dispersion Measure  (\SI{}{\parsec\per\centi\metre\cubed})  &  42.026(9)  &  27.386(10)  &  12.779(12)  &  48.734(6) \\
 Distance (\SI{}{\parsec})  &  2629  &  1803  &  747  &  2546 \\
 \\
 Period (\SI{}{\second})  &  1.38618423347(3)  &  1.240102562457(15)  &  1.97423969309(5)  &  0.36492920980(3) \\
 Period Derivative (\SI{e-15}{\second\per\second})  &  0.136(3)  &  2.8547(14)  &  0.165(4)  &  0.0584(15) \\
 Characteristic Age (\SI{}{\mega\year})  &  162  &  6.89  &  190  &  99.0 \\
 Magnetic Field (\SI{e12}{\gauss})  &  0.439  &  1.9  &  0.578  &  0.148 \\
 \\
 Timing Start (MJD)  &  59395  &  59177  &  59129  &  59401 \\
 Timing End (MJD)  &  60123  &  59982  &  59989  &  60031 \\
 Reference Epoch (MJD)  &  59759  &  59641  &  59500  &  59920 \\
 N\textsubscript{TOAs}  &  37  &  16  &  17  &  32 \\
 RMS Residuals (\SI{}{\micro\second})  &  1683  &  718  &  1409  &  2056 \\
\hline\hline
\\
\hline\hline
 Source  &  J0608+1635 &  J1327+3423  &  J1832+2749  &  J1837+5156 \\
 \\
 Survey  &  PALFA  &  GBNCC/PRAO  &  AO327  &  GBNCC \\
 Solution &  \cite{Parent2022}  &  \cite{Fiore2023}  & \cite{2025ao327} & \href{https://github.com/GBNCC/data}{GBNCC/data} \\
 \hline
 Right Ascension (hms)  &  \hmsangle{06;08;51.654}(18) &  \hmsangle{13;27;7.547}(3)  &  \hmsangle{18;32;19.002}(13)  &  \hmsangle{18;37;51.25}(2) \\
 Declination (dms)  &  \dmsangle{+16;35;2}(3)  &  \dmsangle{+34;23;37.66}(5)  &  \dmsangle{+27;49;36.2}(2)  &  \dmsangle{+51;56;14.6}(2) \\
 Galactic Longitude ($^\circ$)  &  \galangle{193.36751}(8)  &  \galangle{78.605524}(12)  &  \galangle{56.35782}(5)  &  \galangle{80.95498}(10) \\
 Galactic Latitude ($^\circ$)  &  \galangle{-1.5704}(7)  &  \galangle{79.446663}(14)  &  \galangle{16.16734}(6)  &  \galangle{23.04883}(7) \\
 Dispersion Measure  (\SI{}{\parsec\per\centi\metre\cubed})  &  86.146(6) &  4.1850(3)  &  47.387(5)  &  43.819(6) \\
 Distance (\SI{}{\parsec})  &  2107 &  532  &  3440  &  3633 \\
 \\
 Period (\SI{}{\second})  &  0.94584729633(3) &  0.04151271040323(15)  &  0.631706905364(9)  &  0.691906460248(12) \\
 Period Derivative (\SI{e-15}{\second\per\second})  &  13.675(3) &  0.000127(12)  &  1.1249(10)  &  0.2694(16) \\
 Characteristic Age (\SI{}{\mega\year})  &  1.1 &  5180  &  8.9  &  40.7 \\
 Magnetic Field (\SI{e12}{\gauss})  &  3.64 &  0.00232  &  0.853  &  0.437 \\
 \\
 Timing Start (MJD)  &  59395 &  59248  &  59382  &  59382 \\
 Timing End (MJD)  &  60052 &  60101  &  60108  &  60102 \\
 Reference Epoch (MJD)  &  59750 &  59750  &  59750  &  59750 \\
 N\textsubscript{TOAs}  &  41 &  50  &  37  &  38 \\
 RMS Residuals (\SI{}{\micro\second})  &  2287 &  218  &  1368  &  1717 \\
\hline\hline
\\
\hline\hline
 Source  &  J2000+2920  &  J2347+0300 \\
 \\
 Survey  &  PALFA  &  AO327 \\
 Solution &  \cite{Parent2022}  & \cite{2025ao327} \\
 \hline
 Right Ascension (hms)  &  \hmsangle{20;00;16.57}(7)  &  \hmsangle{23;47;44.41}(4) \\
 Declination (dms)  &  \dmsangle{+29;20;7.4}(1.2)  &  \dmsangle{+03;00;14.4}(1.2) \\
 Galactic Longitude ($^\circ$)  &  \galangle{66.5494}(3)  &  \galangle{93.44425}(15) \\
 Galactic Latitude ($^\circ$)  &  \galangle{-0.3527}(3)  &  \galangle{-56.1771}(3) \\
 Dispersion Measure  (\SI{}{\parsec\per\centi\metre\cubed})  &  132.151(19)  &  16.124(9) \\
 Distance (\SI{}{\parsec})  &  6523  &  1154 \\
 \\
 Period (\SI{}{\second})  &  3.0737882818(2)  &  1.386056611589(9) \\
 Period Derivative (\SI{e-15}{\second\per\second})  &  37.40(3)  &  5.2657(12) \\
 Characteristic Age (\SI{}{\mega\year})  &  1.30  &  4.17 \\
 Magnetic Field (\SI{e12}{\gauss})  &  10.8  &  2.73 \\
 \\
 Timing Start (MJD)  &  59381  &  59395 \\
 Timing End (MJD)  &  60102  &  60108 \\
 Reference Epoch (MJD)  &  59750  &  59750 \\
 N\textsubscript{TOAs}  &  32  &  21 \\
 RMS Residuals (\SI{}{\micro\second})  &  6064  &  495 \\
\hline\hline

        \end{tabular}
        \caption{Timing solutions for sources that have previously been announced by work at other telescopes.}
        \label{tab:timingephemeridesprereport}
\end{table*}

\subsection{Previously Timed Sources}\label{sec:timingrespretimed}

From these $\timedmonitored$ sources, $\timedpreeph$ had timing solutions provided at the time of submission of this work. This includes the CHIME PSR J0226+3356~\citep{dongCHIME2022}, LOTAAS PSR J0317+1328~\citep{Wateren2023}, GBT PSR J1327+3423~\citep{Fiore2023}, PALFA PSRs J0608+1635 and J2000+2920~\citep{Parent2022} and AO327 PSRs J1832+2749 and J2347+0300~\citep{2025ao327}. Three further sources, PSRs J0104+6438, J0355+2838 and J1837+5156, appear in the GBNCC GitHub Data repository\footnote{\url{https://github.com/GBNCC/data}} and cite the work of \cite{mcewen2024}.

The ephemerides produced as a part of this work fall within the expectations of the models produced by the cited works. Primarily due to the frequency range of LOFAR, our determined dispersion measure uncertainties are reduced over previous works for the monitored sources, or overlap the uncertainties provided (PSRs J1327+3423, J1832+2749, J2347+0300).

\subsection{Sources of note}

\subsubsection{PSR J0220+3626}
PSR~J0220+3626 has previously been reported to be a pulsar with sub-pulse drifting and giant pulses by the PRAO~\citep{Teplykh2022}. While \ac{ilofar} lacks the sensitivity to probe the sub-pulse properties of the source, 9 single pulses were detected from the source as a part of this work. Our measurement of a \ac{fwhm} of \SI{141}{\milli\second} at \SI{150}{\mega\hertz} is consistent with the findings of \citet{Teplykh2022} who found a $\sim$\SI{220}{\milli\second} main pulse component at \SI{111}{\mega\hertz}.

\subsubsection{PSR J0355+2838}
In the work of \cite{McEwen2020}, PSR J0355+2838 was reported with a significantly modified period that is present in the current version of the pulsar catalogue (v2.6.0), \SI{94.3}{\milli\second}, as opposed to the original survey period, the latter of which was used for this work and is present in the ephemeris, \SI{364.9}{\milli\second}. This does not appear to be an integer multiple of the final rotation period, and we are uncertain how it was obtained. Periodic emission could not be detected from the source from the 4 longest observations of the source (each \SI{89}{\minute}) at \ac{ilofar}, when the data were folded at trials for a topocentric period of \SI{94.3\pm0.13}{\milli\second}. While the pulsar is still represented with this shorter period in v2.6.0 of \textsc{psrcat}, given that the \href{https://github.com/GBNCC/data}{GBNCC/data} GitHub repo (containing work from \cite{mcewen2024}) contains a new ephemeris matching the period we use here, this seems to be an error that will be corrected in a future release of the pulsar catalogue.

\subsubsection{PSR J1327+3423}
PSR J1327+3423 is a $41.5$-ms rotation period pulsar
with a low dispersion measure of \SI{4.185}{\parsec\per\cubic\centi\metre}, with an ephemeris previously reported by~\cite{Fiore2023}, and also announced in the PRAO pulsar and RRAT catalogues. It is suspected to be a recycled pulsar given the timing solution determined it has a rotation derivative of \SI{1.27e-19}{\second\per\second}. It was the fastest rotating source observed as a part of this work, and was initially not going to be monitored due to the relatively low number of time samples per rotation (63) produced with the standard pipeline and strong scattering at low frequencies. 

However, it was noticed that due to its brightness it was visible as an off-axis source when observing RRAT~J1336+3416 as a part of the work described in~\citet{2024MNRAS.527.4397M}, which received a large amount of observing time due to the presence of a secondary off-axis pulse train in the single-pulse timing results. As a result, a sufficient number of \acp{toa} could be generated for the source `for free' using the J1336$+$3416 data, and we tried to form a coherent timing solution for the source given the data was available. These off-axis data were only used to determine \acp{toa} for the source, e.g. not flux densities, and is not included in Table~\ref{tab:timingmonitoredproperties} or~\ref{tab:timingobservedsources} for the source. The uncertainties on the \acp{toa} measurements are relatively high despite the source brightness as a result of being observed off-axis, at a low number of samples per revolution, resulting in a model that can describe the overall basic parameters of the source, but cannot identify higher precision effects on the \acp{toa}, such as the proper-motion parameters, reported by \cite{Fiore2023}.

\subsubsection{PSR J1536+1759}\label{sec:1536}
PSR J1536+1749 was observed to be extremely variable between observations during this work. It was only detected in preview \texttt{prepfold} plots for 55\% of the observations (18 of 33, Table~\ref{tab:timingobservedsources}), with a mean observing time of \SI{1.45}{\hour}. Initially, \acp{toa} were only generated for observations which had detectable emission in the preview plots, which were then used to form the initial timing model for the source. Afterwards, all \acp{toa} were included in the fit, however only an additional 4  \acp{toa} from the 15 non-visual-detections were used for the timing solution as the remaining \acp{toa} had large uncertainties, or large phase offsets. Notably, during an extended local mode window with the station, the source went undetected across 5 observations spread across 6 days. While the previous \SI{59}{\minute} observation of the source had shown a \ac{snr} of 5.5, during this period it was not detected with observations of 89, 89, 119, 149 and \SI{89}{\minute} in length. These observations indicate this source may be a nulling pulsar or an intermittent pulsar.

\subsubsection{PSR J2105+1908}
PSR J2105+1908 was observed following a detection and localisation during observations taken with the LOFAR core (McKenna et al., in prep., LOFAR project code LC16\_016). At the time of that work, the source did not have a known period. and was not initially observed as a part of the \ac{ilofar} RRAT census~\citep{2024MNRAS.527.4397M} due to the low reported peak flux density by the PRAO. Follow-up observations after the core detection and localisation allowed for periodic emission to be detected from the source, however no single pulses were detected during the \SI{17.6}{\hour} of observations performed prior to July 2023. \cite{Tyulbashev2022} later reported a rotation period of \SI{3.5297}{\second} which aligns with the result obtained during the \ac{lofar} core work, and our final timing solution presented in Table~\ref{tab:timingephemerides}.

The nulling behaviour of the source, previously observed as 0.38(11) with the LOFAR \textit{core}, has not yet been characterised for our I-LOFAR observations, as initial attempts to separate the modes demonstrated that the active mode was too quiet to be detected in the typical \SIrange{10}{30}{\second} integration used for this analysis. While increasing the integration time further may allow for the two modes to be separated, the short switching duration observed from the \ac{lofar} core observation, and small $N$ statistics means this is not likely to result in an accurate value, or reduce the uncertainty from the \ac{lofar} core observation.

\section{Conclusions}\label{sec:timingconclusion}

We have presented $\timedmonitored$ pulsar timing ephemerides for pulsars generated from observations using \ac{ilofar} at \SI{150}{\mega\hertz} between 2020 and 2023. From these, $\timedneweph$ sources did not have previous ephemerides published, while the remaining $\timedpreeph$ match recently announced results from other telescopes. Systemic analysis was performed to describe the emission properties of the sources, which were found to have flux densities at \SI{150}{\mega\hertz} ranging from \SIrange{2}{91}{\milli\jansky}, with spectral indices from $-3.0$ to $+0.8$. Duty cycles ranged from 1.0\% to 14\%. 5 of the observed sources showed signs of nulling or highly variable emission.

This work demonstrates the effectiveness of international LOFAR stations for establishing timing solutions for newly discovered pulsars. This has been shown to be feasible even for pulsars where very scant initial information is available, i.e. they can be timed completely `from scratch', and even for pulsars discovered by instruments with much larger gain. The latter leverages the spectra of pulsars which are typically brighter at low frequencies.

As the discovery rate of pulsars is accelerating, thanks to FAST, and looks set to accelerate even more when the Square Kilometre Array telescopes (currently under construction) are operational during this decade, the need for other telescopes to time pulsars is ever growing. This timing procedure is essential for, at the very least, triaging discoveries to identify systems most suitable for dedicated deep monitoring, but also for properly characterising the spin evolution of the pulsar population at large. We will continue and expand upon the work presented here.

\section*{Acknowledgments}

D.J.McK acknowledges funding under the Government of Ireland Postgraduate Scholarship (GOIPG/2019/2798) administered by the Irish Research Council (IRC). D.J.McK and E.F.K. would like to thank Aaron Golden of the University of Galway for hosting them on numerous `busy weeks' which helped immensely in getting this work to completion. The Rosse Observatory is operated by Trinity College Dublin. I-LOFAR infrastructure has benefited from funding from Science Foundation Ireland, a predecessor of Taighde \'{E}ireann --- Research Ireland.

\section*{Data Availability}
Folded profiles and times of arrival per-observation and averaged across the entire set of observations can be found on Zenodo~\citep{mckenna_2025_14938680}. Stokes $I$ filterbanks or pre-time and frequency scrunched archives can be made available on a request to the contact author.

\texttt{tempo2} pulsar timing solutions presented in Tables \ref{tab:timingephemerides} and \ref{tab:timingephemeridesprereport} can be found on Zenodo~\citep{mckenna_2025_14938977} in a machine-readable format.

\bibliographystyle{apsrev4-1}

\bibliography{ie613_pulsar_timing}

\begin{thebibliography}{53}%
\makeatletter
\providecommand \@ifxundefined [1]{%
 \@ifx{#1\undefined}
}%
\providecommand \@ifnum [1]{%
 \ifnum #1\expandafter \@firstoftwo
 \else \expandafter \@secondoftwo
 \fi
}%
\providecommand \@ifx [1]{%
 \ifx #1\expandafter \@firstoftwo
 \else \expandafter \@secondoftwo
 \fi
}%
\providecommand \natexlab [1]{#1}%
\providecommand \enquote  [1]{``#1''}%
\providecommand \bibnamefont  [1]{#1}%
\providecommand \bibfnamefont [1]{#1}%
\providecommand \citenamefont [1]{#1}%
\providecommand \href@noop [0]{\@secondoftwo}%
\providecommand \href [0]{\begingroup \@sanitize@url \@href}%
\providecommand \@href[1]{\@@startlink{#1}\@@href}%
\providecommand \@@href[1]{\endgroup#1\@@endlink}%
\providecommand \@sanitize@url [0]{\catcode `\\12\catcode `\$12\catcode `\&12\catcode `\#12\catcode `\^12\catcode `\_12\catcode `\%12\relax}%
\providecommand \@@startlink[1]{}%
\providecommand \@@endlink[0]{}%
\providecommand \url  [0]{\begingroup\@sanitize@url \@url }%
\providecommand \@url [1]{\endgroup\@href {#1}{\urlprefix }}%
\providecommand \urlprefix  [0]{URL }%
\providecommand \Eprint [0]{\href }%
\providecommand \doibase [0]{http://dx.doi.org/}%
\providecommand \selectlanguage [0]{\@gobble}%
\providecommand \bibinfo  [0]{\@secondoftwo}%
\providecommand \bibfield  [0]{\@secondoftwo}%
\providecommand \translation [1]{[#1]}%
\providecommand \BibitemOpen [0]{}%
\providecommand \bibitemStop [0]{}%
\providecommand \bibitemNoStop [0]{.\EOS\space}%
\providecommand \EOS [0]{\spacefactor3000\relax}%
\providecommand \BibitemShut  [1]{\csname bibitem#1\endcsname}%
\let\auto@bib@innerbib\@empty
\bibitem [{\citenamefont {Lorimer}\ and\ \citenamefont {Kramer}(2005)}]{Lorimer2004}%
  \BibitemOpen
  \bibfield  {author} {\bibinfo {author} {\bibfnamefont {D.~R.}\ \bibnamefont {Lorimer}}\ and\ \bibinfo {author} {\bibfnamefont {M.}~\bibnamefont {Kramer}},\ }\href@noop {} {\emph {\bibinfo {title} {Handbook of {{Pulsar Astronomy}}}}},\ \bibinfo {series} {Cambridge observing handbooks for research astronomers}, Vol.~\bibinfo {volume} {4}\ (\bibinfo  {publisher} {Cambridge University Press},\ \bibinfo {year} {2005})\BibitemShut {NoStop}%
\bibitem [{\citenamefont {{Verbiest}}\ \emph {et~al.}(2021)\citenamefont {{Verbiest}}, \citenamefont {{Os{\l}owski}},\ and\ \citenamefont {{Burke-Spolaor}}}]{2021hgwa.bookE...4V}%
  \BibitemOpen
  \bibfield  {author} {\bibinfo {author} {\bibfnamefont {J.~P.~W.}\ \bibnamefont {{Verbiest}}}, \bibinfo {author} {\bibfnamefont {S.}~\bibnamefont {{Os{\l}owski}}}, \ and\ \bibinfo {author} {\bibfnamefont {S.}~\bibnamefont {{Burke-Spolaor}}},\ }in\ \href {\doibase 10.1007/978-981-15-4702-7_4-1} {\emph {\bibinfo {booktitle} {Handbook of Gravitational Wave Astronomy}}},\ \bibinfo {editor} {edited by\ \bibinfo {editor} {\bibfnamefont {C.}~\bibnamefont {{Bambi}}}, \bibinfo {editor} {\bibfnamefont {S.}~\bibnamefont {{Katsanevas}}}, \ and\ \bibinfo {editor} {\bibfnamefont {K.~D.}\ \bibnamefont {{Kokkotas}}}}\ (\bibinfo {year} {2021})\ p.~\bibinfo {pages} {4}\BibitemShut {NoStop}%
\bibitem [{\citenamefont {van Haarlem}\ \emph {et~al.}(2013)\citenamefont {van Haarlem}, \citenamefont {Wise}, \citenamefont {Gunst}, \citenamefont {Heald}, \citenamefont {McKean}, \citenamefont {Hessels}, \citenamefont {de~Bruyn}, \citenamefont {Nijboer}, \citenamefont {Swinbank}, \citenamefont {Fallows}, \citenamefont {Brentjens}, \citenamefont {Nelles}, \citenamefont {Beck}, \citenamefont {Falcke},\ and\ \citenamefont {et~al.}}]{haarlemLOFARLOwFrequencyARray2013a}%
  \BibitemOpen
  \bibfield  {author} {\bibinfo {author} {\bibfnamefont {M.~P.}\ \bibnamefont {van Haarlem}}, \bibinfo {author} {\bibfnamefont {M.~W.}\ \bibnamefont {Wise}}, \bibinfo {author} {\bibfnamefont {A.~W.}\ \bibnamefont {Gunst}}, \bibinfo {author} {\bibfnamefont {G.}~\bibnamefont {Heald}}, \bibinfo {author} {\bibfnamefont {J.~P.}\ \bibnamefont {McKean}}, \bibinfo {author} {\bibfnamefont {J.~W.~T.}\ \bibnamefont {Hessels}}, \bibinfo {author} {\bibfnamefont {A.~G.}\ \bibnamefont {de~Bruyn}}, \bibinfo {author} {\bibfnamefont {R.}~\bibnamefont {Nijboer}}, \bibinfo {author} {\bibfnamefont {J.}~\bibnamefont {Swinbank}}, \bibinfo {author} {\bibfnamefont {R.}~\bibnamefont {Fallows}}, \bibinfo {author} {\bibfnamefont {M.}~\bibnamefont {Brentjens}}, \bibinfo {author} {\bibfnamefont {A.}~\bibnamefont {Nelles}}, \bibinfo {author} {\bibfnamefont {R.}~\bibnamefont {Beck}}, \bibinfo {author} {\bibfnamefont {H.}~\bibnamefont {Falcke}}, \ and\ \bibinfo {author} {\bibnamefont {et~al.}},\ }\href {\doibase 10.1051/0004-6361/201220873}
  {\bibfield  {journal} {\bibinfo  {journal} {Astronomy \& Astrophysics}\ }\textbf {\bibinfo {volume} {556}},\ \bibinfo {pages} {A2} (\bibinfo {year} {2013})}\BibitemShut {NoStop}%
\bibitem [{\citenamefont {Grießmeier}\ \emph {et~al.}(2021)\citenamefont {Grießmeier}, \citenamefont {Smith}, \citenamefont {Theureau}, \citenamefont {Johnson}, \citenamefont {Kerr}, \citenamefont {Bondonneau}, \citenamefont {Cognard},\ and\ \citenamefont {Serylak}}]{Griessmeier2021}%
  \BibitemOpen
  \bibfield  {author} {\bibinfo {author} {\bibfnamefont {J.-M.}\ \bibnamefont {Grießmeier}}, \bibinfo {author} {\bibfnamefont {D.~A.}\ \bibnamefont {Smith}}, \bibinfo {author} {\bibfnamefont {G.}~\bibnamefont {Theureau}}, \bibinfo {author} {\bibfnamefont {T.~J.}\ \bibnamefont {Johnson}}, \bibinfo {author} {\bibfnamefont {M.}~\bibnamefont {Kerr}}, \bibinfo {author} {\bibfnamefont {L.}~\bibnamefont {Bondonneau}}, \bibinfo {author} {\bibfnamefont {I.}~\bibnamefont {Cognard}}, \ and\ \bibinfo {author} {\bibfnamefont {M.}~\bibnamefont {Serylak}},\ }\href {\doibase 10.1051/0004-6361/202140841} {\bibfield  {journal} {\bibinfo  {journal} {Astronomy \& Astrophysics}\ }\textbf {\bibinfo {volume} {654}},\ \bibinfo {pages} {A43} (\bibinfo {year} {2021})}\BibitemShut {NoStop}%
\bibitem [{\citenamefont {{Donner}}\ \emph {et~al.}(2020)\citenamefont {{Donner}}, \citenamefont {{Verbiest, J. P. W.}}, \citenamefont {{Tiburzi, C.}}, \citenamefont {{Oslowski, S.}}, \citenamefont {{K\"unsem\"oller, J.}}, \citenamefont {{Bak Nielsen, A.-S.}}, \citenamefont {{Grie\ss{}meier, J.-M.}}, \citenamefont {{Serylak, M.}}, \citenamefont {{Kramer, M.}}, \citenamefont {{Anderson, J. M.}}, \citenamefont {{Wucknitz, O.}}, \citenamefont {{Keane, E.}}, \citenamefont {{Kondratiev, V.}}, \citenamefont {{Sobey, C.}}, \citenamefont {{McKee, J. W.}}, \citenamefont {{Bilous, A. V.}}, \citenamefont {{Breton, R. P.}}, \citenamefont {{Br\"uggen, M.}}, \citenamefont {{Ciardi, B.}}, \citenamefont {{Hoeft, M.}}, \citenamefont {{van Leeuwen, J.}},\ and\ \citenamefont {{Vocks, C.}}}]{Donner2020}%
  \BibitemOpen
  \bibfield  {author} {\bibinfo {author} {\bibnamefont {{Donner}}}, \bibinfo {author} {\bibnamefont {{Verbiest, J. P. W.}}}, \bibinfo {author} {\bibnamefont {{Tiburzi, C.}}}, \bibinfo {author} {\bibnamefont {{Oslowski, S.}}}, \bibinfo {author} {\bibnamefont {{K\"unsem\"oller, J.}}}, \bibinfo {author} {\bibnamefont {{Bak Nielsen, A.-S.}}}, \bibinfo {author} {\bibnamefont {{Grie\ss{}meier, J.-M.}}}, \bibinfo {author} {\bibnamefont {{Serylak, M.}}}, \bibinfo {author} {\bibnamefont {{Kramer, M.}}}, \bibinfo {author} {\bibnamefont {{Anderson, J. M.}}}, \bibinfo {author} {\bibnamefont {{Wucknitz, O.}}}, \bibinfo {author} {\bibnamefont {{Keane, E.}}}, \bibinfo {author} {\bibnamefont {{Kondratiev, V.}}}, \bibinfo {author} {\bibnamefont {{Sobey, C.}}}, \bibinfo {author} {\bibnamefont {{McKee, J. W.}}}, \bibinfo {author} {\bibnamefont {{Bilous, A. V.}}}, \bibinfo {author} {\bibnamefont {{Breton, R. P.}}}, \bibinfo {author} {\bibnamefont {{Br\"uggen, M.}}}, \bibinfo {author} {\bibnamefont {{Ciardi, B.}}}, \bibinfo
  {author} {\bibnamefont {{Hoeft, M.}}}, \bibinfo {author} {\bibnamefont {{van Leeuwen, J.}}}, \ and\ \bibinfo {author} {\bibnamefont {{Vocks, C.}}},\ }\href {\doibase 10.1051/0004-6361/202039517} {\bibfield  {journal} {\bibinfo  {journal} {A\&A}\ }\textbf {\bibinfo {volume} {644}},\ \bibinfo {pages} {A153} (\bibinfo {year} {2020})}\BibitemShut {NoStop}%
\bibitem [{\citenamefont {{Susarla}}\ \emph {et~al.}(2024)\citenamefont {{Susarla}}, \citenamefont {{Chalumeau}}, \citenamefont {{Tiburzi}}, \citenamefont {{Keane}}, \citenamefont {{Verbiest}}, \citenamefont {{Hazboun}}, \citenamefont {{Krishnakumar}}, \citenamefont {{Iraci}}, \citenamefont {{Shaifullah}}, \citenamefont {{Golden}}, \citenamefont {{Bak Nielsen}}, \citenamefont {{Donner}}, \citenamefont {{Grie{\ss}meier}}, \citenamefont {{Keith}}, \citenamefont {{Os{\l}owski}}, \citenamefont {{Porayko}}, \citenamefont {{Serylak}}, \citenamefont {{Anderson}}, \citenamefont {{Br{\"u}ggen}}, \citenamefont {{Ciardi}}, \citenamefont {{Dettmar}}, \citenamefont {{Hoeft}}, \citenamefont {{K{\"u}nsem{\"o}ller}}, \citenamefont {{Schwarz}},\ and\ \citenamefont {{Vocks}}}]{susarla2024}%
  \BibitemOpen
  \bibfield  {author} {\bibinfo {author} {\bibfnamefont {S.~C.}\ \bibnamefont {{Susarla}}}, \bibinfo {author} {\bibfnamefont {A.}~\bibnamefont {{Chalumeau}}}, \bibinfo {author} {\bibfnamefont {C.}~\bibnamefont {{Tiburzi}}}, \bibinfo {author} {\bibfnamefont {E.~F.}\ \bibnamefont {{Keane}}}, \bibinfo {author} {\bibfnamefont {J.~P.~W.}\ \bibnamefont {{Verbiest}}}, \bibinfo {author} {\bibfnamefont {J.~S.}\ \bibnamefont {{Hazboun}}}, \bibinfo {author} {\bibfnamefont {M.~A.}\ \bibnamefont {{Krishnakumar}}}, \bibinfo {author} {\bibfnamefont {F.}~\bibnamefont {{Iraci}}}, \bibinfo {author} {\bibfnamefont {G.~M.}\ \bibnamefont {{Shaifullah}}}, \bibinfo {author} {\bibfnamefont {A.}~\bibnamefont {{Golden}}}, \bibinfo {author} {\bibfnamefont {A.~S.}\ \bibnamefont {{Bak Nielsen}}}, \bibinfo {author} {\bibfnamefont {J.}~\bibnamefont {{Donner}}}, \bibinfo {author} {\bibfnamefont {J.~M.}\ \bibnamefont {{Grie{\ss}meier}}}, \bibinfo {author} {\bibfnamefont {M.~J.}\ \bibnamefont {{Keith}}}, \bibinfo {author} {\bibfnamefont
  {S.}~\bibnamefont {{Os{\l}owski}}}, \bibinfo {author} {\bibfnamefont {N.~K.}\ \bibnamefont {{Porayko}}}, \bibinfo {author} {\bibfnamefont {M.}~\bibnamefont {{Serylak}}}, \bibinfo {author} {\bibfnamefont {J.~M.}\ \bibnamefont {{Anderson}}}, \bibinfo {author} {\bibfnamefont {M.}~\bibnamefont {{Br{\"u}ggen}}}, \bibinfo {author} {\bibfnamefont {B.}~\bibnamefont {{Ciardi}}}, \bibinfo {author} {\bibfnamefont {R.~J.}\ \bibnamefont {{Dettmar}}}, \bibinfo {author} {\bibfnamefont {M.}~\bibnamefont {{Hoeft}}}, \bibinfo {author} {\bibfnamefont {J.}~\bibnamefont {{K{\"u}nsem{\"o}ller}}}, \bibinfo {author} {\bibfnamefont {D.}~\bibnamefont {{Schwarz}}}, \ and\ \bibinfo {author} {\bibfnamefont {C.}~\bibnamefont {{Vocks}}},\ }\href {\doibase 10.1051/0004-6361/202450680} {\bibfield  {journal} {\bibinfo  {journal} {\aap}\ }\textbf {\bibinfo {volume} {692}},\ \bibinfo {eid} {A18} (\bibinfo {year} {2024})}\BibitemShut {NoStop}%
\bibitem [{\citenamefont {{Bondonneau}}\ \emph {et~al.}(2020)\citenamefont {{Bondonneau}}, \citenamefont {{Grie{\ss}meier}}, \citenamefont {{Theureau}}, \citenamefont {{Bilous}}, \citenamefont {{Kondratiev}}, \citenamefont {{Serylak}}, \citenamefont {{Keith}},\ and\ \citenamefont {{Lyne}}}]{Bondonneau2020}%
  \BibitemOpen
  \bibfield  {author} {\bibinfo {author} {\bibfnamefont {L.}~\bibnamefont {{Bondonneau}}}, \bibinfo {author} {\bibfnamefont {J.~M.}\ \bibnamefont {{Grie{\ss}meier}}}, \bibinfo {author} {\bibfnamefont {G.}~\bibnamefont {{Theureau}}}, \bibinfo {author} {\bibfnamefont {A.~V.}\ \bibnamefont {{Bilous}}}, \bibinfo {author} {\bibfnamefont {V.~I.}\ \bibnamefont {{Kondratiev}}}, \bibinfo {author} {\bibfnamefont {M.}~\bibnamefont {{Serylak}}}, \bibinfo {author} {\bibfnamefont {M.~J.}\ \bibnamefont {{Keith}}}, \ and\ \bibinfo {author} {\bibfnamefont {A.~G.}\ \bibnamefont {{Lyne}}},\ }\href {\doibase 10.1051/0004-6361/201936829} {\bibfield  {journal} {\bibinfo  {journal} {\aap}\ }\textbf {\bibinfo {volume} {635}},\ \bibinfo {eid} {A76} (\bibinfo {year} {2020})}\BibitemShut {NoStop}%
\bibitem [{\citenamefont {Manchester}\ \emph {et~al.}(2005)\citenamefont {Manchester}, \citenamefont {Hobbs}, \citenamefont {Teoh},\ and\ \citenamefont {Hobbs}}]{manchesterATNFPulsarCatalogue2005}%
  \BibitemOpen
  \bibfield  {author} {\bibinfo {author} {\bibfnamefont {R.~N.}\ \bibnamefont {Manchester}}, \bibinfo {author} {\bibfnamefont {G.~B.}\ \bibnamefont {Hobbs}}, \bibinfo {author} {\bibfnamefont {A.}~\bibnamefont {Teoh}}, \ and\ \bibinfo {author} {\bibfnamefont {M.}~\bibnamefont {Hobbs}},\ }\href {\doibase 10.1086/428488} {\bibfield  {journal} {\bibinfo  {journal} {The Astronomical Journal}\ }\textbf {\bibinfo {volume} {129}},\ \bibinfo {pages} {1993} (\bibinfo {year} {2005})}\BibitemShut {NoStop}%
\bibitem [{\citenamefont {{Han}}\ \emph {et~al.}(2021)\citenamefont {{Han}}, \citenamefont {{Wang}}, \citenamefont {{Wang}}, \citenamefont {{Wang}}, \citenamefont {{Zhou}}, \citenamefont {{Sun}}, \citenamefont {{Yan}}, \citenamefont {{Su}}, \citenamefont {{Jing}}, \citenamefont {{Chen}}, \citenamefont {{Gao}}, \citenamefont {{Hou}}, \citenamefont {{Xu}}, \citenamefont {{Lee}}, \citenamefont {{Wang}}, \citenamefont {{Jiang}}, \citenamefont {{Xu}}, \citenamefont {{Yan}}, \citenamefont {{Gan}}, \citenamefont {{Guan}}, \citenamefont {{Huang}}, \citenamefont {{Jiang}}, \citenamefont {{Li}}, \citenamefont {{Men}}, \citenamefont {{Sun}}, \citenamefont {{Wang}}, \citenamefont {{Wang}}, \citenamefont {{Wang}}, \citenamefont {{Xie}}, \citenamefont {{Xu}}, \citenamefont {{Yao}}, \citenamefont {{You}}, \citenamefont {{Yu}}, \citenamefont {{Yuan}}, \citenamefont {{Yuen}}, \citenamefont {{Zhang}},\ and\ \citenamefont {{Zhu}}}]{2021RAA....21..107H}%
  \BibitemOpen
  \bibfield  {author} {\bibinfo {author} {\bibfnamefont {J.~L.}\ \bibnamefont {{Han}}}, \bibinfo {author} {\bibfnamefont {C.}~\bibnamefont {{Wang}}}, \bibinfo {author} {\bibfnamefont {P.~F.}\ \bibnamefont {{Wang}}}, \bibinfo {author} {\bibfnamefont {T.}~\bibnamefont {{Wang}}}, \bibinfo {author} {\bibfnamefont {D.~J.}\ \bibnamefont {{Zhou}}}, \bibinfo {author} {\bibfnamefont {J.-H.}\ \bibnamefont {{Sun}}}, \bibinfo {author} {\bibfnamefont {Y.}~\bibnamefont {{Yan}}}, \bibinfo {author} {\bibfnamefont {W.-Q.}\ \bibnamefont {{Su}}}, \bibinfo {author} {\bibfnamefont {W.-C.}\ \bibnamefont {{Jing}}}, \bibinfo {author} {\bibfnamefont {X.}~\bibnamefont {{Chen}}}, \bibinfo {author} {\bibfnamefont {X.~Y.}\ \bibnamefont {{Gao}}}, \bibinfo {author} {\bibfnamefont {L.-G.}\ \bibnamefont {{Hou}}}, \bibinfo {author} {\bibfnamefont {J.}~\bibnamefont {{Xu}}}, \bibinfo {author} {\bibfnamefont {K.~J.}\ \bibnamefont {{Lee}}}, \bibinfo {author} {\bibfnamefont {N.}~\bibnamefont {{Wang}}}, \bibinfo {author} {\bibfnamefont
  {P.}~\bibnamefont {{Jiang}}}, \bibinfo {author} {\bibfnamefont {R.-X.}\ \bibnamefont {{Xu}}}, \bibinfo {author} {\bibfnamefont {J.}~\bibnamefont {{Yan}}}, \bibinfo {author} {\bibfnamefont {H.-Q.}\ \bibnamefont {{Gan}}}, \bibinfo {author} {\bibfnamefont {X.}~\bibnamefont {{Guan}}}, \bibinfo {author} {\bibfnamefont {W.-J.}\ \bibnamefont {{Huang}}}, \bibinfo {author} {\bibfnamefont {J.-C.}\ \bibnamefont {{Jiang}}}, \bibinfo {author} {\bibfnamefont {H.}~\bibnamefont {{Li}}}, \bibinfo {author} {\bibfnamefont {Y.-P.}\ \bibnamefont {{Men}}}, \bibinfo {author} {\bibfnamefont {C.}~\bibnamefont {{Sun}}}, \bibinfo {author} {\bibfnamefont {B.-J.}\ \bibnamefont {{Wang}}}, \bibinfo {author} {\bibfnamefont {H.~G.}\ \bibnamefont {{Wang}}}, \bibinfo {author} {\bibfnamefont {S.-Q.}\ \bibnamefont {{Wang}}}, \bibinfo {author} {\bibfnamefont {J.-T.}\ \bibnamefont {{Xie}}}, \bibinfo {author} {\bibfnamefont {H.}~\bibnamefont {{Xu}}}, \bibinfo {author} {\bibfnamefont {R.}~\bibnamefont {{Yao}}}, \bibinfo {author} {\bibfnamefont
  {X.-P.}\ \bibnamefont {{You}}}, \bibinfo {author} {\bibfnamefont {D.~J.}\ \bibnamefont {{Yu}}}, \bibinfo {author} {\bibfnamefont {J.-P.}\ \bibnamefont {{Yuan}}}, \bibinfo {author} {\bibfnamefont {R.}~\bibnamefont {{Yuen}}}, \bibinfo {author} {\bibfnamefont {C.-F.}\ \bibnamefont {{Zhang}}}, \ and\ \bibinfo {author} {\bibfnamefont {Y.}~\bibnamefont {{Zhu}}},\ }\href {\doibase 10.1088/1674-4527/21/5/107} {\bibfield  {journal} {\bibinfo  {journal} {Research in Astronomy and Astrophysics}\ }\textbf {\bibinfo {volume} {21}},\ \bibinfo {eid} {107} (\bibinfo {year} {2021})}\BibitemShut {NoStop}%
\bibitem [{\citenamefont {{Zhou}}\ \emph {et~al.}(2023)\citenamefont {{Zhou}}, \citenamefont {{Han}}, \citenamefont {{Xu}}, \citenamefont {{Wang}}, \citenamefont {{Wang}}, \citenamefont {{Wang}}, \citenamefont {{Jing}}, \citenamefont {{Chen}}, \citenamefont {{Yan}}, \citenamefont {{Su}}, \citenamefont {{Gan}}, \citenamefont {{Jiang}}, \citenamefont {{Sun}}, \citenamefont {{Wang}}, \citenamefont {{Wang}}, \citenamefont {{Wang}}, \citenamefont {{Xu}},\ and\ \citenamefont {{You}}}]{2023RAA....23j4001Z}%
  \BibitemOpen
  \bibfield  {author} {\bibinfo {author} {\bibfnamefont {D.~J.}\ \bibnamefont {{Zhou}}}, \bibinfo {author} {\bibfnamefont {J.~L.}\ \bibnamefont {{Han}}}, \bibinfo {author} {\bibfnamefont {J.}~\bibnamefont {{Xu}}}, \bibinfo {author} {\bibfnamefont {C.}~\bibnamefont {{Wang}}}, \bibinfo {author} {\bibfnamefont {P.~F.}\ \bibnamefont {{Wang}}}, \bibinfo {author} {\bibfnamefont {T.}~\bibnamefont {{Wang}}}, \bibinfo {author} {\bibfnamefont {W.-C.}\ \bibnamefont {{Jing}}}, \bibinfo {author} {\bibfnamefont {X.}~\bibnamefont {{Chen}}}, \bibinfo {author} {\bibfnamefont {Y.}~\bibnamefont {{Yan}}}, \bibinfo {author} {\bibfnamefont {W.-Q.}\ \bibnamefont {{Su}}}, \bibinfo {author} {\bibfnamefont {H.-Q.}\ \bibnamefont {{Gan}}}, \bibinfo {author} {\bibfnamefont {P.}~\bibnamefont {{Jiang}}}, \bibinfo {author} {\bibfnamefont {J.-H.}\ \bibnamefont {{Sun}}}, \bibinfo {author} {\bibfnamefont {H.-G.}\ \bibnamefont {{Wang}}}, \bibinfo {author} {\bibfnamefont {N.}~\bibnamefont {{Wang}}}, \bibinfo {author} {\bibfnamefont {S.-Q.}\
  \bibnamefont {{Wang}}}, \bibinfo {author} {\bibfnamefont {R.-X.}\ \bibnamefont {{Xu}}}, \ and\ \bibinfo {author} {\bibfnamefont {X.-P.}\ \bibnamefont {{You}}},\ }\href {\doibase 10.1088/1674-4527/accc76} {\bibfield  {journal} {\bibinfo  {journal} {Research in Astronomy and Astrophysics}\ }\textbf {\bibinfo {volume} {23}},\ \bibinfo {eid} {104001} (\bibinfo {year} {2023})}\BibitemShut {NoStop}%
\bibitem [{\citenamefont {{Han}}\ \emph {et~al.}(2025)\citenamefont {{Han}}, \citenamefont {{Zhou}}, \citenamefont {{Wang}}, \citenamefont {{Su}}, \citenamefont {{Yan}}, \citenamefont {{Jing}}, \citenamefont {{Yang}}, \citenamefont {{Wang}}, \citenamefont {{Wang}}, \citenamefont {{Xu}}, \citenamefont {{Cai}}, \citenamefont {{Sun}}, \citenamefont {{Yang}}, \citenamefont {{Xu}}, \citenamefont {{Wang}},\ and\ \citenamefont {{You}}}]{2025RAA....25a4001H}%
  \BibitemOpen
  \bibfield  {author} {\bibinfo {author} {\bibfnamefont {J.~L.}\ \bibnamefont {{Han}}}, \bibinfo {author} {\bibfnamefont {D.~J.}\ \bibnamefont {{Zhou}}}, \bibinfo {author} {\bibfnamefont {C.}~\bibnamefont {{Wang}}}, \bibinfo {author} {\bibfnamefont {W.~Q.}\ \bibnamefont {{Su}}}, \bibinfo {author} {\bibfnamefont {Y.}~\bibnamefont {{Yan}}}, \bibinfo {author} {\bibfnamefont {W.~C.}\ \bibnamefont {{Jing}}}, \bibinfo {author} {\bibfnamefont {Z.~L.}\ \bibnamefont {{Yang}}}, \bibinfo {author} {\bibfnamefont {P.~F.}\ \bibnamefont {{Wang}}}, \bibinfo {author} {\bibfnamefont {T.}~\bibnamefont {{Wang}}}, \bibinfo {author} {\bibfnamefont {J.}~\bibnamefont {{Xu}}}, \bibinfo {author} {\bibfnamefont {N.~N.}\ \bibnamefont {{Cai}}}, \bibinfo {author} {\bibfnamefont {J.~H.}\ \bibnamefont {{Sun}}}, \bibinfo {author} {\bibfnamefont {Q.~L.}\ \bibnamefont {{Yang}}}, \bibinfo {author} {\bibfnamefont {R.~X.}\ \bibnamefont {{Xu}}}, \bibinfo {author} {\bibfnamefont {H.~G.}\ \bibnamefont {{Wang}}}, \ and\ \bibinfo {author}
  {\bibfnamefont {X.~P.}\ \bibnamefont {{You}}},\ }\href {\doibase 10.1088/1674-4527/ada3b7} {\bibfield  {journal} {\bibinfo  {journal} {Research in Astronomy and Astrophysics}\ }\textbf {\bibinfo {volume} {25}},\ \bibinfo {eid} {014001} (\bibinfo {year} {2025})}\BibitemShut {NoStop}%
\bibitem [{\citenamefont {{Jankowski}}\ \emph {et~al.}(2018)\citenamefont {{Jankowski}}, \citenamefont {{van Straten}}, \citenamefont {{Keane}}, \citenamefont {{Bailes}}, \citenamefont {{Barr}}, \citenamefont {{Johnston}},\ and\ \citenamefont {{Kerr}}}]{2018MNRAS.473.4436J}%
  \BibitemOpen
  \bibfield  {author} {\bibinfo {author} {\bibfnamefont {F.}~\bibnamefont {{Jankowski}}}, \bibinfo {author} {\bibfnamefont {W.}~\bibnamefont {{van Straten}}}, \bibinfo {author} {\bibfnamefont {E.~F.}\ \bibnamefont {{Keane}}}, \bibinfo {author} {\bibfnamefont {M.}~\bibnamefont {{Bailes}}}, \bibinfo {author} {\bibfnamefont {E.~D.}\ \bibnamefont {{Barr}}}, \bibinfo {author} {\bibfnamefont {S.}~\bibnamefont {{Johnston}}}, \ and\ \bibinfo {author} {\bibfnamefont {M.}~\bibnamefont {{Kerr}}},\ }\href {\doibase 10.1093/mnras/stx2476} {\bibfield  {journal} {\bibinfo  {journal} {\mnras}\ }\textbf {\bibinfo {volume} {473}},\ \bibinfo {pages} {4436} (\bibinfo {year} {2018})}\BibitemShut {NoStop}%
\bibitem [{\citenamefont {Sanidas}\ \emph {et~al.}(2019)\citenamefont {Sanidas}, \citenamefont {Cooper}, \citenamefont {Bassa}, \citenamefont {Hessels}, \citenamefont {Kondratiev}, \citenamefont {Michilli}, \citenamefont {Stappers}, \citenamefont {Tan}, \citenamefont {van Leeuwen}, \citenamefont {Cerrigone}, \citenamefont {Fallows}, \citenamefont {Iacobelli}, \citenamefont {Orr{\'u}}, \citenamefont {Pizzo}, \citenamefont {Shulevski}, \citenamefont {Toribio}, \citenamefont {ter Veen}, \citenamefont {Zucca}, \citenamefont {Bondonneau}, \citenamefont {Grie{\ss}meier}, \citenamefont {Karastergiou}, \citenamefont {Kramer},\ and\ \citenamefont {Sobey}}]{sanidasLOFARTiedArrayAllSky2019b}%
  \BibitemOpen
  \bibfield  {author} {\bibinfo {author} {\bibfnamefont {S.}~\bibnamefont {Sanidas}}, \bibinfo {author} {\bibfnamefont {S.}~\bibnamefont {Cooper}}, \bibinfo {author} {\bibfnamefont {C.~G.}\ \bibnamefont {Bassa}}, \bibinfo {author} {\bibfnamefont {J.~W.~T.}\ \bibnamefont {Hessels}}, \bibinfo {author} {\bibfnamefont {V.~I.}\ \bibnamefont {Kondratiev}}, \bibinfo {author} {\bibfnamefont {D.}~\bibnamefont {Michilli}}, \bibinfo {author} {\bibfnamefont {B.~W.}\ \bibnamefont {Stappers}}, \bibinfo {author} {\bibfnamefont {C.~M.}\ \bibnamefont {Tan}}, \bibinfo {author} {\bibfnamefont {J.}~\bibnamefont {van Leeuwen}}, \bibinfo {author} {\bibfnamefont {L.}~\bibnamefont {Cerrigone}}, \bibinfo {author} {\bibfnamefont {R.~A.}\ \bibnamefont {Fallows}}, \bibinfo {author} {\bibfnamefont {M.}~\bibnamefont {Iacobelli}}, \bibinfo {author} {\bibfnamefont {E.}~\bibnamefont {Orr{\'u}}}, \bibinfo {author} {\bibfnamefont {R.~F.}\ \bibnamefont {Pizzo}}, \bibinfo {author} {\bibfnamefont {A.}~\bibnamefont {Shulevski}}, \bibinfo {author}
  {\bibfnamefont {M.~C.}\ \bibnamefont {Toribio}}, \bibinfo {author} {\bibfnamefont {S.}~\bibnamefont {ter Veen}}, \bibinfo {author} {\bibfnamefont {P.}~\bibnamefont {Zucca}}, \bibinfo {author} {\bibfnamefont {L.}~\bibnamefont {Bondonneau}}, \bibinfo {author} {\bibfnamefont {J.-M.}\ \bibnamefont {Grie{\ss}meier}}, \bibinfo {author} {\bibfnamefont {A.}~\bibnamefont {Karastergiou}}, \bibinfo {author} {\bibfnamefont {M.}~\bibnamefont {Kramer}}, \ and\ \bibinfo {author} {\bibfnamefont {C.}~\bibnamefont {Sobey}},\ }\href {\doibase 10.1051/0004-6361/201935609} {\bibfield  {journal} {\bibinfo  {journal} {Astronomy \& Astrophysics}\ }\textbf {\bibinfo {volume} {626}},\ \bibinfo {pages} {A104} (\bibinfo {year} {2019})}\BibitemShut {NoStop}%
\bibitem [{\citenamefont {McLaughlin}\ \emph {et~al.}(2006)\citenamefont {McLaughlin}, \citenamefont {Lyne}, \citenamefont {Lorimer}, \citenamefont {Kramer}, \citenamefont {Faulkner}, \citenamefont {Manchester}, \citenamefont {Cordes}, \citenamefont {Camilo}, \citenamefont {Possenti}, \citenamefont {Stairs}, \citenamefont {Hobbs}, \citenamefont {D'Amico}, \citenamefont {Burgay},\ and\ \citenamefont {O'Brien}}]{mclaughlinTransientRadioBursts2006a}%
  \BibitemOpen
  \bibfield  {author} {\bibinfo {author} {\bibfnamefont {M.~A.}\ \bibnamefont {McLaughlin}}, \bibinfo {author} {\bibfnamefont {A.~G.}\ \bibnamefont {Lyne}}, \bibinfo {author} {\bibfnamefont {D.~R.}\ \bibnamefont {Lorimer}}, \bibinfo {author} {\bibfnamefont {M.}~\bibnamefont {Kramer}}, \bibinfo {author} {\bibfnamefont {A.~J.}\ \bibnamefont {Faulkner}}, \bibinfo {author} {\bibfnamefont {R.~N.}\ \bibnamefont {Manchester}}, \bibinfo {author} {\bibfnamefont {J.~M.}\ \bibnamefont {Cordes}}, \bibinfo {author} {\bibfnamefont {F.}~\bibnamefont {Camilo}}, \bibinfo {author} {\bibfnamefont {A.}~\bibnamefont {Possenti}}, \bibinfo {author} {\bibfnamefont {I.~H.}\ \bibnamefont {Stairs}}, \bibinfo {author} {\bibfnamefont {G.}~\bibnamefont {Hobbs}}, \bibinfo {author} {\bibfnamefont {N.}~\bibnamefont {D'Amico}}, \bibinfo {author} {\bibfnamefont {M.}~\bibnamefont {Burgay}}, \ and\ \bibinfo {author} {\bibfnamefont {J.~T.}\ \bibnamefont {O'Brien}},\ }\href {\doibase 10.1038/nature04440} {\bibfield  {journal} {\bibinfo  {journal}
  {nat}\ }\textbf {\bibinfo {volume} {439}},\ \bibinfo {pages} {817} (\bibinfo {year} {2006})}\BibitemShut {NoStop}%
\bibitem [{\citenamefont {{McKenna}}\ \emph {et~al.}(2024{\natexlab{a}})\citenamefont {{McKenna}}, \citenamefont {{Keane}}, \citenamefont {{Gallagher}},\ and\ \citenamefont {{McCauley}}}]{2024MNRAS.527.4397M}%
  \BibitemOpen
  \bibfield  {author} {\bibinfo {author} {\bibfnamefont {D.~J.}\ \bibnamefont {{McKenna}}}, \bibinfo {author} {\bibfnamefont {E.~F.}\ \bibnamefont {{Keane}}}, \bibinfo {author} {\bibfnamefont {P.~T.}\ \bibnamefont {{Gallagher}}}, \ and\ \bibinfo {author} {\bibfnamefont {J.}~\bibnamefont {{McCauley}}},\ }\href {\doibase 10.1093/mnras/stad2900} {\bibfield  {journal} {\bibinfo  {journal} {\mnras}\ }\textbf {\bibinfo {volume} {527}},\ \bibinfo {pages} {4397} (\bibinfo {year} {2024}{\natexlab{a}})}\BibitemShut {NoStop}%
\bibitem [{\citenamefont {{Tyul'bashev}}\ \emph {et~al.}(2016)\citenamefont {{Tyul'bashev}}, \citenamefont {{Tyul'bashev}}, \citenamefont {{Oreshko}},\ and\ \citenamefont {{Logvinenko}}}]{2016ARep...60..220T}%
  \BibitemOpen
  \bibfield  {author} {\bibinfo {author} {\bibfnamefont {S.~A.}\ \bibnamefont {{Tyul'bashev}}}, \bibinfo {author} {\bibfnamefont {V.~S.}\ \bibnamefont {{Tyul'bashev}}}, \bibinfo {author} {\bibfnamefont {V.~V.}\ \bibnamefont {{Oreshko}}}, \ and\ \bibinfo {author} {\bibfnamefont {S.~V.}\ \bibnamefont {{Logvinenko}}},\ }\href {\doibase 10.1134/S1063772916020128} {\bibfield  {journal} {\bibinfo  {journal} {Astronomy Reports}\ }\textbf {\bibinfo {volume} {60}},\ \bibinfo {pages} {220} (\bibinfo {year} {2016})}\BibitemShut {NoStop}%
\bibitem [{\citenamefont {Tyul’bashev}\ \emph {et~al.}(2024)\citenamefont {Tyul’bashev}, \citenamefont {Tyul’basheva}, \citenamefont {Kitaeva}, \citenamefont {Ovchinnikov}, \citenamefont {Oreshko},\ and\ \citenamefont {Logvinenko}}]{10.1093/mnras/stae070}%
  \BibitemOpen
  \bibfield  {author} {\bibinfo {author} {\bibfnamefont {S.~A.}\ \bibnamefont {Tyul’bashev}}, \bibinfo {author} {\bibfnamefont {G.~E.}\ \bibnamefont {Tyul’basheva}}, \bibinfo {author} {\bibfnamefont {M.~A.}\ \bibnamefont {Kitaeva}}, \bibinfo {author} {\bibfnamefont {I.~L.}\ \bibnamefont {Ovchinnikov}}, \bibinfo {author} {\bibfnamefont {V.~V.}\ \bibnamefont {Oreshko}}, \ and\ \bibinfo {author} {\bibfnamefont {S.~V.}\ \bibnamefont {Logvinenko}},\ }\href {\doibase 10.1093/mnras/stae070} {\bibfield  {journal} {\bibinfo  {journal} {Monthly Notices of the Royal Astronomical Society}\ }\textbf {\bibinfo {volume} {528}},\ \bibinfo {pages} {2220} (\bibinfo {year} {2024})}\BibitemShut {NoStop}%
\bibitem [{\citenamefont {{Stovall}}\ \emph {et~al.}(2014)\citenamefont {{Stovall}}, \citenamefont {{Lynch}}, \citenamefont {{Ransom}}, \citenamefont {{Archibald}}, \citenamefont {{Banaszak}}, \citenamefont {{Biwer}}, \citenamefont {{Boyles}}, \citenamefont {{Dartez}}, \citenamefont {{Day}}, \citenamefont {{Ford}}, \citenamefont {{Flanigan}}, \citenamefont {{Garcia}}, \citenamefont {{Hessels}}, \citenamefont {{Hinojosa}}, \citenamefont {{Jenet}}, \citenamefont {{Kaplan}}, \citenamefont {{Karako-Argaman}}, \citenamefont {{Kaspi}}, \citenamefont {{Kondratiev}}, \citenamefont {{Leake}}, \citenamefont {{Lorimer}}, \citenamefont {{Lunsford}}, \citenamefont {{Martinez}}, \citenamefont {{Mata}}, \citenamefont {{McLaughlin}}, \citenamefont {{Roberts}}, \citenamefont {{Rohr}}, \citenamefont {{Siemens}}, \citenamefont {{Stairs}}, \citenamefont {{van Leeuwen}}, \citenamefont {{Walker}},\ and\ \citenamefont {{Wells}}}]{2014ApJ...791...67S}%
  \BibitemOpen
  \bibfield  {author} {\bibinfo {author} {\bibfnamefont {K.}~\bibnamefont {{Stovall}}}, \bibinfo {author} {\bibfnamefont {R.~S.}\ \bibnamefont {{Lynch}}}, \bibinfo {author} {\bibfnamefont {S.~M.}\ \bibnamefont {{Ransom}}}, \bibinfo {author} {\bibfnamefont {A.~M.}\ \bibnamefont {{Archibald}}}, \bibinfo {author} {\bibfnamefont {S.}~\bibnamefont {{Banaszak}}}, \bibinfo {author} {\bibfnamefont {C.~M.}\ \bibnamefont {{Biwer}}}, \bibinfo {author} {\bibfnamefont {J.}~\bibnamefont {{Boyles}}}, \bibinfo {author} {\bibfnamefont {L.~P.}\ \bibnamefont {{Dartez}}}, \bibinfo {author} {\bibfnamefont {D.}~\bibnamefont {{Day}}}, \bibinfo {author} {\bibfnamefont {A.~J.}\ \bibnamefont {{Ford}}}, \bibinfo {author} {\bibfnamefont {J.}~\bibnamefont {{Flanigan}}}, \bibinfo {author} {\bibfnamefont {A.}~\bibnamefont {{Garcia}}}, \bibinfo {author} {\bibfnamefont {J.~W.~T.}\ \bibnamefont {{Hessels}}}, \bibinfo {author} {\bibfnamefont {J.}~\bibnamefont {{Hinojosa}}}, \bibinfo {author} {\bibfnamefont {F.~A.}\ \bibnamefont {{Jenet}}},
  \bibinfo {author} {\bibfnamefont {D.~L.}\ \bibnamefont {{Kaplan}}}, \bibinfo {author} {\bibfnamefont {C.}~\bibnamefont {{Karako-Argaman}}}, \bibinfo {author} {\bibfnamefont {V.~M.}\ \bibnamefont {{Kaspi}}}, \bibinfo {author} {\bibfnamefont {V.~I.}\ \bibnamefont {{Kondratiev}}}, \bibinfo {author} {\bibfnamefont {S.}~\bibnamefont {{Leake}}}, \bibinfo {author} {\bibfnamefont {D.~R.}\ \bibnamefont {{Lorimer}}}, \bibinfo {author} {\bibfnamefont {G.}~\bibnamefont {{Lunsford}}}, \bibinfo {author} {\bibfnamefont {J.~G.}\ \bibnamefont {{Martinez}}}, \bibinfo {author} {\bibfnamefont {A.}~\bibnamefont {{Mata}}}, \bibinfo {author} {\bibfnamefont {M.~A.}\ \bibnamefont {{McLaughlin}}}, \bibinfo {author} {\bibfnamefont {M.~S.~E.}\ \bibnamefont {{Roberts}}}, \bibinfo {author} {\bibfnamefont {M.~D.}\ \bibnamefont {{Rohr}}}, \bibinfo {author} {\bibfnamefont {X.}~\bibnamefont {{Siemens}}}, \bibinfo {author} {\bibfnamefont {I.~H.}\ \bibnamefont {{Stairs}}}, \bibinfo {author} {\bibfnamefont {J.}~\bibnamefont {{van Leeuwen}}},
  \bibinfo {author} {\bibfnamefont {A.~N.}\ \bibnamefont {{Walker}}}, \ and\ \bibinfo {author} {\bibfnamefont {B.~L.}\ \bibnamefont {{Wells}}},\ }\href {\doibase 10.1088/0004-637X/791/1/67} {\bibfield  {journal} {\bibinfo  {journal} {\apj}\ }\textbf {\bibinfo {volume} {791}},\ \bibinfo {eid} {67} (\bibinfo {year} {2014})}\BibitemShut {NoStop}%
\bibitem [{\citenamefont {{Cordes}}\ \emph {et~al.}(2006)\citenamefont {{Cordes}}, \citenamefont {{Freire}}, \citenamefont {{Lorimer}}, \citenamefont {{Camilo}}, \citenamefont {{Champion}}, \citenamefont {{Nice}}, \citenamefont {{Ramachandran}}, \citenamefont {{Hessels}}, \citenamefont {{Vlemmings}}, \citenamefont {{van Leeuwen}}, \citenamefont {{Ransom}}, \citenamefont {{Bhat}}, \citenamefont {{Arzoumanian}}, \citenamefont {{McLaughlin}}, \citenamefont {{Kaspi}}, \citenamefont {{Kasian}}, \citenamefont {{Deneva}}, \citenamefont {{Reid}}, \citenamefont {{Chatterjee}}, \citenamefont {{Han}}, \citenamefont {{Backer}}, \citenamefont {{Stairs}}, \citenamefont {{Deshpande}},\ and\ \citenamefont {{Faucher-Gigu{\`e}re}}}]{2006ApJ...637..446C}%
  \BibitemOpen
  \bibfield  {author} {\bibinfo {author} {\bibfnamefont {J.~M.}\ \bibnamefont {{Cordes}}}, \bibinfo {author} {\bibfnamefont {P.~C.~C.}\ \bibnamefont {{Freire}}}, \bibinfo {author} {\bibfnamefont {D.~R.}\ \bibnamefont {{Lorimer}}}, \bibinfo {author} {\bibfnamefont {F.}~\bibnamefont {{Camilo}}}, \bibinfo {author} {\bibfnamefont {D.~J.}\ \bibnamefont {{Champion}}}, \bibinfo {author} {\bibfnamefont {D.~J.}\ \bibnamefont {{Nice}}}, \bibinfo {author} {\bibfnamefont {R.}~\bibnamefont {{Ramachandran}}}, \bibinfo {author} {\bibfnamefont {J.~W.~T.}\ \bibnamefont {{Hessels}}}, \bibinfo {author} {\bibfnamefont {W.}~\bibnamefont {{Vlemmings}}}, \bibinfo {author} {\bibfnamefont {J.}~\bibnamefont {{van Leeuwen}}}, \bibinfo {author} {\bibfnamefont {S.~M.}\ \bibnamefont {{Ransom}}}, \bibinfo {author} {\bibfnamefont {N.~D.~R.}\ \bibnamefont {{Bhat}}}, \bibinfo {author} {\bibfnamefont {Z.}~\bibnamefont {{Arzoumanian}}}, \bibinfo {author} {\bibfnamefont {M.~A.}\ \bibnamefont {{McLaughlin}}}, \bibinfo {author} {\bibfnamefont
  {V.~M.}\ \bibnamefont {{Kaspi}}}, \bibinfo {author} {\bibfnamefont {L.}~\bibnamefont {{Kasian}}}, \bibinfo {author} {\bibfnamefont {J.~S.}\ \bibnamefont {{Deneva}}}, \bibinfo {author} {\bibfnamefont {B.}~\bibnamefont {{Reid}}}, \bibinfo {author} {\bibfnamefont {S.}~\bibnamefont {{Chatterjee}}}, \bibinfo {author} {\bibfnamefont {J.~L.}\ \bibnamefont {{Han}}}, \bibinfo {author} {\bibfnamefont {D.~C.}\ \bibnamefont {{Backer}}}, \bibinfo {author} {\bibfnamefont {I.~H.}\ \bibnamefont {{Stairs}}}, \bibinfo {author} {\bibfnamefont {A.~A.}\ \bibnamefont {{Deshpande}}}, \ and\ \bibinfo {author} {\bibfnamefont {C.~A.}\ \bibnamefont {{Faucher-Gigu{\`e}re}}},\ }\href {\doibase 10.1086/498335} {\bibfield  {journal} {\bibinfo  {journal} {\apj}\ }\textbf {\bibinfo {volume} {637}},\ \bibinfo {pages} {446} (\bibinfo {year} {2006})}\BibitemShut {NoStop}%
\bibitem [{\citenamefont {{Deneva}}\ \emph {et~al.}(2013)\citenamefont {{Deneva}}, \citenamefont {{Stovall}}, \citenamefont {{McLaughlin}}, \citenamefont {{Bates}}, \citenamefont {{Freire}}, \citenamefont {{Martinez}}, \citenamefont {{Jenet}},\ and\ \citenamefont {{Bagchi}}}]{2013ApJ...775...51D}%
  \BibitemOpen
  \bibfield  {author} {\bibinfo {author} {\bibfnamefont {J.~S.}\ \bibnamefont {{Deneva}}}, \bibinfo {author} {\bibfnamefont {K.}~\bibnamefont {{Stovall}}}, \bibinfo {author} {\bibfnamefont {M.~A.}\ \bibnamefont {{McLaughlin}}}, \bibinfo {author} {\bibfnamefont {S.~D.}\ \bibnamefont {{Bates}}}, \bibinfo {author} {\bibfnamefont {P.~C.~C.}\ \bibnamefont {{Freire}}}, \bibinfo {author} {\bibfnamefont {J.~G.}\ \bibnamefont {{Martinez}}}, \bibinfo {author} {\bibfnamefont {F.}~\bibnamefont {{Jenet}}}, \ and\ \bibinfo {author} {\bibfnamefont {M.}~\bibnamefont {{Bagchi}}},\ }\href {\doibase 10.1088/0004-637X/775/1/51} {\bibfield  {journal} {\bibinfo  {journal} {\apj}\ }\textbf {\bibinfo {volume} {775}},\ \bibinfo {eid} {51} (\bibinfo {year} {2013})}\BibitemShut {NoStop}%
\bibitem [{\citenamefont {{Sobey}}\ \emph {et~al.}(2022)\citenamefont {{Sobey}}, \citenamefont {{Bassa}}, \citenamefont {{O'Sullivan}}, \citenamefont {{Callingham}}, \citenamefont {{Tan}}, \citenamefont {{Hessels}}, \citenamefont {{Kondratiev}}, \citenamefont {{Stappers}}, \citenamefont {{Tiburzi}}, \citenamefont {{Heald}}, \citenamefont {{Shimwell}}, \citenamefont {{Breton}}, \citenamefont {{Kirwan}}, \citenamefont {{Vedantham}}, \citenamefont {{Carretti}}, \citenamefont {{Grie{\ss}meier}}, \citenamefont {{Haverkorn}},\ and\ \citenamefont {{Karastergiou}}}]{2022A&A...661A..87S}%
  \BibitemOpen
  \bibfield  {author} {\bibinfo {author} {\bibfnamefont {C.}~\bibnamefont {{Sobey}}}, \bibinfo {author} {\bibfnamefont {C.~G.}\ \bibnamefont {{Bassa}}}, \bibinfo {author} {\bibfnamefont {S.~P.}\ \bibnamefont {{O'Sullivan}}}, \bibinfo {author} {\bibfnamefont {J.~R.}\ \bibnamefont {{Callingham}}}, \bibinfo {author} {\bibfnamefont {C.~M.}\ \bibnamefont {{Tan}}}, \bibinfo {author} {\bibfnamefont {J.~W.~T.}\ \bibnamefont {{Hessels}}}, \bibinfo {author} {\bibfnamefont {V.~I.}\ \bibnamefont {{Kondratiev}}}, \bibinfo {author} {\bibfnamefont {B.~W.}\ \bibnamefont {{Stappers}}}, \bibinfo {author} {\bibfnamefont {C.}~\bibnamefont {{Tiburzi}}}, \bibinfo {author} {\bibfnamefont {G.}~\bibnamefont {{Heald}}}, \bibinfo {author} {\bibfnamefont {T.}~\bibnamefont {{Shimwell}}}, \bibinfo {author} {\bibfnamefont {R.~P.}\ \bibnamefont {{Breton}}}, \bibinfo {author} {\bibfnamefont {M.}~\bibnamefont {{Kirwan}}}, \bibinfo {author} {\bibfnamefont {H.~K.}\ \bibnamefont {{Vedantham}}}, \bibinfo {author} {\bibfnamefont {E.}~\bibnamefont
  {{Carretti}}}, \bibinfo {author} {\bibfnamefont {J.~M.}\ \bibnamefont {{Grie{\ss}meier}}}, \bibinfo {author} {\bibfnamefont {M.}~\bibnamefont {{Haverkorn}}}, \ and\ \bibinfo {author} {\bibfnamefont {A.}~\bibnamefont {{Karastergiou}}},\ }\href {\doibase 10.1051/0004-6361/202142636} {\bibfield  {journal} {\bibinfo  {journal} {\aap}\ }\textbf {\bibinfo {volume} {661}},\ \bibinfo {eid} {A87} (\bibinfo {year} {2022})}\BibitemShut {NoStop}%
\bibitem [{\citenamefont {Ransom}(2011)}]{ransomPRESTOPulsaRExploration2011}%
  \BibitemOpen
  \bibfield  {author} {\bibinfo {author} {\bibfnamefont {S.}~\bibnamefont {Ransom}},\ }\href@noop {} {\bibfield  {journal} {\bibinfo  {journal} {Astrophysics Source Code Library}\ ,\ \bibinfo {pages} {ascl:1107.017}} (\bibinfo {year} {2011})}\BibitemShut {NoStop}%
\bibitem [{\citenamefont {Bassa}\ \emph {et~al.}(2017)\citenamefont {Bassa}, \citenamefont {Pleunis},\ and\ \citenamefont {Hessels}}]{bassaEnablingPulsarFast2017}%
  \BibitemOpen
  \bibfield  {author} {\bibinfo {author} {\bibfnamefont {C.~G.}\ \bibnamefont {Bassa}}, \bibinfo {author} {\bibfnamefont {Z.}~\bibnamefont {Pleunis}}, \ and\ \bibinfo {author} {\bibfnamefont {J.~W.~T.}\ \bibnamefont {Hessels}},\ }\href {\doibase 10.1016/j.ascom.2017.01.004} {\bibfield  {journal} {\bibinfo  {journal} {Astronomy and Computing}\ }\textbf {\bibinfo {volume} {18}},\ \bibinfo {pages} {40} (\bibinfo {year} {2017})}\BibitemShut {NoStop}%
\bibitem [{\citenamefont {{McKenna}}\ \emph {et~al.}(2024{\natexlab{b}})\citenamefont {{McKenna}}, \citenamefont {{Keane}}, \citenamefont {{Gallagher}},\ and\ \citenamefont {{McCauley}}}]{2024JOSS....9.5517M}%
  \BibitemOpen
  \bibfield  {author} {\bibinfo {author} {\bibfnamefont {D.}~\bibnamefont {{McKenna}}}, \bibinfo {author} {\bibfnamefont {E.}~\bibnamefont {{Keane}}}, \bibinfo {author} {\bibfnamefont {P.}~\bibnamefont {{Gallagher}}}, \ and\ \bibinfo {author} {\bibfnamefont {J.}~\bibnamefont {{McCauley}}},\ }\href {\doibase 10.21105/joss.05517} {\bibfield  {journal} {\bibinfo  {journal} {The Journal of Open Source Software}\ }\textbf {\bibinfo {volume} {9}},\ \bibinfo {eid} {5517} (\bibinfo {year} {2024}{\natexlab{b}})}\BibitemShut {NoStop}%
\bibitem [{\citenamefont {{Murphy}}\ \emph {et~al.}(2021)\citenamefont {{Murphy}}, \citenamefont {{Callanan, P.}}, \citenamefont {{McCauley, J.}}, \citenamefont {{McKenna, D. J.}}, \citenamefont {{Fionnag\'ain, D. \'O}}, \citenamefont {{Louis, C. K.}}, \citenamefont {{Redman, M. P.}}, \citenamefont {{Ca\~nizares, L. A.}}, \citenamefont {{Carley, E. P.}}, \citenamefont {{Maloney, S. A.}}, \citenamefont {{Coghlan, B.}}, \citenamefont {{Daly, M.}}, \citenamefont {{Scully, J.}}, \citenamefont {{Dooley, J.}}, \citenamefont {{Gajjar, V.}}, \citenamefont {{Giese, C.}}, \citenamefont {{Brennan, A.}}, \citenamefont {{Keane, E. F.}}, \citenamefont {{Maguire, C. A.}}, \citenamefont {{Quinn, J.}}, \citenamefont {{Mooney, S.}}, \citenamefont {{Ryan, A. M.}}, \citenamefont {{Walsh, J.}}, \citenamefont {{Jackman, C. M.}}, \citenamefont {{Golden, A.}}, \citenamefont {{Ray, T. P.}}, \citenamefont {{Doyle, J. G.}}, \citenamefont {{Rigney, J.}}, \citenamefont {{Burton, M.}},\ and\ \citenamefont {{Gallagher, P.
  T.}}}]{murphyFirstResultsREALtime2021}%
  \BibitemOpen
  \bibfield  {author} {\bibinfo {author} {\bibfnamefont {P.~C.}\ \bibnamefont {{Murphy}}}, \bibinfo {author} {\bibnamefont {{Callanan, P.}}}, \bibinfo {author} {\bibnamefont {{McCauley, J.}}}, \bibinfo {author} {\bibnamefont {{McKenna, D. J.}}}, \bibinfo {author} {\bibnamefont {{Fionnag\'ain, D. \'O}}}, \bibinfo {author} {\bibnamefont {{Louis, C. K.}}}, \bibinfo {author} {\bibnamefont {{Redman, M. P.}}}, \bibinfo {author} {\bibnamefont {{Ca\~nizares, L. A.}}}, \bibinfo {author} {\bibnamefont {{Carley, E. P.}}}, \bibinfo {author} {\bibnamefont {{Maloney, S. A.}}}, \bibinfo {author} {\bibnamefont {{Coghlan, B.}}}, \bibinfo {author} {\bibnamefont {{Daly, M.}}}, \bibinfo {author} {\bibnamefont {{Scully, J.}}}, \bibinfo {author} {\bibnamefont {{Dooley, J.}}}, \bibinfo {author} {\bibnamefont {{Gajjar, V.}}}, \bibinfo {author} {\bibnamefont {{Giese, C.}}}, \bibinfo {author} {\bibnamefont {{Brennan, A.}}}, \bibinfo {author} {\bibnamefont {{Keane, E. F.}}}, \bibinfo {author} {\bibnamefont {{Maguire, C. A.}}}, \bibinfo
  {author} {\bibnamefont {{Quinn, J.}}}, \bibinfo {author} {\bibnamefont {{Mooney, S.}}}, \bibinfo {author} {\bibnamefont {{Ryan, A. M.}}}, \bibinfo {author} {\bibnamefont {{Walsh, J.}}}, \bibinfo {author} {\bibnamefont {{Jackman, C. M.}}}, \bibinfo {author} {\bibnamefont {{Golden, A.}}}, \bibinfo {author} {\bibnamefont {{Ray, T. P.}}}, \bibinfo {author} {\bibnamefont {{Doyle, J. G.}}}, \bibinfo {author} {\bibnamefont {{Rigney, J.}}}, \bibinfo {author} {\bibnamefont {{Burton, M.}}}, \ and\ \bibinfo {author} {\bibnamefont {{Gallagher, P. T.}}},\ }\href {\doibase 10.1051/0004-6361/202140415} {\bibfield  {journal} {\bibinfo  {journal} {A\&A}\ }\textbf {\bibinfo {volume} {655}},\ \bibinfo {pages} {A16} (\bibinfo {year} {2021})}\BibitemShut {NoStop}%
\bibitem [{\citenamefont {{van Straten}}\ and\ \citenamefont {Bailes}(2011)}]{vanstratenDSPSRDigitalSignal2011}%
  \BibitemOpen
  \bibfield  {author} {\bibinfo {author} {\bibfnamefont {W.}~\bibnamefont {{van Straten}}}\ and\ \bibinfo {author} {\bibfnamefont {M.}~\bibnamefont {Bailes}},\ }\href {\doibase 10.1071/AS10021} {\bibfield  {journal} {\bibinfo  {journal} {Publications of the Astronomical Society of Australia}\ }\textbf {\bibinfo {volume} {28}},\ \bibinfo {pages} {1} (\bibinfo {year} {2011})}\BibitemShut {NoStop}%
\bibitem [{\citenamefont {Collet}\ and\ \citenamefont {Kucherawy}(2018)}]{Collet2018}%
  \BibitemOpen
  \bibfield  {author} {\bibinfo {author} {\bibfnamefont {Y.}~\bibnamefont {Collet}}\ and\ \bibinfo {author} {\bibfnamefont {M.}~\bibnamefont {Kucherawy}},\ }\href {\doibase 10.17487/RFC8478} {\emph {\bibinfo {title} {Zstandard {Compression} and the application/zstd {Media} {Type}}}},\ \bibinfo {type} {Tech. Rep.}\ \bibinfo {number} {RFC 8478}\ (\bibinfo  {institution} {Internet Engineering Task Force},\ \bibinfo {year} {2018})\BibitemShut {NoStop}%
\bibitem [{\citenamefont {Hotan}\ \emph {et~al.}(2004)\citenamefont {Hotan}, \citenamefont {{van Straten}},\ and\ \citenamefont {Manchester}}]{hotanPSRCHIVEPSRFITSOpen2004}%
  \BibitemOpen
  \bibfield  {author} {\bibinfo {author} {\bibfnamefont {A.~W.}\ \bibnamefont {Hotan}}, \bibinfo {author} {\bibfnamefont {W.}~\bibnamefont {{van Straten}}}, \ and\ \bibinfo {author} {\bibfnamefont {R.~N.}\ \bibnamefont {Manchester}},\ }\href {\doibase 10.1071/AS04022} {\bibfield  {journal} {\bibinfo  {journal} {Publications of the Astronomical Society of Australia}\ }\textbf {\bibinfo {volume} {21}},\ \bibinfo {pages} {302} (\bibinfo {year} {2004})}\BibitemShut {NoStop}%
\bibitem [{\citenamefont {Morello}\ \emph {et~al.}(2019)\citenamefont {Morello}, \citenamefont {Barr}, \citenamefont {Cooper}, \citenamefont {Bailes}, \citenamefont {Bates}, \citenamefont {Bhat}, \citenamefont {Burgay}, \citenamefont {{Burke-Spolaor}}, \citenamefont {Cameron}, \citenamefont {Champion}, \citenamefont {Eatough}, \citenamefont {Flynn}, \citenamefont {Jameson}, \citenamefont {Johnston}, \citenamefont {Keith}, \citenamefont {Keane}, \citenamefont {Kramer}, \citenamefont {Levin}, \citenamefont {Ng}, \citenamefont {Petroff}, \citenamefont {Possenti}, \citenamefont {Stappers}, \citenamefont {{van Straten}},\ and\ \citenamefont {Tiburzi}}]{morelloHighTimeResolution2019}%
  \BibitemOpen
  \bibfield  {author} {\bibinfo {author} {\bibfnamefont {V.}~\bibnamefont {Morello}}, \bibinfo {author} {\bibfnamefont {E.~D.}\ \bibnamefont {Barr}}, \bibinfo {author} {\bibfnamefont {S.}~\bibnamefont {Cooper}}, \bibinfo {author} {\bibfnamefont {M.}~\bibnamefont {Bailes}}, \bibinfo {author} {\bibfnamefont {S.}~\bibnamefont {Bates}}, \bibinfo {author} {\bibfnamefont {N.~D.~R.}\ \bibnamefont {Bhat}}, \bibinfo {author} {\bibfnamefont {M.}~\bibnamefont {Burgay}}, \bibinfo {author} {\bibfnamefont {S.}~\bibnamefont {{Burke-Spolaor}}}, \bibinfo {author} {\bibfnamefont {A.~D.}\ \bibnamefont {Cameron}}, \bibinfo {author} {\bibfnamefont {D.~J.}\ \bibnamefont {Champion}}, \bibinfo {author} {\bibfnamefont {R.~P.}\ \bibnamefont {Eatough}}, \bibinfo {author} {\bibfnamefont {C.~M.~L.}\ \bibnamefont {Flynn}}, \bibinfo {author} {\bibfnamefont {A.}~\bibnamefont {Jameson}}, \bibinfo {author} {\bibfnamefont {S.}~\bibnamefont {Johnston}}, \bibinfo {author} {\bibfnamefont {M.~J.}\ \bibnamefont {Keith}}, \bibinfo {author}
  {\bibfnamefont {E.~F.}\ \bibnamefont {Keane}}, \bibinfo {author} {\bibfnamefont {M.}~\bibnamefont {Kramer}}, \bibinfo {author} {\bibfnamefont {L.}~\bibnamefont {Levin}}, \bibinfo {author} {\bibfnamefont {C.}~\bibnamefont {Ng}}, \bibinfo {author} {\bibfnamefont {E.}~\bibnamefont {Petroff}}, \bibinfo {author} {\bibfnamefont {A.}~\bibnamefont {Possenti}}, \bibinfo {author} {\bibfnamefont {B.~W.}\ \bibnamefont {Stappers}}, \bibinfo {author} {\bibfnamefont {W.}~\bibnamefont {{van Straten}}}, \ and\ \bibinfo {author} {\bibfnamefont {C.}~\bibnamefont {Tiburzi}},\ }\href {\doibase 10.1093/mnras/sty3328} {\bibfield  {journal} {\bibinfo  {journal} {Monthly Notices of the Royal Astronomical Society}\ }\textbf {\bibinfo {volume} {483}},\ \bibinfo {pages} {3673} (\bibinfo {year} {2019})}\BibitemShut {NoStop}%
\bibitem [{\citenamefont {Hobbs}\ \emph {et~al.}(2006)\citenamefont {Hobbs}, \citenamefont {Edwards},\ and\ \citenamefont {Manchester}}]{hobbsTEMPO2NewPulsartiming2006}%
  \BibitemOpen
  \bibfield  {author} {\bibinfo {author} {\bibfnamefont {G.~B.}\ \bibnamefont {Hobbs}}, \bibinfo {author} {\bibfnamefont {R.~T.}\ \bibnamefont {Edwards}}, \ and\ \bibinfo {author} {\bibfnamefont {R.~N.}\ \bibnamefont {Manchester}},\ }\href {\doibase 10.1111/j.1365-2966.2006.10302.x} {\bibfield  {journal} {\bibinfo  {journal} {Monthly Notices of the Royal Astronomical Society}\ }\textbf {\bibinfo {volume} {369}},\ \bibinfo {pages} {655} (\bibinfo {year} {2006})}\BibitemShut {NoStop}%
\bibitem [{\citenamefont {Kondratiev}\ \emph {et~al.}(2016)\citenamefont {Kondratiev}, \citenamefont {Verbiest}, \citenamefont {Hessels}, \citenamefont {Bilous}, \citenamefont {Stappers}, \citenamefont {Kramer}, \citenamefont {Keane}, \citenamefont {Noutsos}, \citenamefont {Os{\l}owski}, \citenamefont {Breton}, \citenamefont {Hassall}, \citenamefont {Alexov}, \citenamefont {Cooper}, \citenamefont {Falcke}, \citenamefont {Grie{\ss}meier}, \citenamefont {Karastergiou}, \citenamefont {Kuniyoshi}, \citenamefont {Pilia}, \citenamefont {Sobey}, \citenamefont {ter Veen}, \citenamefont {van Leeuwen}, \citenamefont {Weltevrede}, \citenamefont {Bell}, \citenamefont {Broderick}, \citenamefont {Corbel}, \citenamefont {Eisl{\"o}ffel}, \citenamefont {Markoff}, \citenamefont {Rowlinson}, \citenamefont {Swinbank}, \citenamefont {Wijers}, \citenamefont {Wijnands},\ and\ \citenamefont {Zarka}}]{kondratievLOFARCensusMillisecond2016a}%
  \BibitemOpen
  \bibfield  {author} {\bibinfo {author} {\bibfnamefont {V.~I.}\ \bibnamefont {Kondratiev}}, \bibinfo {author} {\bibfnamefont {J.~P.~W.}\ \bibnamefont {Verbiest}}, \bibinfo {author} {\bibfnamefont {J.~W.~T.}\ \bibnamefont {Hessels}}, \bibinfo {author} {\bibfnamefont {A.~V.}\ \bibnamefont {Bilous}}, \bibinfo {author} {\bibfnamefont {B.~W.}\ \bibnamefont {Stappers}}, \bibinfo {author} {\bibfnamefont {M.}~\bibnamefont {Kramer}}, \bibinfo {author} {\bibfnamefont {E.~F.}\ \bibnamefont {Keane}}, \bibinfo {author} {\bibfnamefont {A.}~\bibnamefont {Noutsos}}, \bibinfo {author} {\bibfnamefont {S.}~\bibnamefont {Os{\l}owski}}, \bibinfo {author} {\bibfnamefont {R.~P.}\ \bibnamefont {Breton}}, \bibinfo {author} {\bibfnamefont {T.~E.}\ \bibnamefont {Hassall}}, \bibinfo {author} {\bibfnamefont {A.}~\bibnamefont {Alexov}}, \bibinfo {author} {\bibfnamefont {S.}~\bibnamefont {Cooper}}, \bibinfo {author} {\bibfnamefont {H.}~\bibnamefont {Falcke}}, \bibinfo {author} {\bibfnamefont {J.-M.}\ \bibnamefont {Grie{\ss}meier}},
  \bibinfo {author} {\bibfnamefont {A.}~\bibnamefont {Karastergiou}}, \bibinfo {author} {\bibfnamefont {M.}~\bibnamefont {Kuniyoshi}}, \bibinfo {author} {\bibfnamefont {M.}~\bibnamefont {Pilia}}, \bibinfo {author} {\bibfnamefont {C.}~\bibnamefont {Sobey}}, \bibinfo {author} {\bibfnamefont {S.}~\bibnamefont {ter Veen}}, \bibinfo {author} {\bibfnamefont {J.}~\bibnamefont {van Leeuwen}}, \bibinfo {author} {\bibfnamefont {P.}~\bibnamefont {Weltevrede}}, \bibinfo {author} {\bibfnamefont {M.~E.}\ \bibnamefont {Bell}}, \bibinfo {author} {\bibfnamefont {J.~W.}\ \bibnamefont {Broderick}}, \bibinfo {author} {\bibfnamefont {S.}~\bibnamefont {Corbel}}, \bibinfo {author} {\bibfnamefont {J.}~\bibnamefont {Eisl{\"o}ffel}}, \bibinfo {author} {\bibfnamefont {S.}~\bibnamefont {Markoff}}, \bibinfo {author} {\bibfnamefont {A.}~\bibnamefont {Rowlinson}}, \bibinfo {author} {\bibfnamefont {J.~D.}\ \bibnamefont {Swinbank}}, \bibinfo {author} {\bibfnamefont {R.~a. M.~J.}\ \bibnamefont {Wijers}}, \bibinfo {author} {\bibfnamefont
  {R.}~\bibnamefont {Wijnands}}, \ and\ \bibinfo {author} {\bibfnamefont {P.}~\bibnamefont {Zarka}},\ }\href {\doibase 10.1051/0004-6361/201527178} {\bibfield  {journal} {\bibinfo  {journal} {Astronomy \& Astrophysics}\ }\textbf {\bibinfo {volume} {585}},\ \bibinfo {pages} {A128} (\bibinfo {year} {2016})}\BibitemShut {NoStop}%
\bibitem [{\citenamefont {{Dowell}}\ \emph {et~al.}(2017)\citenamefont {{Dowell}}, \citenamefont {{Taylor}}, \citenamefont {{Schinzel}}, \citenamefont {{Kassim}},\ and\ \citenamefont {{Stovall}}}]{2017MNRAS.469.4537D}%
  \BibitemOpen
  \bibfield  {author} {\bibinfo {author} {\bibfnamefont {J.}~\bibnamefont {{Dowell}}}, \bibinfo {author} {\bibfnamefont {G.~B.}\ \bibnamefont {{Taylor}}}, \bibinfo {author} {\bibfnamefont {F.~K.}\ \bibnamefont {{Schinzel}}}, \bibinfo {author} {\bibfnamefont {N.~E.}\ \bibnamefont {{Kassim}}}, \ and\ \bibinfo {author} {\bibfnamefont {K.}~\bibnamefont {{Stovall}}},\ }\href {\doibase 10.1093/mnras/stx1136} {\bibfield  {journal} {\bibinfo  {journal} {\mnras}\ }\textbf {\bibinfo {volume} {469}},\ \bibinfo {pages} {4537} (\bibinfo {year} {2017})}\BibitemShut {NoStop}%
\bibitem [{\citenamefont {Price}(2021)}]{priceGlobalSkyModels2021}%
  \BibitemOpen
  \bibfield  {author} {\bibinfo {author} {\bibfnamefont {D.~C.}\ \bibnamefont {Price}},\ }\href {\doibase 10.3847/2515-5172/ac332c} {\bibfield  {journal} {\bibinfo  {journal} {Research Notes of the AAS}\ }\textbf {\bibinfo {volume} {5}},\ \bibinfo {pages} {246} (\bibinfo {year} {2021})}\BibitemShut {NoStop}%
\bibitem [{\citenamefont {Wijnholds}\ and\ \citenamefont {{van Cappellen}}(2011)}]{wijnholdsSituAntennaPerformance2011}%
  \BibitemOpen
  \bibfield  {author} {\bibinfo {author} {\bibfnamefont {S.~J.}\ \bibnamefont {Wijnholds}}\ and\ \bibinfo {author} {\bibfnamefont {W.~A.}\ \bibnamefont {{van Cappellen}}},\ }\href {\doibase 10.1109/TAP.2011.2122225} {\bibfield  {journal} {\bibinfo  {journal} {IEEE Transactions on Antennas and Propagation}\ }\textbf {\bibinfo {volume} {59}},\ \bibinfo {pages} {1981} (\bibinfo {year} {2011})}\BibitemShut {NoStop}%
\bibitem [{\citenamefont {Carozzi}(2020)}]{carozzi2baOrNot2baDreamBeam2020}%
  \BibitemOpen
  \bibfield  {author} {\bibinfo {author} {\bibfnamefont {T.}~\bibnamefont {Carozzi}},\ }\href {https://github.com/2baOrNot2ba/dreamBeam} {\enquote {\bibinfo {title} {{dreamBeam}},}\ } (\bibinfo {year} {2020})\BibitemShut {NoStop}%
\bibitem [{\citenamefont {Newville}\ \emph {et~al.}(2021)\citenamefont {Newville}, \citenamefont {Otten}, \citenamefont {Nelson}, \citenamefont {Ingargiola}, \citenamefont {Stensitzki}, \citenamefont {Allan}, \citenamefont {Fox}, \citenamefont {Carter}, \citenamefont {Micha{\l}}, \citenamefont {Osborn}, \citenamefont {Pustakhod}, \citenamefont {{lneuhaus}}, \citenamefont {Weigand}, \citenamefont {Glenn}, \citenamefont {Deil}, \citenamefont {Mark}, \citenamefont {Hansen}, \citenamefont {Pasquevich}, \citenamefont {Foks}, \citenamefont {Zobrist}, \citenamefont {Frost}, \citenamefont {Beelen}, \citenamefont {Stuermer}, \citenamefont {{azelcer}}, \citenamefont {Hannum}, \citenamefont {Polloreno}, \citenamefont {Nielsen}, \citenamefont {Caldwell}, \citenamefont {Almarza},\ and\ \citenamefont {Persaud}}]{newvilleLmfitLmfitpy2021}%
  \BibitemOpen
  \bibfield  {author} {\bibinfo {author} {\bibfnamefont {M.}~\bibnamefont {Newville}}, \bibinfo {author} {\bibfnamefont {R.}~\bibnamefont {Otten}}, \bibinfo {author} {\bibfnamefont {A.}~\bibnamefont {Nelson}}, \bibinfo {author} {\bibfnamefont {A.}~\bibnamefont {Ingargiola}}, \bibinfo {author} {\bibfnamefont {T.}~\bibnamefont {Stensitzki}}, \bibinfo {author} {\bibfnamefont {D.}~\bibnamefont {Allan}}, \bibinfo {author} {\bibfnamefont {A.}~\bibnamefont {Fox}}, \bibinfo {author} {\bibfnamefont {F.}~\bibnamefont {Carter}}, \bibinfo {author} {\bibnamefont {Micha{\l}}}, \bibinfo {author} {\bibfnamefont {R.}~\bibnamefont {Osborn}}, \bibinfo {author} {\bibfnamefont {D.}~\bibnamefont {Pustakhod}}, \bibinfo {author} {\bibnamefont {{lneuhaus}}}, \bibinfo {author} {\bibfnamefont {S.}~\bibnamefont {Weigand}}, \bibinfo {author} {\bibnamefont {Glenn}}, \bibinfo {author} {\bibfnamefont {C.}~\bibnamefont {Deil}}, \bibinfo {author} {\bibnamefont {Mark}}, \bibinfo {author} {\bibfnamefont {A.~L.~R.}\ \bibnamefont {Hansen}},
  \bibinfo {author} {\bibfnamefont {G.}~\bibnamefont {Pasquevich}}, \bibinfo {author} {\bibfnamefont {L.}~\bibnamefont {Foks}}, \bibinfo {author} {\bibfnamefont {N.}~\bibnamefont {Zobrist}}, \bibinfo {author} {\bibfnamefont {O.}~\bibnamefont {Frost}}, \bibinfo {author} {\bibfnamefont {A.}~\bibnamefont {Beelen}}, \bibinfo {author} {\bibnamefont {Stuermer}}, \bibinfo {author} {\bibnamefont {{azelcer}}}, \bibinfo {author} {\bibfnamefont {A.}~\bibnamefont {Hannum}}, \bibinfo {author} {\bibfnamefont {A.}~\bibnamefont {Polloreno}}, \bibinfo {author} {\bibfnamefont {J.~H.}\ \bibnamefont {Nielsen}}, \bibinfo {author} {\bibfnamefont {S.}~\bibnamefont {Caldwell}}, \bibinfo {author} {\bibfnamefont {A.}~\bibnamefont {Almarza}}, \ and\ \bibinfo {author} {\bibfnamefont {A.}~\bibnamefont {Persaud}},\ }\href {\doibase 10.5281/zenodo.5570790} {\enquote {\bibinfo {title} {Lmfit/lmfit-py: 1.0.3},}\ }\bibinfo {howpublished} {Zenodo} (\bibinfo {year} {2021})\BibitemShut {NoStop}%
\bibitem [{\citenamefont {{Bilous}}\ \emph {et~al.}(2016)\citenamefont {{Bilous}}, \citenamefont {{Kondratiev}}, \citenamefont {{Kramer}}, \citenamefont {{Keane}}, \citenamefont {{Hessels}}, \citenamefont {{Stappers}}, \citenamefont {{Malofeev}}, \citenamefont {{Sobey}}, \citenamefont {{Breton}}, \citenamefont {{Cooper}}, \citenamefont {{Falcke}}, \citenamefont {{Karastergiou}}, \citenamefont {{Michilli}}, \citenamefont {{Os{\l}owski}}, \citenamefont {{Sanidas}}, \citenamefont {{ter Veen}}, \citenamefont {{van Leeuwen}}, \citenamefont {{Verbiest}}, \citenamefont {{Weltevrede}}, \citenamefont {{Zarka}}, \citenamefont {{Grie{\ss}meier}}, \citenamefont {{Serylak}}, \citenamefont {{Bell}}, \citenamefont {{Broderick}}, \citenamefont {{Eisl{\"o}ffel}}, \citenamefont {{Markoff}},\ and\ \citenamefont {{Rowlinson}}}]{2016A&A...591A.134B}%
  \BibitemOpen
  \bibfield  {author} {\bibinfo {author} {\bibfnamefont {A.~V.}\ \bibnamefont {{Bilous}}}, \bibinfo {author} {\bibfnamefont {V.~I.}\ \bibnamefont {{Kondratiev}}}, \bibinfo {author} {\bibfnamefont {M.}~\bibnamefont {{Kramer}}}, \bibinfo {author} {\bibfnamefont {E.~F.}\ \bibnamefont {{Keane}}}, \bibinfo {author} {\bibfnamefont {J.~W.~T.}\ \bibnamefont {{Hessels}}}, \bibinfo {author} {\bibfnamefont {B.~W.}\ \bibnamefont {{Stappers}}}, \bibinfo {author} {\bibfnamefont {V.~M.}\ \bibnamefont {{Malofeev}}}, \bibinfo {author} {\bibfnamefont {C.}~\bibnamefont {{Sobey}}}, \bibinfo {author} {\bibfnamefont {R.~P.}\ \bibnamefont {{Breton}}}, \bibinfo {author} {\bibfnamefont {S.}~\bibnamefont {{Cooper}}}, \bibinfo {author} {\bibfnamefont {H.}~\bibnamefont {{Falcke}}}, \bibinfo {author} {\bibfnamefont {A.}~\bibnamefont {{Karastergiou}}}, \bibinfo {author} {\bibfnamefont {D.}~\bibnamefont {{Michilli}}}, \bibinfo {author} {\bibfnamefont {S.}~\bibnamefont {{Os{\l}owski}}}, \bibinfo {author} {\bibfnamefont {S.}~\bibnamefont
  {{Sanidas}}}, \bibinfo {author} {\bibfnamefont {S.}~\bibnamefont {{ter Veen}}}, \bibinfo {author} {\bibfnamefont {J.}~\bibnamefont {{van Leeuwen}}}, \bibinfo {author} {\bibfnamefont {J.~P.~W.}\ \bibnamefont {{Verbiest}}}, \bibinfo {author} {\bibfnamefont {P.}~\bibnamefont {{Weltevrede}}}, \bibinfo {author} {\bibfnamefont {P.}~\bibnamefont {{Zarka}}}, \bibinfo {author} {\bibfnamefont {J.~M.}\ \bibnamefont {{Grie{\ss}meier}}}, \bibinfo {author} {\bibfnamefont {M.}~\bibnamefont {{Serylak}}}, \bibinfo {author} {\bibfnamefont {M.~E.}\ \bibnamefont {{Bell}}}, \bibinfo {author} {\bibfnamefont {J.~W.}\ \bibnamefont {{Broderick}}}, \bibinfo {author} {\bibfnamefont {J.}~\bibnamefont {{Eisl{\"o}ffel}}}, \bibinfo {author} {\bibfnamefont {S.}~\bibnamefont {{Markoff}}}, \ and\ \bibinfo {author} {\bibfnamefont {A.}~\bibnamefont {{Rowlinson}}},\ }\href {\doibase 10.1051/0004-6361/201527702} {\bibfield  {journal} {\bibinfo  {journal} {\aap}\ }\textbf {\bibinfo {volume} {591}},\ \bibinfo {eid} {A134} (\bibinfo {year}
  {2016})}\BibitemShut {NoStop}%
\bibitem [{\citenamefont {Tyul'bashev}\ \emph {et~al.}(2018)\citenamefont {Tyul'bashev}, \citenamefont {Tyul'bashev},\ and\ \citenamefont {Malofeev}}]{tyulbashevDetection25New2018}%
  \BibitemOpen
  \bibfield  {author} {\bibinfo {author} {\bibfnamefont {S.~A.}\ \bibnamefont {Tyul'bashev}}, \bibinfo {author} {\bibfnamefont {V.~S.}\ \bibnamefont {Tyul'bashev}}, \ and\ \bibinfo {author} {\bibfnamefont {V.~M.}\ \bibnamefont {Malofeev}},\ }\href {\doibase 10.1051/0004-6361/201833102} {\bibfield  {journal} {\bibinfo  {journal} {Astronomy and Astrophysics}\ }\textbf {\bibinfo {volume} {618}},\ \bibinfo {pages} {A70} (\bibinfo {year} {2018})}\BibitemShut {NoStop}%
\bibitem [{\citenamefont {Seymour}\ \emph {et~al.}(2019)\citenamefont {Seymour}, \citenamefont {Michilli},\ and\ \citenamefont {Pleunis}}]{Seymour2019}%
  \BibitemOpen
  \bibfield  {author} {\bibinfo {author} {\bibfnamefont {A.}~\bibnamefont {Seymour}}, \bibinfo {author} {\bibfnamefont {D.}~\bibnamefont {Michilli}}, \ and\ \bibinfo {author} {\bibfnamefont {Z.}~\bibnamefont {Pleunis}},\ }\href {https://ui.adsabs.harvard.edu/abs/2019ascl.soft10004S} {\bibfield  {journal} {\bibinfo  {journal} {Astrophysics Source Code Library}\ ,\ \bibinfo {pages} {ascl:1910.004}} (\bibinfo {year} {2019})},\ \bibinfo {note} {aDS Bibcode: 2019ascl.soft10004S}\BibitemShut {NoStop}%
\bibitem [{\citenamefont {{CHIME/FRB Collaboration}}\ \emph {et~al.}(2019)\citenamefont {{CHIME/FRB Collaboration}}, \citenamefont {{Andersen}}, \citenamefont {{Bandura}}, \citenamefont {{Bhardwaj}}, \citenamefont {{Boubel}}, \citenamefont {{Boyce}}, \citenamefont {{Boyle}}, \citenamefont {{Brar}}, \citenamefont {{Cassanelli}}, \citenamefont {{Chawla}}, \citenamefont {{Cubranic}}, \citenamefont {{Deng}}, \citenamefont {{Dobbs}}, \citenamefont {{Fandino}}, \citenamefont {{Fonseca}}, \citenamefont {{Gaensler}}, \citenamefont {{Gilbert}}, \citenamefont {{Giri}}, \citenamefont {{Good}}, \citenamefont {{Halpern}}, \citenamefont {{Hill}}, \citenamefont {{Hinshaw}}, \citenamefont {{H{\"o}fer}}, \citenamefont {{Josephy}}, \citenamefont {{Kaspi}}, \citenamefont {{Kothes}}, \citenamefont {{Landecker}}, \citenamefont {{Lang}}, \citenamefont {{Li}}, \citenamefont {{Lin}}, \citenamefont {{Masui}}, \citenamefont {{Mena-Parra}}, \citenamefont {{Merryfield}}, \citenamefont {{Mckinven}}, \citenamefont {{Michilli}},
  \citenamefont {{Milutinovic}}, \citenamefont {{Naidu}}, \citenamefont {{Newburgh}}, \citenamefont {{Ng}}, \citenamefont {{Patel}}, \citenamefont {{Pen}}, \citenamefont {{Pinsonneault-Marotte}}, \citenamefont {{Pleunis}}, \citenamefont {{Rafiei-Ravandi}}, \citenamefont {{Rahman}}, \citenamefont {{Ransom}}, \citenamefont {{Renard}}, \citenamefont {{Scholz}}, \citenamefont {{Siegel}}, \citenamefont {{Singh}}, \citenamefont {{Smith}}, \citenamefont {{Stairs}}, \citenamefont {{Tendulkar}}, \citenamefont {{Tretyakov}}, \citenamefont {{Vanderlinde}}, \citenamefont {{Yadav}},\ and\ \citenamefont {{Zwaniga}}}]{2019ApJ...885L..24C}%
  \BibitemOpen
  \bibfield  {author} {\bibinfo {author} {\bibnamefont {{CHIME/FRB Collaboration}}}, \bibinfo {author} {\bibfnamefont {B.~C.}\ \bibnamefont {{Andersen}}}, \bibinfo {author} {\bibfnamefont {K.}~\bibnamefont {{Bandura}}}, \bibinfo {author} {\bibfnamefont {M.}~\bibnamefont {{Bhardwaj}}}, \bibinfo {author} {\bibfnamefont {P.}~\bibnamefont {{Boubel}}}, \bibinfo {author} {\bibfnamefont {M.~M.}\ \bibnamefont {{Boyce}}}, \bibinfo {author} {\bibfnamefont {P.~J.}\ \bibnamefont {{Boyle}}}, \bibinfo {author} {\bibfnamefont {C.}~\bibnamefont {{Brar}}}, \bibinfo {author} {\bibfnamefont {T.}~\bibnamefont {{Cassanelli}}}, \bibinfo {author} {\bibfnamefont {P.}~\bibnamefont {{Chawla}}}, \bibinfo {author} {\bibfnamefont {D.}~\bibnamefont {{Cubranic}}}, \bibinfo {author} {\bibfnamefont {M.}~\bibnamefont {{Deng}}}, \bibinfo {author} {\bibfnamefont {M.}~\bibnamefont {{Dobbs}}}, \bibinfo {author} {\bibfnamefont {M.}~\bibnamefont {{Fandino}}}, \bibinfo {author} {\bibfnamefont {E.}~\bibnamefont {{Fonseca}}}, \bibinfo {author}
  {\bibfnamefont {B.~M.}\ \bibnamefont {{Gaensler}}}, \bibinfo {author} {\bibfnamefont {A.~J.}\ \bibnamefont {{Gilbert}}}, \bibinfo {author} {\bibfnamefont {U.}~\bibnamefont {{Giri}}}, \bibinfo {author} {\bibfnamefont {D.~C.}\ \bibnamefont {{Good}}}, \bibinfo {author} {\bibfnamefont {M.}~\bibnamefont {{Halpern}}}, \bibinfo {author} {\bibfnamefont {A.~S.}\ \bibnamefont {{Hill}}}, \bibinfo {author} {\bibfnamefont {G.}~\bibnamefont {{Hinshaw}}}, \bibinfo {author} {\bibfnamefont {C.}~\bibnamefont {{H{\"o}fer}}}, \bibinfo {author} {\bibfnamefont {A.}~\bibnamefont {{Josephy}}}, \bibinfo {author} {\bibfnamefont {V.~M.}\ \bibnamefont {{Kaspi}}}, \bibinfo {author} {\bibfnamefont {R.}~\bibnamefont {{Kothes}}}, \bibinfo {author} {\bibfnamefont {T.~L.}\ \bibnamefont {{Landecker}}}, \bibinfo {author} {\bibfnamefont {D.~A.}\ \bibnamefont {{Lang}}}, \bibinfo {author} {\bibfnamefont {D.~Z.}\ \bibnamefont {{Li}}}, \bibinfo {author} {\bibfnamefont {H.~H.}\ \bibnamefont {{Lin}}}, \bibinfo {author} {\bibfnamefont {K.~W.}\
  \bibnamefont {{Masui}}}, \bibinfo {author} {\bibfnamefont {J.}~\bibnamefont {{Mena-Parra}}}, \bibinfo {author} {\bibfnamefont {M.}~\bibnamefont {{Merryfield}}}, \bibinfo {author} {\bibfnamefont {R.}~\bibnamefont {{Mckinven}}}, \bibinfo {author} {\bibfnamefont {D.}~\bibnamefont {{Michilli}}}, \bibinfo {author} {\bibfnamefont {N.}~\bibnamefont {{Milutinovic}}}, \bibinfo {author} {\bibfnamefont {A.}~\bibnamefont {{Naidu}}}, \bibinfo {author} {\bibfnamefont {L.~B.}\ \bibnamefont {{Newburgh}}}, \bibinfo {author} {\bibfnamefont {C.}~\bibnamefont {{Ng}}}, \bibinfo {author} {\bibfnamefont {C.}~\bibnamefont {{Patel}}}, \bibinfo {author} {\bibfnamefont {U.}~\bibnamefont {{Pen}}}, \bibinfo {author} {\bibfnamefont {T.}~\bibnamefont {{Pinsonneault-Marotte}}}, \bibinfo {author} {\bibfnamefont {Z.}~\bibnamefont {{Pleunis}}}, \bibinfo {author} {\bibfnamefont {M.}~\bibnamefont {{Rafiei-Ravandi}}}, \bibinfo {author} {\bibfnamefont {M.}~\bibnamefont {{Rahman}}}, \bibinfo {author} {\bibfnamefont {S.~M.}\ \bibnamefont
  {{Ransom}}}, \bibinfo {author} {\bibfnamefont {A.}~\bibnamefont {{Renard}}}, \bibinfo {author} {\bibfnamefont {P.}~\bibnamefont {{Scholz}}}, \bibinfo {author} {\bibfnamefont {S.~R.}\ \bibnamefont {{Siegel}}}, \bibinfo {author} {\bibfnamefont {S.}~\bibnamefont {{Singh}}}, \bibinfo {author} {\bibfnamefont {K.~M.}\ \bibnamefont {{Smith}}}, \bibinfo {author} {\bibfnamefont {I.~H.}\ \bibnamefont {{Stairs}}}, \bibinfo {author} {\bibfnamefont {S.~P.}\ \bibnamefont {{Tendulkar}}}, \bibinfo {author} {\bibfnamefont {I.}~\bibnamefont {{Tretyakov}}}, \bibinfo {author} {\bibfnamefont {K.}~\bibnamefont {{Vanderlinde}}}, \bibinfo {author} {\bibfnamefont {P.}~\bibnamefont {{Yadav}}}, \ and\ \bibinfo {author} {\bibfnamefont {A.~V.}\ \bibnamefont {{Zwaniga}}},\ }\href {\doibase 10.3847/2041-8213/ab4a80} {\bibfield  {journal} {\bibinfo  {journal} {\apjl}\ }\textbf {\bibinfo {volume} {885}},\ \bibinfo {eid} {L24} (\bibinfo {year} {2019})}\BibitemShut {NoStop}%
\bibitem [{\citenamefont {Ocker}\ and\ \citenamefont {Cordes}(2026)}]{Ocker_2026}%
  \BibitemOpen
  \bibfield  {author} {\bibinfo {author} {\bibfnamefont {S.~K.}\ \bibnamefont {Ocker}}\ and\ \bibinfo {author} {\bibfnamefont {J.~M.}\ \bibnamefont {Cordes}},\ }\href {\doibase 10.3847/1538-4357/ae5825} {\bibfield  {journal} {\bibinfo  {journal} {The Astrophysical Journal}\ }\textbf {\bibinfo {volume} {1002}},\ \bibinfo {pages} {3} (\bibinfo {year} {2026})}\BibitemShut {NoStop}%
\bibitem [{\citenamefont {{Pitkin}}(2018)}]{psrqpy}%
  \BibitemOpen
  \bibfield  {author} {\bibinfo {author} {\bibfnamefont {M.}~\bibnamefont {{Pitkin}}},\ }\href {\doibase 10.21105/joss.00538} {\bibfield  {journal} {\bibinfo  {journal} {{Journal of Open Source Software}}\ }\textbf {\bibinfo {volume} {3}},\ \bibinfo {pages} {538} (\bibinfo {year} {2018})}\BibitemShut {NoStop}%
\bibitem [{\citenamefont {{Dong}}\ \emph {et~al.}(2023)\citenamefont {{Dong}}, \citenamefont {{Crowter}}, \citenamefont {{Meyers}}, \citenamefont {{Pleunis}}, \citenamefont {{Stairs}}, \citenamefont {{Tan}}, \citenamefont {{Yu}}, \citenamefont {{Boyle}}, \citenamefont {{Cook}}, \citenamefont {{Fonseca}}, \citenamefont {{Gaensler}}, \citenamefont {{Good}}, \citenamefont {{Kaspi}}, \citenamefont {{McKee}}, \citenamefont {{Patel}},\ and\ \citenamefont {{Pearlman}}}]{dongCHIME2022}%
  \BibitemOpen
  \bibfield  {author} {\bibinfo {author} {\bibfnamefont {F.~A.}\ \bibnamefont {{Dong}}}, \bibinfo {author} {\bibfnamefont {K.}~\bibnamefont {{Crowter}}}, \bibinfo {author} {\bibfnamefont {B.~W.}\ \bibnamefont {{Meyers}}}, \bibinfo {author} {\bibfnamefont {Z.}~\bibnamefont {{Pleunis}}}, \bibinfo {author} {\bibfnamefont {I.}~\bibnamefont {{Stairs}}}, \bibinfo {author} {\bibfnamefont {C.~M.}\ \bibnamefont {{Tan}}}, \bibinfo {author} {\bibfnamefont {T.~T.}\ \bibnamefont {{Yu}}}, \bibinfo {author} {\bibfnamefont {P.~J.}\ \bibnamefont {{Boyle}}}, \bibinfo {author} {\bibfnamefont {A.~M.}\ \bibnamefont {{Cook}}}, \bibinfo {author} {\bibfnamefont {E.}~\bibnamefont {{Fonseca}}}, \bibinfo {author} {\bibfnamefont {B.~M.}\ \bibnamefont {{Gaensler}}}, \bibinfo {author} {\bibfnamefont {D.~C.}\ \bibnamefont {{Good}}}, \bibinfo {author} {\bibfnamefont {V.}~\bibnamefont {{Kaspi}}}, \bibinfo {author} {\bibfnamefont {J.~W.}\ \bibnamefont {{McKee}}}, \bibinfo {author} {\bibfnamefont {C.}~\bibnamefont {{Patel}}}, \ and\ \bibinfo
  {author} {\bibfnamefont {A.~B.}\ \bibnamefont {{Pearlman}}},\ }\href {\doibase 10.1093/mnras/stad2012} {\bibfield  {journal} {\bibinfo  {journal} {\mnras}\ }\textbf {\bibinfo {volume} {524}},\ \bibinfo {pages} {5132} (\bibinfo {year} {2023})}\BibitemShut {NoStop}%
\bibitem [{\citenamefont {Wateren}\ \emph {et~al.}(2023)\citenamefont {Wateren}, \citenamefont {Bassa}, \citenamefont {Cooper}, \citenamefont {Grießmeier}, \citenamefont {Stappers}, \citenamefont {Hessels}, \citenamefont {Kondratiev}, \citenamefont {Michilli}, \citenamefont {Tan}, \citenamefont {Tiburzi}, \citenamefont {Weltevrede}, \citenamefont {Nielsen}, \citenamefont {Carozzi}, \citenamefont {Ciardi}, \citenamefont {Cognard}, \citenamefont {Dettmar}, \citenamefont {Karastergiou}, \citenamefont {Kramer}, \citenamefont {Künsemöller}, \citenamefont {Osłowski}, \citenamefont {Serylak}, \citenamefont {Vocks},\ and\ \citenamefont {Wucknitz}}]{Wateren2023}%
  \BibitemOpen
  \bibfield  {author} {\bibinfo {author} {\bibfnamefont {E.~v.~d.}\ \bibnamefont {Wateren}}, \bibinfo {author} {\bibfnamefont {C.~G.}\ \bibnamefont {Bassa}}, \bibinfo {author} {\bibfnamefont {S.}~\bibnamefont {Cooper}}, \bibinfo {author} {\bibfnamefont {J.-M.}\ \bibnamefont {Grießmeier}}, \bibinfo {author} {\bibfnamefont {B.~W.}\ \bibnamefont {Stappers}}, \bibinfo {author} {\bibfnamefont {J.~W.~T.}\ \bibnamefont {Hessels}}, \bibinfo {author} {\bibfnamefont {V.~I.}\ \bibnamefont {Kondratiev}}, \bibinfo {author} {\bibfnamefont {D.}~\bibnamefont {Michilli}}, \bibinfo {author} {\bibfnamefont {C.~M.}\ \bibnamefont {Tan}}, \bibinfo {author} {\bibfnamefont {C.}~\bibnamefont {Tiburzi}}, \bibinfo {author} {\bibfnamefont {P.}~\bibnamefont {Weltevrede}}, \bibinfo {author} {\bibfnamefont {A.-S.~B.}\ \bibnamefont {Nielsen}}, \bibinfo {author} {\bibfnamefont {T.~D.}\ \bibnamefont {Carozzi}}, \bibinfo {author} {\bibfnamefont {B.}~\bibnamefont {Ciardi}}, \bibinfo {author} {\bibfnamefont {I.}~\bibnamefont {Cognard}},
  \bibinfo {author} {\bibfnamefont {R.-J.}\ \bibnamefont {Dettmar}}, \bibinfo {author} {\bibfnamefont {A.}~\bibnamefont {Karastergiou}}, \bibinfo {author} {\bibfnamefont {M.}~\bibnamefont {Kramer}}, \bibinfo {author} {\bibfnamefont {J.}~\bibnamefont {Künsemöller}}, \bibinfo {author} {\bibfnamefont {S.}~\bibnamefont {Osłowski}}, \bibinfo {author} {\bibfnamefont {M.}~\bibnamefont {Serylak}}, \bibinfo {author} {\bibfnamefont {C.}~\bibnamefont {Vocks}}, \ and\ \bibinfo {author} {\bibfnamefont {O.}~\bibnamefont {Wucknitz}},\ }\href {\doibase 10.1051/0004-6361/202245122} {\bibfield  {journal} {\bibinfo  {journal} {Astronomy \& Astrophysics}\ }\textbf {\bibinfo {volume} {669}},\ \bibinfo {pages} {A160} (\bibinfo {year} {2023})}\BibitemShut {NoStop}%
\bibitem [{\citenamefont {Parent}\ \emph {et~al.}(2022)\citenamefont {Parent}, \citenamefont {Sewalls}, \citenamefont {Freire}, \citenamefont {Matheny}, \citenamefont {Lyne}, \citenamefont {Perera}, \citenamefont {Cardoso}, \citenamefont {McLaughlin}, \citenamefont {Allen}, \citenamefont {Brazier}, \citenamefont {Camilo}, \citenamefont {Chatterjee}, \citenamefont {Cordes}, \citenamefont {Crawford}, \citenamefont {Deneva}, \citenamefont {Dong}, \citenamefont {Ferdman}, \citenamefont {Fonseca}, \citenamefont {Hessels}, \citenamefont {Kaspi}, \citenamefont {Knispel}, \citenamefont {Leeuwen}, \citenamefont {Lynch}, \citenamefont {Meyers}, \citenamefont {McKee}, \citenamefont {Mickaliger}, \citenamefont {Patel}, \citenamefont {Ransom}, \citenamefont {Rochon}, \citenamefont {Scholz}, \citenamefont {Stairs}, \citenamefont {Stappers}, \citenamefont {Tan},\ and\ \citenamefont {Zhu}}]{Parent2022}%
  \BibitemOpen
  \bibfield  {author} {\bibinfo {author} {\bibfnamefont {E.}~\bibnamefont {Parent}}, \bibinfo {author} {\bibfnamefont {H.}~\bibnamefont {Sewalls}}, \bibinfo {author} {\bibfnamefont {P.~C.~C.}\ \bibnamefont {Freire}}, \bibinfo {author} {\bibfnamefont {T.}~\bibnamefont {Matheny}}, \bibinfo {author} {\bibfnamefont {A.~G.}\ \bibnamefont {Lyne}}, \bibinfo {author} {\bibfnamefont {B.~B.~P.}\ \bibnamefont {Perera}}, \bibinfo {author} {\bibfnamefont {F.}~\bibnamefont {Cardoso}}, \bibinfo {author} {\bibfnamefont {M.~A.}\ \bibnamefont {McLaughlin}}, \bibinfo {author} {\bibfnamefont {B.}~\bibnamefont {Allen}}, \bibinfo {author} {\bibfnamefont {A.}~\bibnamefont {Brazier}}, \bibinfo {author} {\bibfnamefont {F.}~\bibnamefont {Camilo}}, \bibinfo {author} {\bibfnamefont {S.}~\bibnamefont {Chatterjee}}, \bibinfo {author} {\bibfnamefont {J.~M.}\ \bibnamefont {Cordes}}, \bibinfo {author} {\bibfnamefont {F.}~\bibnamefont {Crawford}}, \bibinfo {author} {\bibfnamefont {J.~S.}\ \bibnamefont {Deneva}}, \bibinfo {author}
  {\bibfnamefont {F.~A.}\ \bibnamefont {Dong}}, \bibinfo {author} {\bibfnamefont {R.~D.}\ \bibnamefont {Ferdman}}, \bibinfo {author} {\bibfnamefont {E.}~\bibnamefont {Fonseca}}, \bibinfo {author} {\bibfnamefont {J.~W.~T.}\ \bibnamefont {Hessels}}, \bibinfo {author} {\bibfnamefont {V.~M.}\ \bibnamefont {Kaspi}}, \bibinfo {author} {\bibfnamefont {B.}~\bibnamefont {Knispel}}, \bibinfo {author} {\bibfnamefont {J.~v.}\ \bibnamefont {Leeuwen}}, \bibinfo {author} {\bibfnamefont {R.~S.}\ \bibnamefont {Lynch}}, \bibinfo {author} {\bibfnamefont {B.~M.}\ \bibnamefont {Meyers}}, \bibinfo {author} {\bibfnamefont {J.~W.}\ \bibnamefont {McKee}}, \bibinfo {author} {\bibfnamefont {M.~B.}\ \bibnamefont {Mickaliger}}, \bibinfo {author} {\bibfnamefont {C.}~\bibnamefont {Patel}}, \bibinfo {author} {\bibfnamefont {S.~M.}\ \bibnamefont {Ransom}}, \bibinfo {author} {\bibfnamefont {A.}~\bibnamefont {Rochon}}, \bibinfo {author} {\bibfnamefont {P.}~\bibnamefont {Scholz}}, \bibinfo {author} {\bibfnamefont {I.~H.}\ \bibnamefont
  {Stairs}}, \bibinfo {author} {\bibfnamefont {B.~W.}\ \bibnamefont {Stappers}}, \bibinfo {author} {\bibfnamefont {C.~M.}\ \bibnamefont {Tan}}, \ and\ \bibinfo {author} {\bibfnamefont {W.~W.}\ \bibnamefont {Zhu}},\ }\href {\doibase 10.3847/1538-4357/ac375d} {\bibfield  {journal} {\bibinfo  {journal} {The Astrophysical Journal}\ }\textbf {\bibinfo {volume} {924}},\ \bibinfo {pages} {135} (\bibinfo {year} {2022})}\BibitemShut {NoStop}%
\bibitem [{\citenamefont {{Fiore}}\ \emph {et~al.}(2023)\citenamefont {{Fiore}}, \citenamefont {{Levin}}, \citenamefont {{McLaughlin}}, \citenamefont {{Anumarlapudi}}, \citenamefont {{Kaplan}}, \citenamefont {{Swiggum}}, \citenamefont {{Agazie}}, \citenamefont {{Bavisotto}}, \citenamefont {{Chawla}}, \citenamefont {{DeCesar}}, \citenamefont {{Dolch}}, \citenamefont {{Fonseca}}, \citenamefont {{Kaspi}}, \citenamefont {{Komassa}}, \citenamefont {{Kondratiev}}, \citenamefont {{van Leeuwen}}, \citenamefont {{Lewis}}, \citenamefont {{Lynch}}, \citenamefont {{McEwen}}, \citenamefont {{Mundorf}}, \citenamefont {{Al Noori}}, \citenamefont {{Parent}}, \citenamefont {{Pleunis}}, \citenamefont {{Ransom}}, \citenamefont {{Siemens}}, \citenamefont {{Spiewak}}, \citenamefont {{Stairs}}, \citenamefont {{Surnis}},\ and\ \citenamefont {{Tobin}}}]{Fiore2023}%
  \BibitemOpen
  \bibfield  {author} {\bibinfo {author} {\bibfnamefont {W.}~\bibnamefont {{Fiore}}}, \bibinfo {author} {\bibfnamefont {L.}~\bibnamefont {{Levin}}}, \bibinfo {author} {\bibfnamefont {M.~A.}\ \bibnamefont {{McLaughlin}}}, \bibinfo {author} {\bibfnamefont {A.}~\bibnamefont {{Anumarlapudi}}}, \bibinfo {author} {\bibfnamefont {D.~L.}\ \bibnamefont {{Kaplan}}}, \bibinfo {author} {\bibfnamefont {J.~K.}\ \bibnamefont {{Swiggum}}}, \bibinfo {author} {\bibfnamefont {G.~Y.}\ \bibnamefont {{Agazie}}}, \bibinfo {author} {\bibfnamefont {R.}~\bibnamefont {{Bavisotto}}}, \bibinfo {author} {\bibfnamefont {P.}~\bibnamefont {{Chawla}}}, \bibinfo {author} {\bibfnamefont {M.~E.}\ \bibnamefont {{DeCesar}}}, \bibinfo {author} {\bibfnamefont {T.}~\bibnamefont {{Dolch}}}, \bibinfo {author} {\bibfnamefont {E.}~\bibnamefont {{Fonseca}}}, \bibinfo {author} {\bibfnamefont {V.~M.}\ \bibnamefont {{Kaspi}}}, \bibinfo {author} {\bibfnamefont {Z.}~\bibnamefont {{Komassa}}}, \bibinfo {author} {\bibfnamefont {V.~I.}\ \bibnamefont
  {{Kondratiev}}}, \bibinfo {author} {\bibfnamefont {J.}~\bibnamefont {{van Leeuwen}}}, \bibinfo {author} {\bibfnamefont {E.~F.}\ \bibnamefont {{Lewis}}}, \bibinfo {author} {\bibfnamefont {R.~S.}\ \bibnamefont {{Lynch}}}, \bibinfo {author} {\bibfnamefont {A.~E.}\ \bibnamefont {{McEwen}}}, \bibinfo {author} {\bibfnamefont {R.}~\bibnamefont {{Mundorf}}}, \bibinfo {author} {\bibfnamefont {H.}~\bibnamefont {{Al Noori}}}, \bibinfo {author} {\bibfnamefont {E.}~\bibnamefont {{Parent}}}, \bibinfo {author} {\bibfnamefont {Z.}~\bibnamefont {{Pleunis}}}, \bibinfo {author} {\bibfnamefont {S.~M.}\ \bibnamefont {{Ransom}}}, \bibinfo {author} {\bibfnamefont {X.}~\bibnamefont {{Siemens}}}, \bibinfo {author} {\bibfnamefont {R.}~\bibnamefont {{Spiewak}}}, \bibinfo {author} {\bibfnamefont {I.~H.}\ \bibnamefont {{Stairs}}}, \bibinfo {author} {\bibfnamefont {M.}~\bibnamefont {{Surnis}}}, \ and\ \bibinfo {author} {\bibfnamefont {T.~J.}\ \bibnamefont {{Tobin}}},\ }\href {\doibase 10.3847/1538-4357/aceef7} {\bibfield  {journal}
  {\bibinfo  {journal} {\apj}\ }\textbf {\bibinfo {volume} {956}},\ \bibinfo {eid} {40} (\bibinfo {year} {2023})}\BibitemShut {NoStop}%
\bibitem [{\citenamefont {{Olszanski}}\ \emph {et~al.}(2025)\citenamefont {{Olszanski}}, \citenamefont {{Lewis}}, \citenamefont {{Deneva}}, \citenamefont {{McLaughlin}}, \citenamefont {{Stovall}}, \citenamefont {{Freire}}, \citenamefont {{Perera}}, \citenamefont {{Bagchi}},\ and\ \citenamefont {{Martinez}}}]{2025ao327}%
  \BibitemOpen
  \bibfield  {author} {\bibinfo {author} {\bibfnamefont {T.~E.~E.}\ \bibnamefont {{Olszanski}}}, \bibinfo {author} {\bibfnamefont {E.~F.}\ \bibnamefont {{Lewis}}}, \bibinfo {author} {\bibfnamefont {J.~S.}\ \bibnamefont {{Deneva}}}, \bibinfo {author} {\bibfnamefont {M.~A.}\ \bibnamefont {{McLaughlin}}}, \bibinfo {author} {\bibfnamefont {K.}~\bibnamefont {{Stovall}}}, \bibinfo {author} {\bibfnamefont {P.~C.~C.}\ \bibnamefont {{Freire}}}, \bibinfo {author} {\bibfnamefont {B.~B.~P.}\ \bibnamefont {{Perera}}}, \bibinfo {author} {\bibfnamefont {M.}~\bibnamefont {{Bagchi}}}, \ and\ \bibinfo {author} {\bibfnamefont {J.~G.}\ \bibnamefont {{Martinez}}},\ }\href {\doibase 10.48550/arXiv.2502.04571} {\bibfield  {journal} {\bibinfo  {journal} {arXiv e-prints}\ ,\ \bibinfo {eid} {arXiv:2502.04571}} (\bibinfo {year} {2025})}\BibitemShut {NoStop}%
\bibitem [{\citenamefont {{McEwen}}\ \emph {et~al.}(2024)\citenamefont {{McEwen}}, \citenamefont {{Swiggum}}, \citenamefont {{Kaplan}}, \citenamefont {{Tan}}, \citenamefont {{Meyers}}, \citenamefont {{Fonseca}}, \citenamefont {{Agazie}}, \citenamefont {{Chawla}}, \citenamefont {{Crowter}}, \citenamefont {{DeCesar}}, \citenamefont {{Dolch}}, \citenamefont {{Dong}}, \citenamefont {{Fiore}}, \citenamefont {{Fonseca}}, \citenamefont {{Good}}, \citenamefont {{Istrate}}, \citenamefont {{Kaspi}}, \citenamefont {{Kondratiev}}, \citenamefont {{van Leeuwen}}, \citenamefont {{Levin}}, \citenamefont {{Lewis}}, \citenamefont {{Lynch}}, \citenamefont {{Masui}}, \citenamefont {{McKee}}, \citenamefont {{McLaughlin}}, \citenamefont {{Al Noori}}, \citenamefont {{Parent}}, \citenamefont {{Ransom}}, \citenamefont {{Siemens}}, \citenamefont {{Spiewak}},\ and\ \citenamefont {{Stairs}}}]{mcewen2024}%
  \BibitemOpen
  \bibfield  {author} {\bibinfo {author} {\bibfnamefont {A.~E.}\ \bibnamefont {{McEwen}}}, \bibinfo {author} {\bibfnamefont {J.~K.}\ \bibnamefont {{Swiggum}}}, \bibinfo {author} {\bibfnamefont {D.~L.}\ \bibnamefont {{Kaplan}}}, \bibinfo {author} {\bibfnamefont {C.~M.}\ \bibnamefont {{Tan}}}, \bibinfo {author} {\bibfnamefont {B.~W.}\ \bibnamefont {{Meyers}}}, \bibinfo {author} {\bibfnamefont {E.}~\bibnamefont {{Fonseca}}}, \bibinfo {author} {\bibfnamefont {G.~Y.}\ \bibnamefont {{Agazie}}}, \bibinfo {author} {\bibfnamefont {P.}~\bibnamefont {{Chawla}}}, \bibinfo {author} {\bibfnamefont {K.}~\bibnamefont {{Crowter}}}, \bibinfo {author} {\bibfnamefont {M.~E.}\ \bibnamefont {{DeCesar}}}, \bibinfo {author} {\bibfnamefont {T.}~\bibnamefont {{Dolch}}}, \bibinfo {author} {\bibfnamefont {F.~A.}\ \bibnamefont {{Dong}}}, \bibinfo {author} {\bibfnamefont {W.}~\bibnamefont {{Fiore}}}, \bibinfo {author} {\bibfnamefont {E.}~\bibnamefont {{Fonseca}}}, \bibinfo {author} {\bibfnamefont {D.~C.}\ \bibnamefont {{Good}}}, \bibinfo
  {author} {\bibfnamefont {A.~G.}\ \bibnamefont {{Istrate}}}, \bibinfo {author} {\bibfnamefont {V.~M.}\ \bibnamefont {{Kaspi}}}, \bibinfo {author} {\bibfnamefont {V.~I.}\ \bibnamefont {{Kondratiev}}}, \bibinfo {author} {\bibfnamefont {J.}~\bibnamefont {{van Leeuwen}}}, \bibinfo {author} {\bibfnamefont {L.}~\bibnamefont {{Levin}}}, \bibinfo {author} {\bibfnamefont {E.~F.}\ \bibnamefont {{Lewis}}}, \bibinfo {author} {\bibfnamefont {R.~S.}\ \bibnamefont {{Lynch}}}, \bibinfo {author} {\bibfnamefont {K.~W.}\ \bibnamefont {{Masui}}}, \bibinfo {author} {\bibfnamefont {J.~W.}\ \bibnamefont {{McKee}}}, \bibinfo {author} {\bibfnamefont {M.~A.}\ \bibnamefont {{McLaughlin}}}, \bibinfo {author} {\bibfnamefont {H.}~\bibnamefont {{Al Noori}}}, \bibinfo {author} {\bibfnamefont {E.}~\bibnamefont {{Parent}}}, \bibinfo {author} {\bibfnamefont {S.~M.}\ \bibnamefont {{Ransom}}}, \bibinfo {author} {\bibfnamefont {X.}~\bibnamefont {{Siemens}}}, \bibinfo {author} {\bibfnamefont {R.}~\bibnamefont {{Spiewak}}}, \ and\ \bibinfo
  {author} {\bibfnamefont {I.~H.}\ \bibnamefont {{Stairs}}},\ }\href {\doibase 10.3847/1538-4357/ad11f0} {\bibfield  {journal} {\bibinfo  {journal} {\apj}\ }\textbf {\bibinfo {volume} {962}},\ \bibinfo {eid} {167} (\bibinfo {year} {2024})}\BibitemShut {NoStop}%
\bibitem [{\citenamefont {{Teplykh}}\ \emph {et~al.}(2022)\citenamefont {{Teplykh}}, \citenamefont {{Malofeev}}, \citenamefont {{Malov}},\ and\ \citenamefont {{Tyul'bashev}}}]{Teplykh2022}%
  \BibitemOpen
  \bibfield  {author} {\bibinfo {author} {\bibfnamefont {D.}~\bibnamefont {{Teplykh}}}, \bibinfo {author} {\bibfnamefont {V.}~\bibnamefont {{Malofeev}}}, \bibinfo {author} {\bibfnamefont {O.}~\bibnamefont {{Malov}}}, \ and\ \bibinfo {author} {\bibfnamefont {S.}~\bibnamefont {{Tyul'bashev}}},\ }\href {\doibase 10.1515/astro-2022-0019} {\bibfield  {journal} {\bibinfo  {journal} {Open Astronomy}\ }\textbf {\bibinfo {volume} {31}},\ \bibinfo {pages} {166} (\bibinfo {year} {2022})}\BibitemShut {NoStop}%
\bibitem [{\citenamefont {McEwen}\ \emph {et~al.}(2020)\citenamefont {McEwen}, \citenamefont {Spiewak}, \citenamefont {Swiggum}, \citenamefont {Kaplan}, \citenamefont {Fiore}, \citenamefont {Agazie}, \citenamefont {Blumer}, \citenamefont {Chawla}, \citenamefont {DeCesar}, \citenamefont {Kaspi}, \citenamefont {Kondratiev}, \citenamefont {LaRose}, \citenamefont {Levin}, \citenamefont {Lynch}, \citenamefont {McLaughlin}, \citenamefont {Mingyar}, \citenamefont {Noori}, \citenamefont {Ransom}, \citenamefont {Roberts}, \citenamefont {Schmiedekamp}, \citenamefont {Schmiedekamp}, \citenamefont {Siemens}, \citenamefont {Stairs}, \citenamefont {Stovall}, \citenamefont {Surnis},\ and\ \citenamefont {Leeuwen}}]{McEwen2020}%
  \BibitemOpen
  \bibfield  {author} {\bibinfo {author} {\bibfnamefont {A.~E.}\ \bibnamefont {McEwen}}, \bibinfo {author} {\bibfnamefont {R.}~\bibnamefont {Spiewak}}, \bibinfo {author} {\bibfnamefont {J.~K.}\ \bibnamefont {Swiggum}}, \bibinfo {author} {\bibfnamefont {D.~L.}\ \bibnamefont {Kaplan}}, \bibinfo {author} {\bibfnamefont {W.}~\bibnamefont {Fiore}}, \bibinfo {author} {\bibfnamefont {G.~Y.}\ \bibnamefont {Agazie}}, \bibinfo {author} {\bibfnamefont {H.}~\bibnamefont {Blumer}}, \bibinfo {author} {\bibfnamefont {P.}~\bibnamefont {Chawla}}, \bibinfo {author} {\bibfnamefont {M.}~\bibnamefont {DeCesar}}, \bibinfo {author} {\bibfnamefont {V.~M.}\ \bibnamefont {Kaspi}}, \bibinfo {author} {\bibfnamefont {V.~I.}\ \bibnamefont {Kondratiev}}, \bibinfo {author} {\bibfnamefont {M.}~\bibnamefont {LaRose}}, \bibinfo {author} {\bibfnamefont {L.}~\bibnamefont {Levin}}, \bibinfo {author} {\bibfnamefont {R.~S.}\ \bibnamefont {Lynch}}, \bibinfo {author} {\bibfnamefont {M.}~\bibnamefont {McLaughlin}}, \bibinfo {author} {\bibfnamefont
  {M.}~\bibnamefont {Mingyar}}, \bibinfo {author} {\bibfnamefont {H.~A.}\ \bibnamefont {Noori}}, \bibinfo {author} {\bibfnamefont {S.~M.}\ \bibnamefont {Ransom}}, \bibinfo {author} {\bibfnamefont {M.~S.~E.}\ \bibnamefont {Roberts}}, \bibinfo {author} {\bibfnamefont {A.}~\bibnamefont {Schmiedekamp}}, \bibinfo {author} {\bibfnamefont {C.}~\bibnamefont {Schmiedekamp}}, \bibinfo {author} {\bibfnamefont {X.}~\bibnamefont {Siemens}}, \bibinfo {author} {\bibfnamefont {I.}~\bibnamefont {Stairs}}, \bibinfo {author} {\bibfnamefont {K.}~\bibnamefont {Stovall}}, \bibinfo {author} {\bibfnamefont {M.}~\bibnamefont {Surnis}}, \ and\ \bibinfo {author} {\bibfnamefont {J.~v.}\ \bibnamefont {Leeuwen}},\ }\href {\doibase 10.3847/1538-4357/ab75e2} {\bibfield  {journal} {\bibinfo  {journal} {The Astrophysical Journal}\ }\textbf {\bibinfo {volume} {892}},\ \bibinfo {pages} {76} (\bibinfo {year} {2020})}\BibitemShut {NoStop}%
\bibitem [{\citenamefont {Tyul'bashev}\ \emph {et~al.}(2022)\citenamefont {Tyul'bashev}, \citenamefont {Tyul'basheva},\ and\ \citenamefont {Kitaeva}}]{Tyulbashev2022}%
  \BibitemOpen
  \bibfield  {author} {\bibinfo {author} {\bibfnamefont {S.~A.}\ \bibnamefont {Tyul'bashev}}, \bibinfo {author} {\bibfnamefont {G.~E.}\ \bibnamefont {Tyul'basheva}}, \ and\ \bibinfo {author} {\bibfnamefont {M.~A.}\ \bibnamefont {Kitaeva}},\ }in\ \href {\doibase 10.22323/1.425.0043} {\emph {\bibinfo {booktitle} {Proceedings of {The} {Multifaceted} {Universe}: {Theory} and {Observations} - 2022 — {PoS}({MUTO2022})}}},\ Vol.\ \bibinfo {volume} {425}\ (\bibinfo  {publisher} {SISSA Medialab},\ \bibinfo {year} {2022})\ p.\ \bibinfo {pages} {043}\BibitemShut {NoStop}%
\bibitem [{\citenamefont {McKenna}(2025{\natexlab{a}})}]{mckenna_2025_14938680}%
  \BibitemOpen
  \bibfield  {author} {\bibinfo {author} {\bibfnamefont {D.}~\bibnamefont {McKenna}},\ }\href {\doibase 10.5281/zenodo.14938680} {\enquote {\bibinfo {title} {Data produced as a part of a pulsar timing campaign with the irish lofar station between 2020 and 2023 at 150 mhz},}\ } (\bibinfo {year} {2025}{\natexlab{a}})\BibitemShut {NoStop}%
\bibitem [{\citenamefont {McKenna}(2025{\natexlab{b}})}]{mckenna_2025_14938977}%
  \BibitemOpen
  \bibfield  {author} {\bibinfo {author} {\bibfnamefont {D.}~\bibnamefont {McKenna}},\ }\href {\doibase 10.5281/zenodo.14938977} {\enquote {\bibinfo {title} {Pulsar timing ephemerides produced as a part of an observing campaign with the irish lofar station between 2020 and 2023 at 150 mhz},}\ } (\bibinfo {year} {2025}{\natexlab{b}})\BibitemShut {NoStop}%
\end{thebibliography}%

\begin{appendix}
\section{Observed Sources}\label{app:timingobservations}
Table~\ref{tab:timingobservedsources} contains a list of the sources observed as a part of this work if they were detected, while Table~\ref{tab:timingobservedsourcesnondetected} contains sources that were not detected. Each source lists the number of times they were observed and detected, alongside the total time spent observing each source, and the minimum and maximum time spent observing the source in one observation.

\begin{table}[!t]
\centering
    \begin{tabular}{lccccc}
    \hline
    \multicolumn{6}{c}{Detected Sources} \\
Source & Cat. & N\textsubscript{obs} & N\textsubscript{det} & T\textsubscript{obs} & T\textsubscript{max} \\
& & & & (\SI{}{\hour}) & (\SI{}{\minute}) \\
    \hline\hline

J0104+6438 & GL & 39 & 34 & 20.4 & 153 \\
J0146+3055 & P & 23 & 22 & 24.6 & 149 \\
J0220+3626 & PL & 22 & 21 & 19.1 & 120 \\
J0226+3356 & C & 16 & 15 & 26.0 & 178 \\
J0317+1328 & LP & 17 & 10 & 12.6 & 154 \\
J0350+2341 & P & 1 & 1 & 1.0 & 59 \\
J0355+2838 & GL & 33 & 29 & 28.6 & 89 \\
J0608+1635 & AL & 42 & 40 & 18.3 & 59 \\
J0928+3037 & P & 1 & 1 & 1.0 & 59 \\
J1110+58 & G & 1 & 1 & 1.0 & 59 \\
J1132+2514 & LP & 33 & 31 & 23.7 & 89 \\
J1239+32 & G & 2 & 2 & 2.0 & 59 \\
J1243+3946 & LP & 26 & 25 & 19.8 & 104 \\
J1327+3423 & GLP & 15 & 15 & 6.1 & 59 \\
J1536+1749 & P & 33 & 18 & 47.7 & 179 \\
J1832+2749 & AL & 40 & 38 & 34.2 & 120 \\
J1836+5156 & GL & 41 & 36 & 32.9 & 149 \\
J1844+21 & P & 1 & 1 & 1.0 & 59 \\
J1958+2213 & P & 1 & 1 & 1.0 & 59 \\
J2000+2920 & GL & 36 & 34 & 38.4 & 119 \\
J2105+1908 & P & 15 & 7 & 17.6 & 119 \\
J2202+2134 & P & 36 & 34 & 28.5 & 134 \\
J2347+0300 & AL & 24 & 22 & 18.1 & 119 \\
\hline\hline\hline
    \end{tabular}

    \caption[Sources observed for timing with I-LOFAR.]{Sources observed for this work, with a distinction between detected and non-detected sources at the Irish LOFAR station. The columns contain the number of observations, the number of detections, the total  observing time on the source and the longest single pointing on the source. For catalogues, ``A'' represents surveys conducted with the Arecibo telescope, ``G'' is the GBNCC survey, ``L'' is the LOTAAS survey, and ``P'' are the PRAO catalogues.}
    \label{tab:timingobservedsources}
\end{table}
\begin{table}[!t]
\centering
        \begin{tabular}{lccccc}
        \hline
\multicolumn{6}{c}{Non-detected Sources} \\
Source & Cat. & N\textsubscript{obs} & N\textsubscript{det} & T\textsubscript{obs} & T\textsubscript{max} \\
& & & & (\SI{}{\hour}) & (\SI{}{\minute}) \\
    \hline\hline
J0109+11 & P & 2 & 0 & 2.0 & 59 \\
J0305+1127 & P & 2 & 0 & 3.7 & 164 \\
J0357-05 & G & 1 & 0 & 1.0 & 61 \\
J0509+37 & P & 2 & 0 & 3.0 & 119 \\
J0933+3245 & P & 1 & 0 & 1.0 & 59 \\
J1243+1752 & P & 3 & 0 & 3.3 & 78 \\
J1354+24 & G & 6 & 0 & 5.4 & 59 \\
J1430+22 & G & 5 & 0 & 5.2 & 109 \\
J1439+76 & G & 7 & 0 & 6.4 & 85 \\
J1651+1422 & P & 3 & 0 & 4.9 & 119 \\
J1843+21 & P & 2 & 0 & 2.9 & 114 \\
J1844+4117 & P & 1 & 0 & 1.0 & 59 \\
J1921+34 & P & 1 & 0 & 1.0 & 62 \\
J2029+34 & P & 1 & 0 & 1.2 & 74 \\
J2253+1237 & P & 2 & 0 & 2.7 & 104 \\
J2333+20 & P & 1 & 0 & 1.0 & 59 \\
\hline\hline
    \end{tabular}
    \caption[Sources observed for timing with I-LOFAR.]{Sources observed for this work, but were not detected. Columns and catalogue legend matches the information provided in Table~\ref{tab:timingobservedsources}}
    \label{tab:timingobservedsourcesnondetected}
\end{table}
\section{Pulsar Profiles}

\begin{figure*}[!p]
\centering
    \begin{tabular}{ccc}
\includegraphics[width=0.3\textwidth]{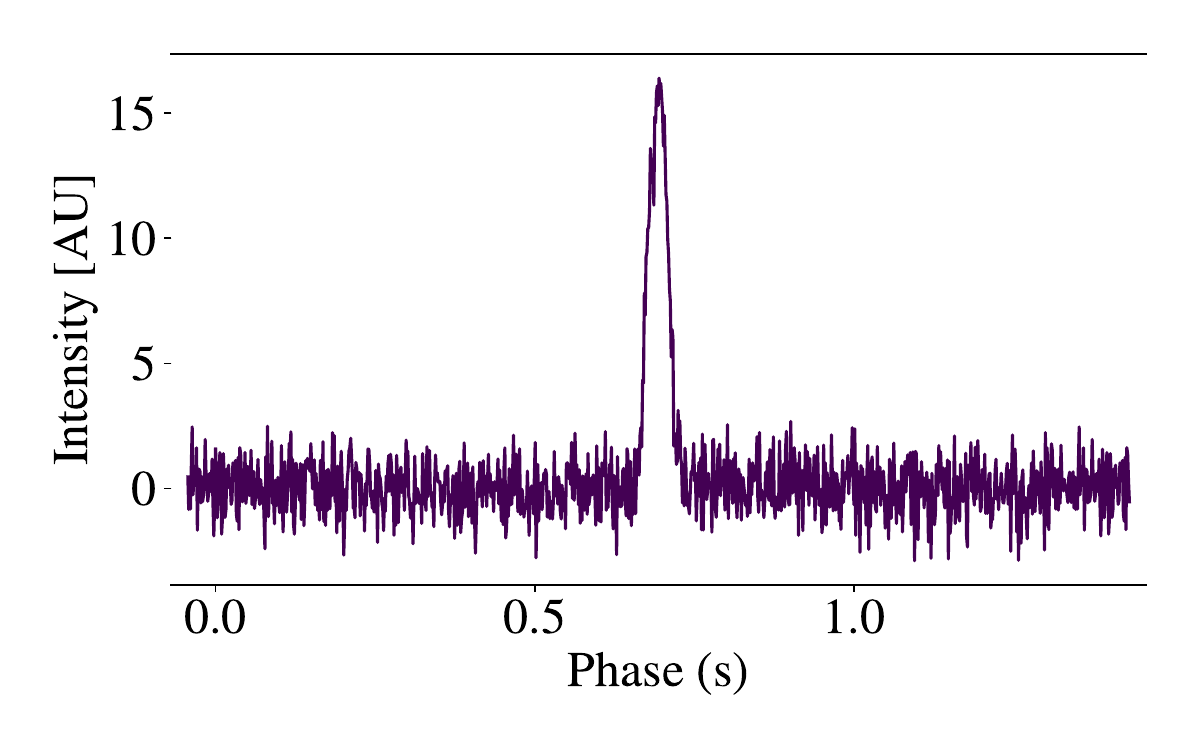} & 
\includegraphics[width=0.3\textwidth]{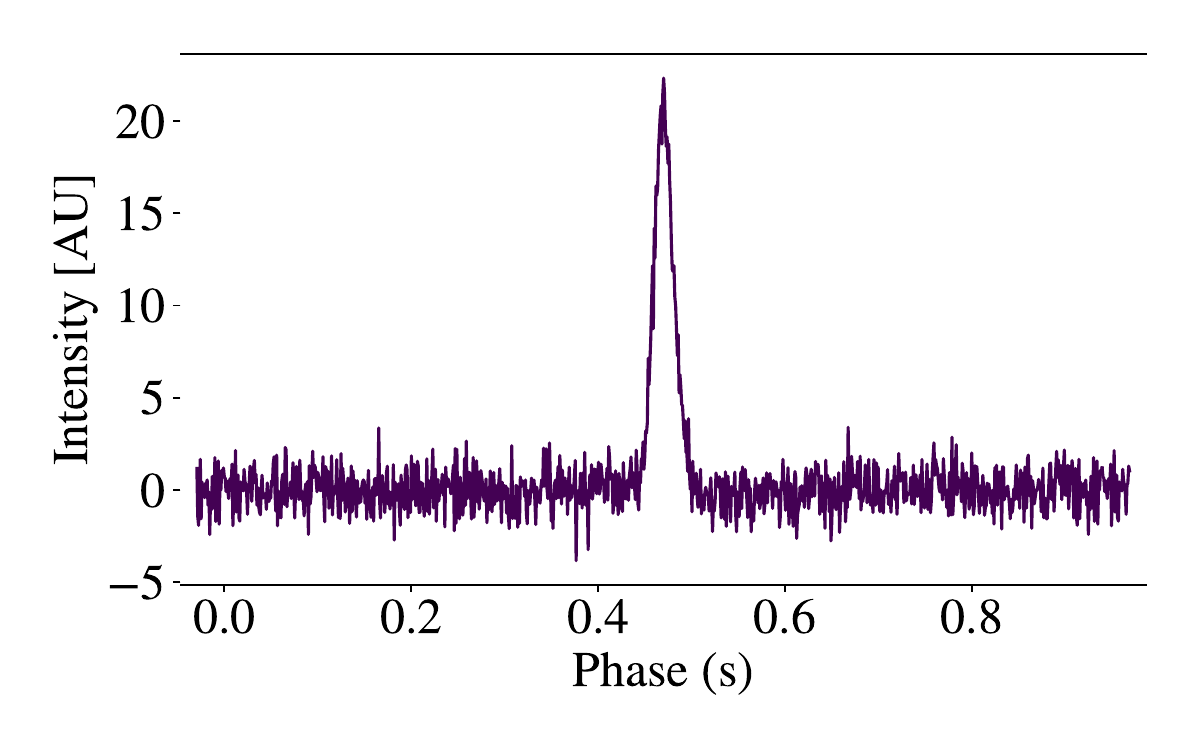} & 
\includegraphics[width=0.3\textwidth]{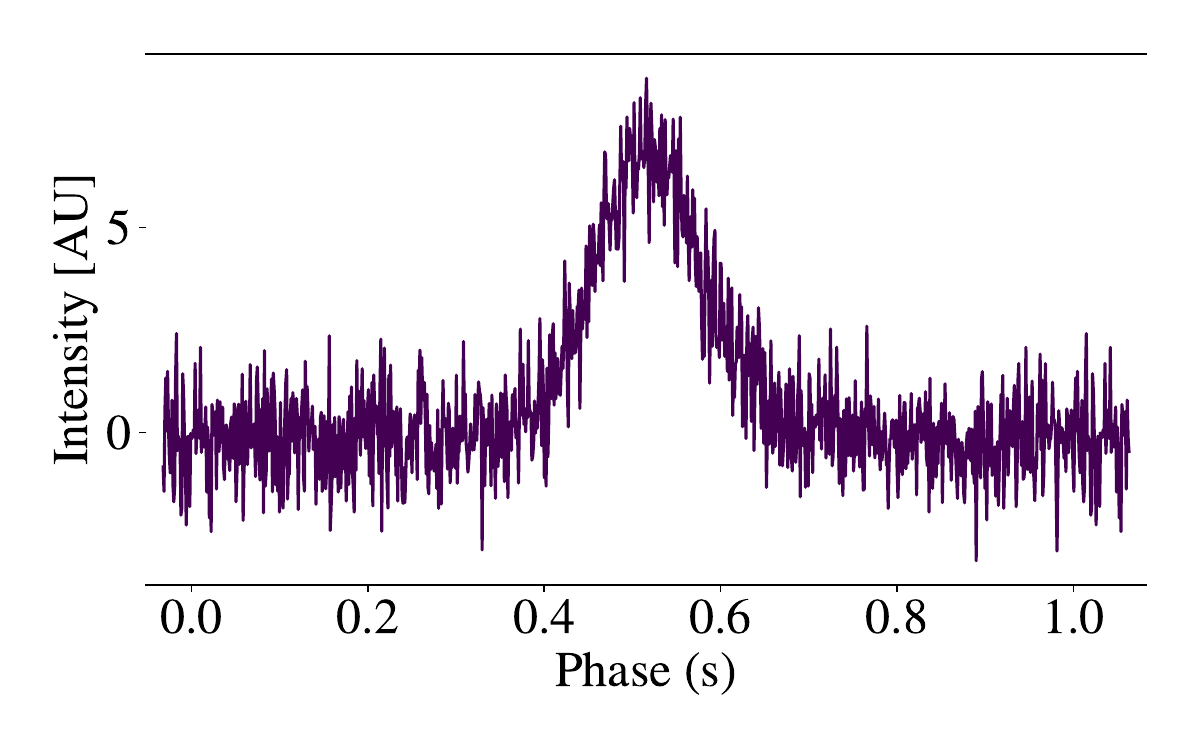} \\
 (a) J0104+6438 & (b) J0146+3055 & (c) J0220+3626 \\
 \includegraphics[width=0.3\textwidth]{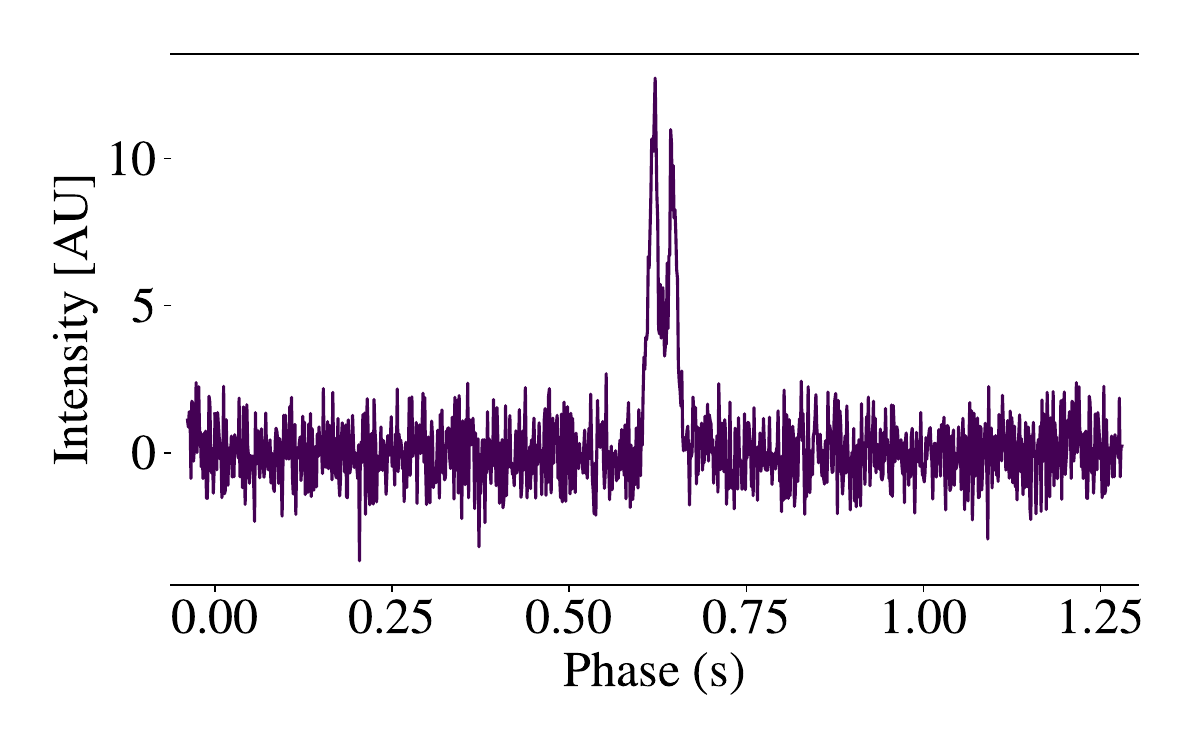} & 
 \includegraphics[width=0.3\textwidth]{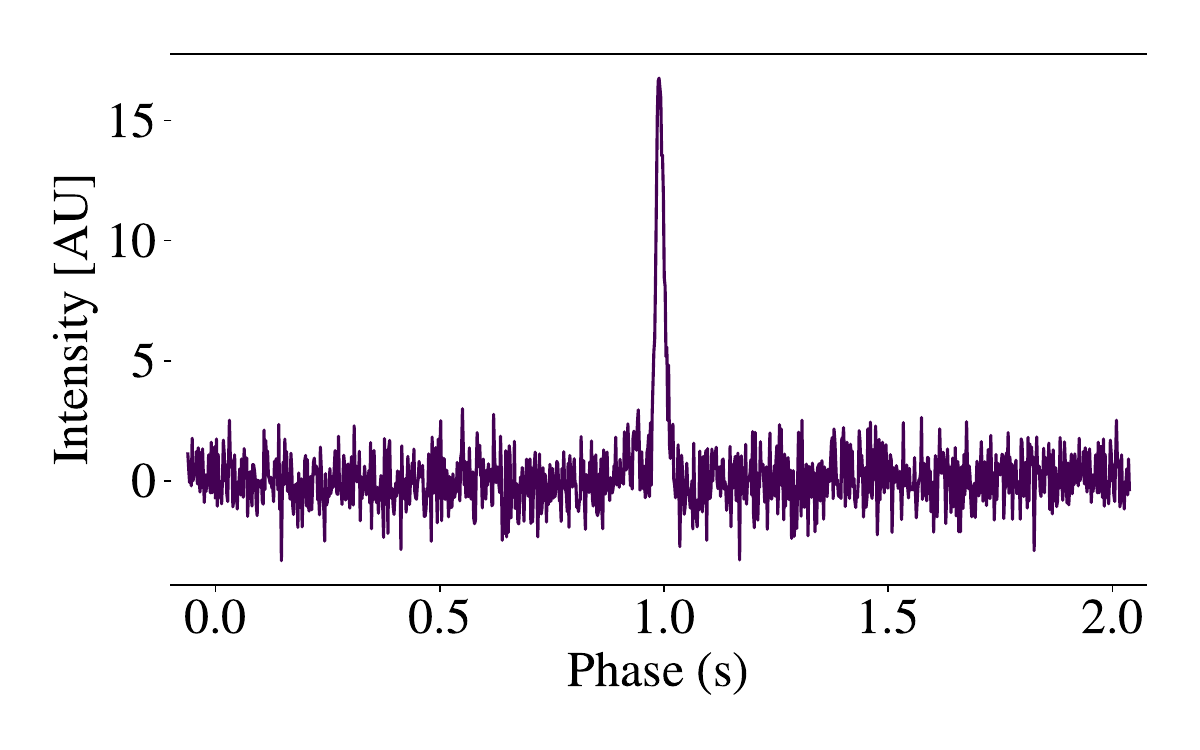} & \includegraphics[width=0.3\textwidth]{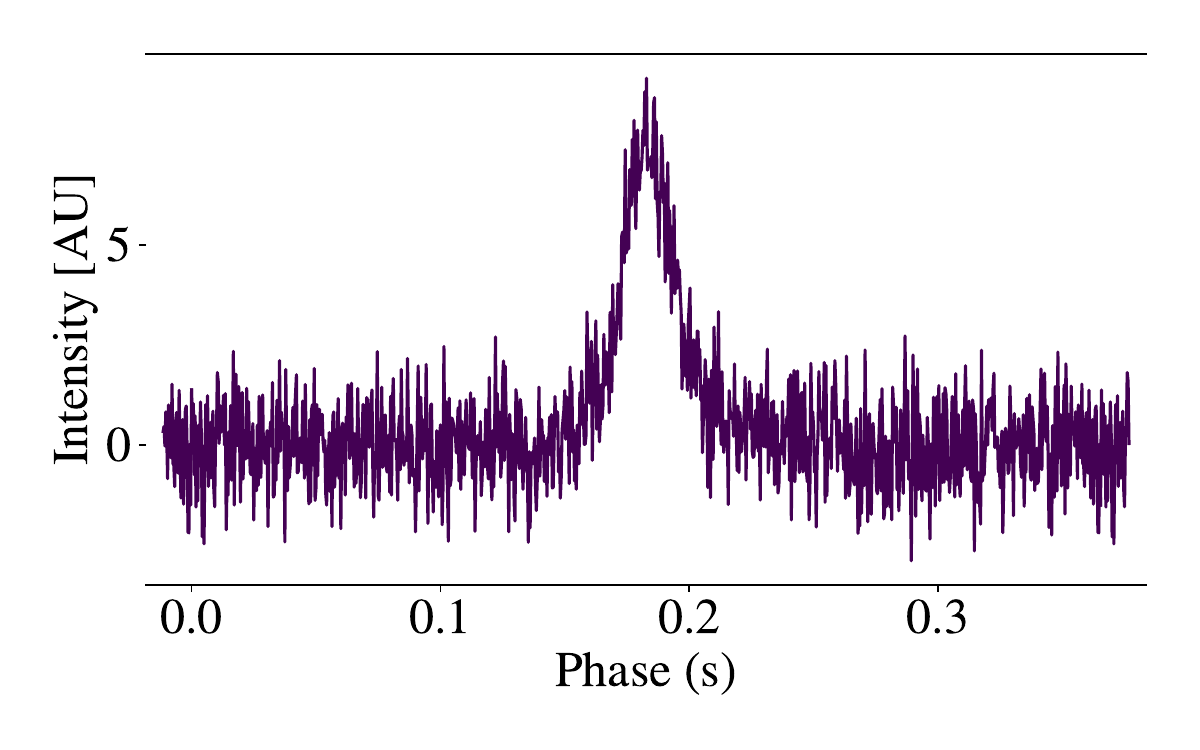} \\
 (d) J0226+3356 & (e) J0317+1328 & (f) J0355+2838 \\
\includegraphics[width=0.3\textwidth]{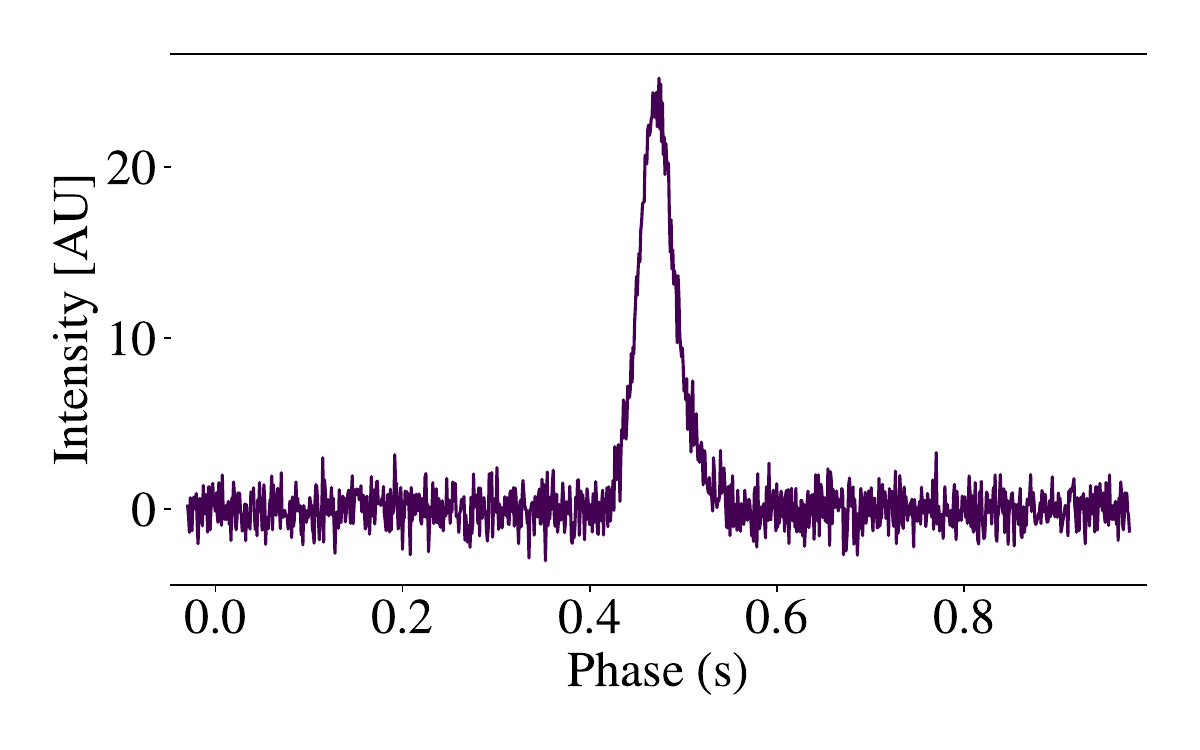} &
\includegraphics[width=0.3\textwidth]{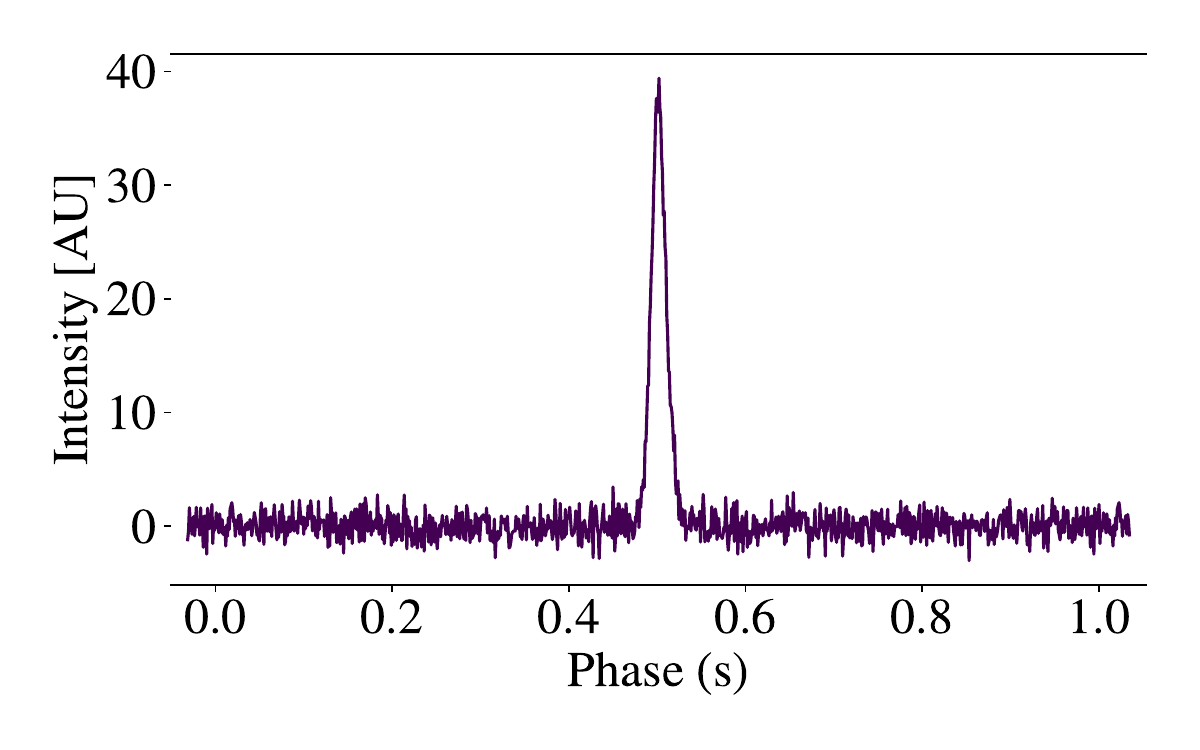} &
 \includegraphics[width=0.3\textwidth]{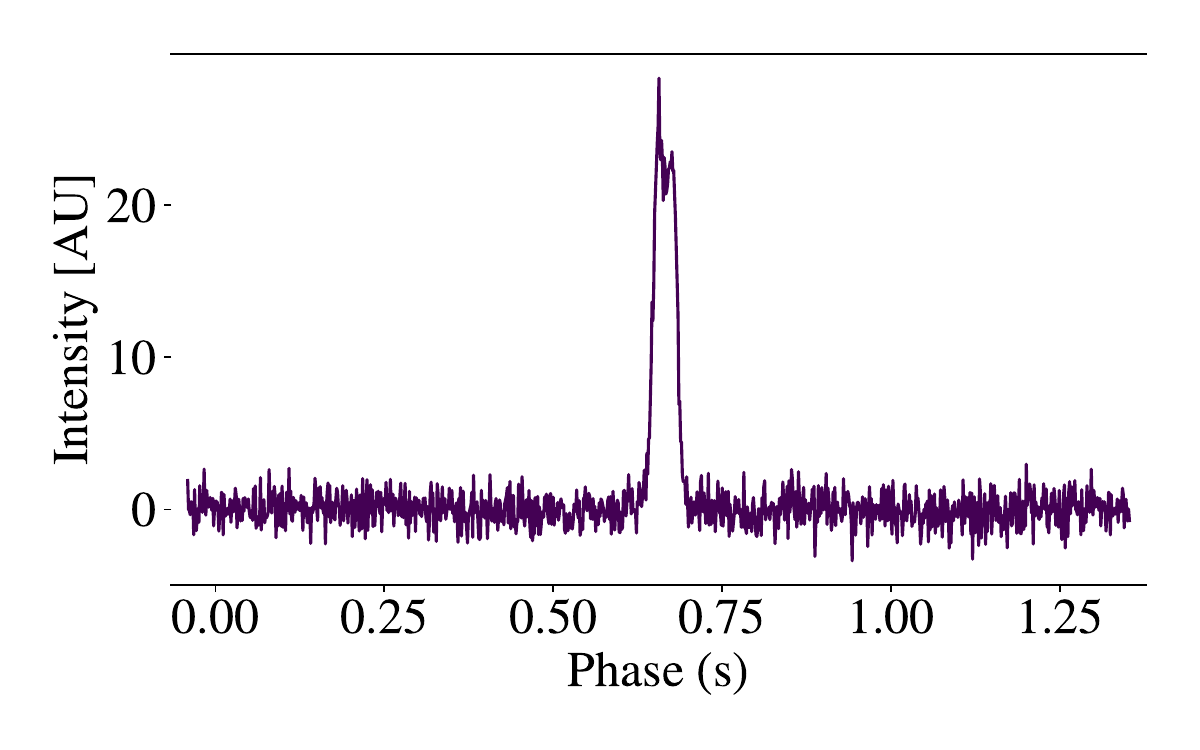} \\
(g) J0608+1635 & (h) J1132+2513 & (i) J1242+3946 \\
\includegraphics[width=0.3\textwidth]{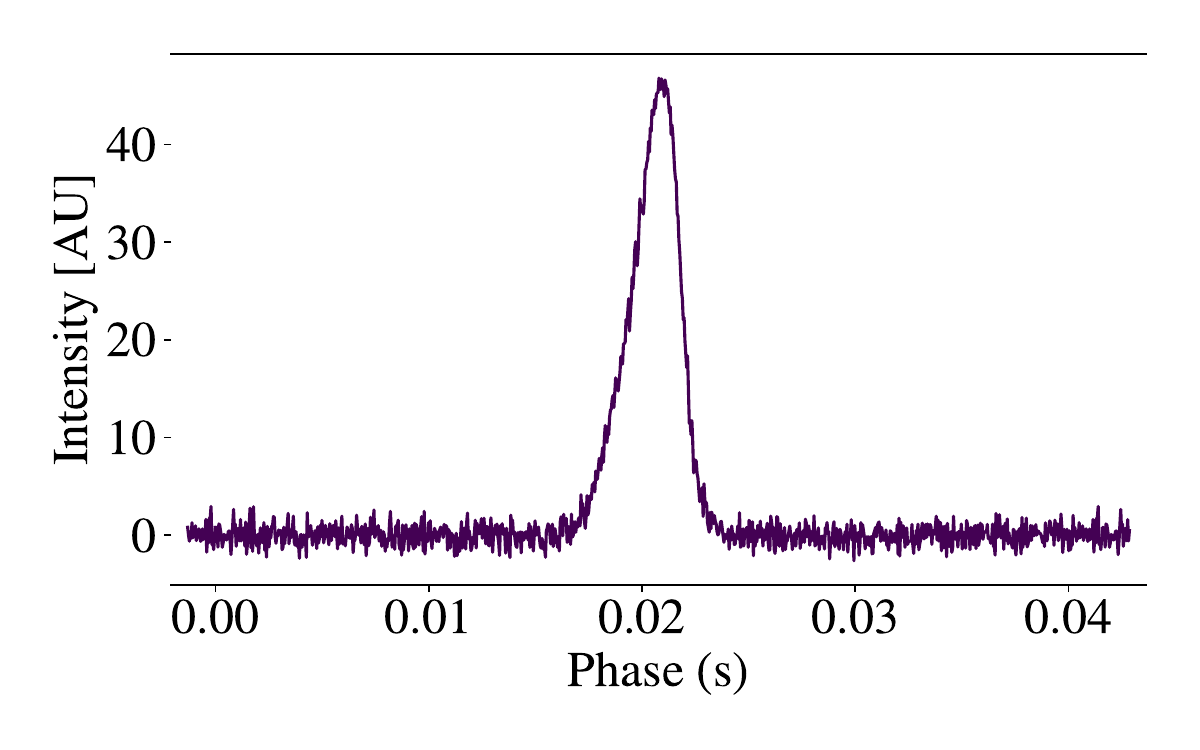} & 
\includegraphics[width=0.3\textwidth]{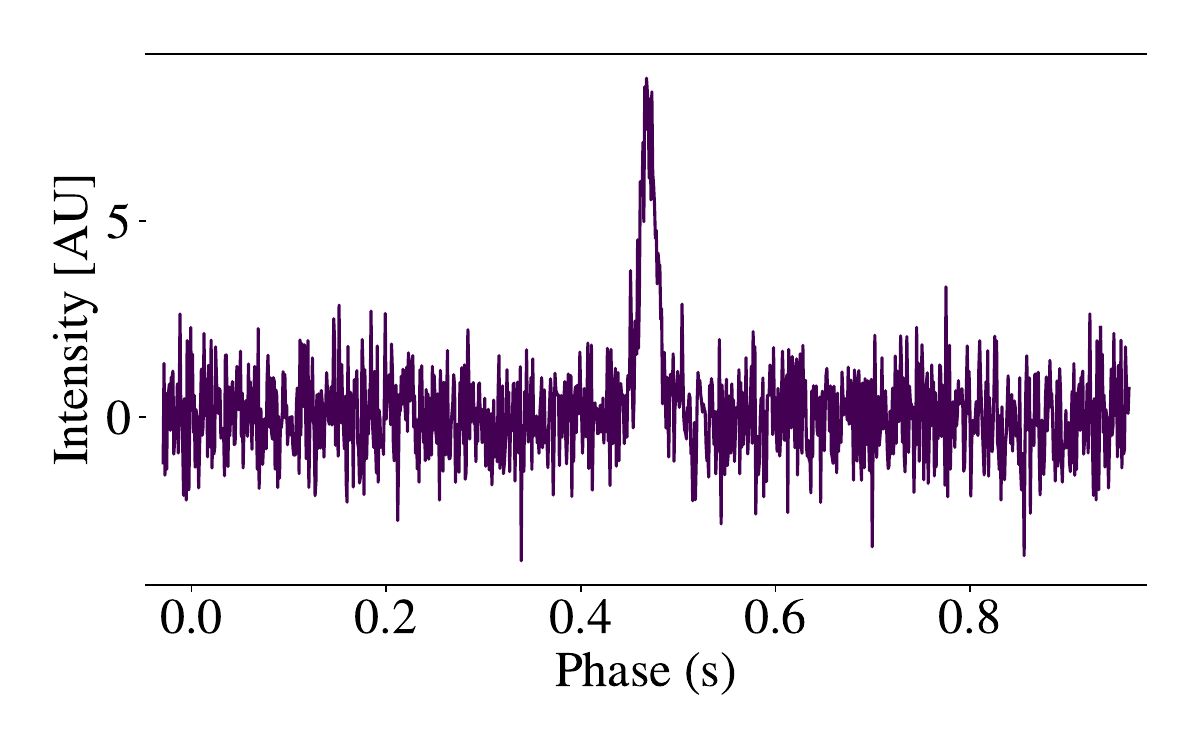} &
\includegraphics[width=0.3\textwidth]{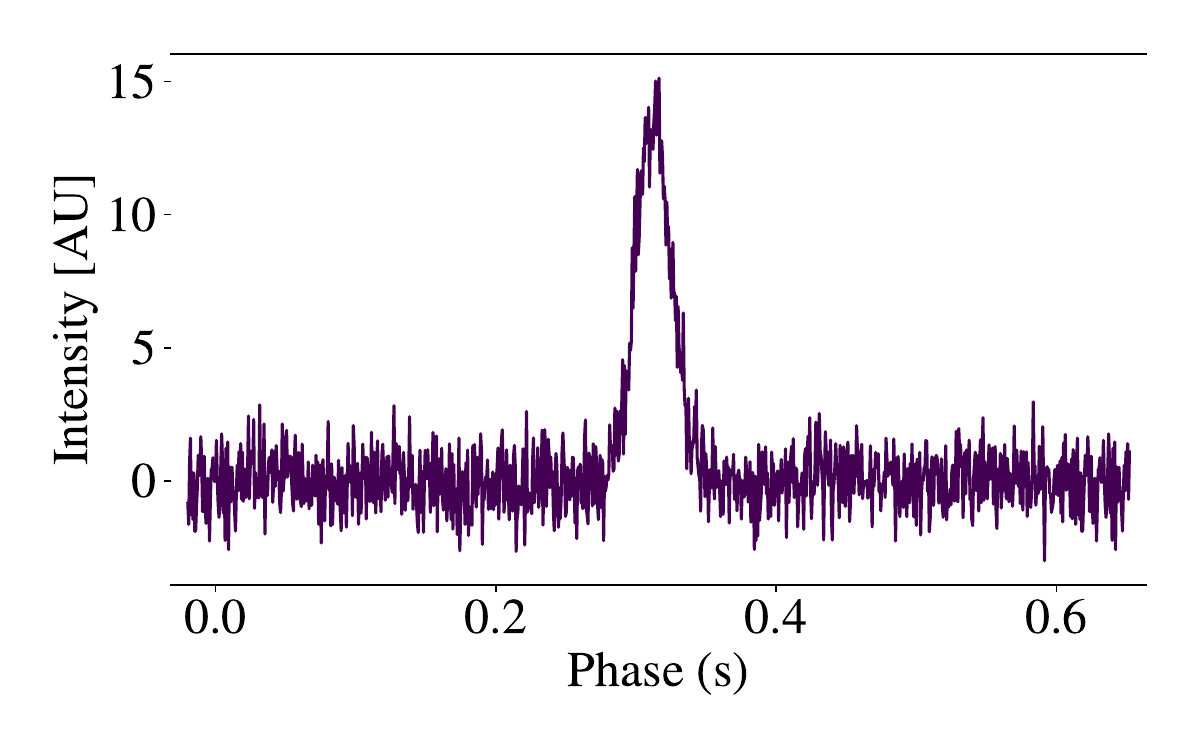} \\
(j) J1327+3423 & (k) J1536+1749 & (l) J1832+2749 \\
 \includegraphics[width=0.3\textwidth]{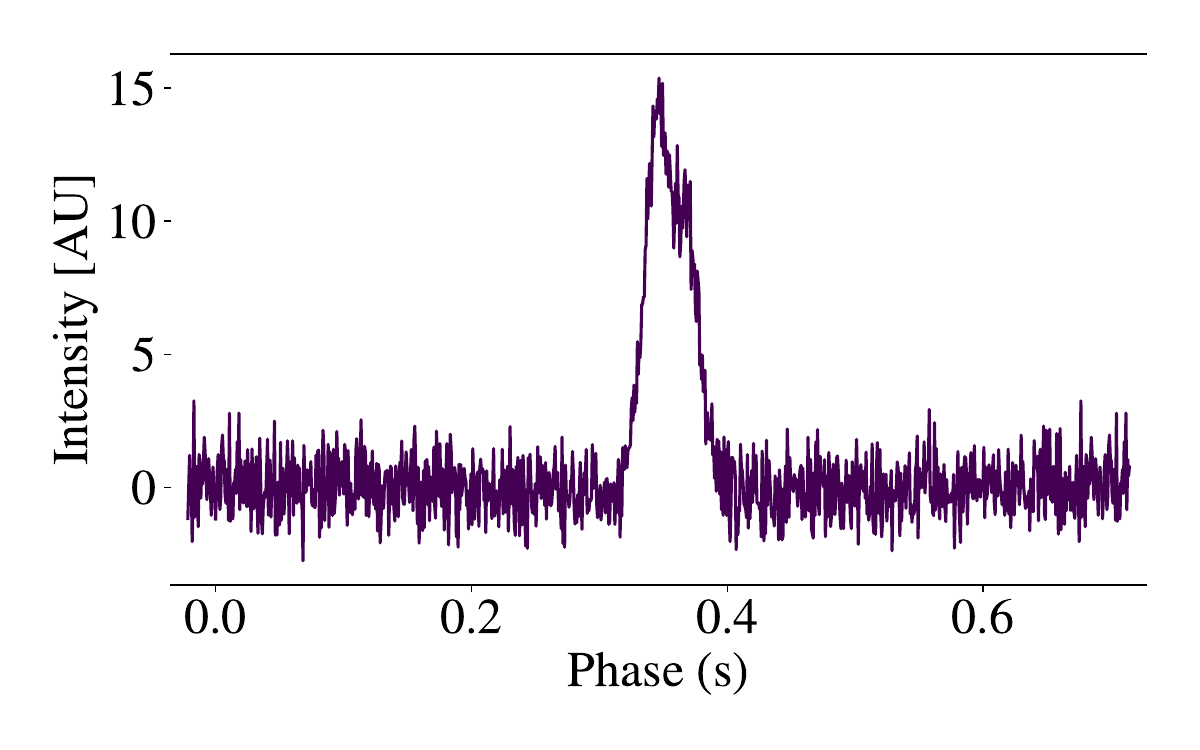} & \includegraphics[width=0.3\textwidth]{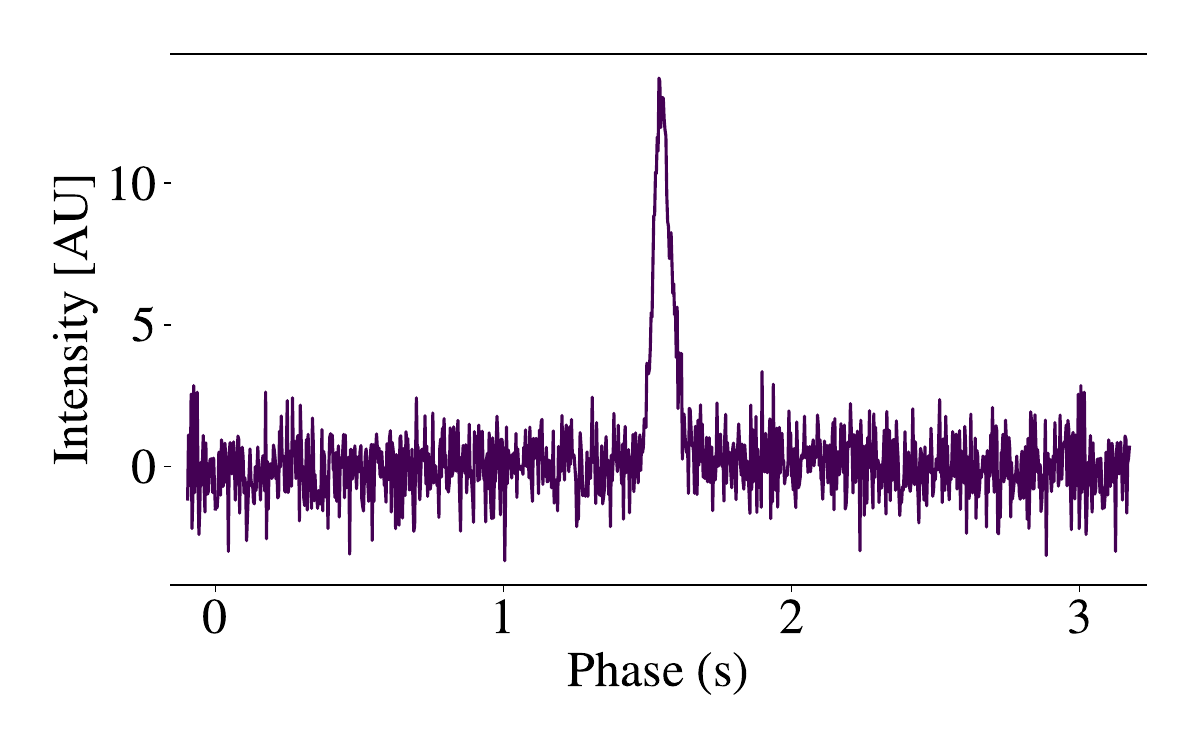} &
 \includegraphics[width=0.3\textwidth]{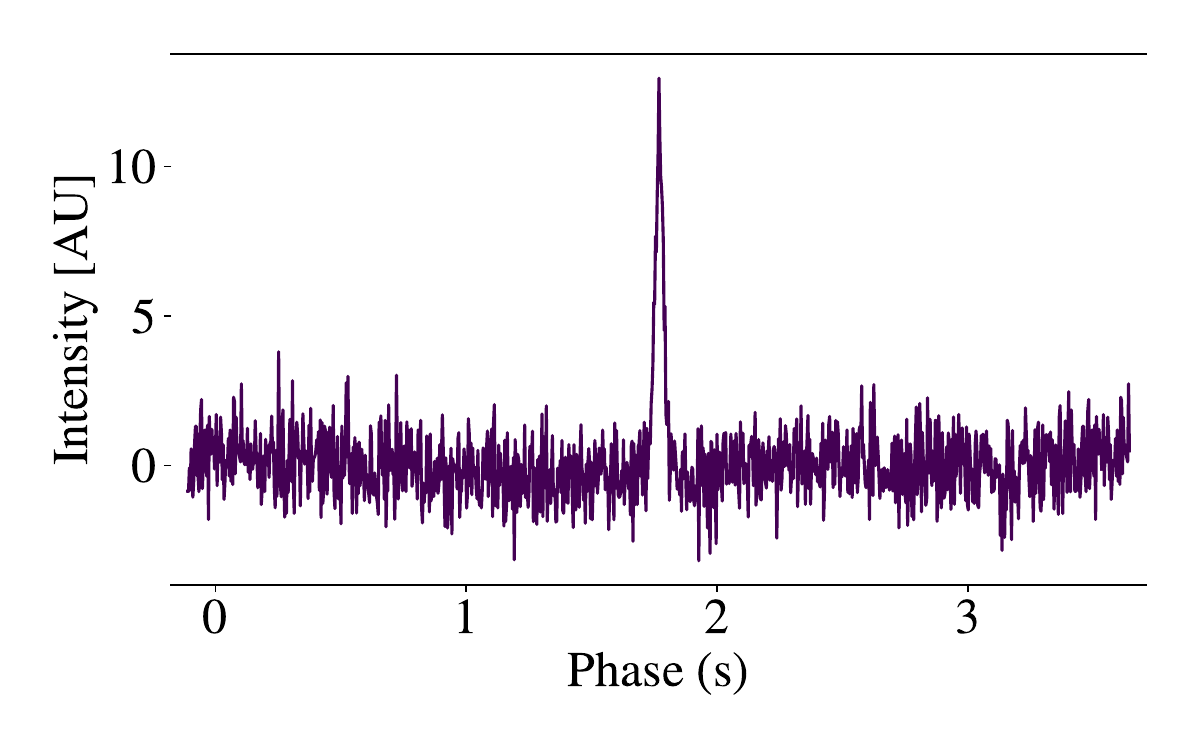} \\
 (m) J1836+5156 & (n) J2000+2920 & (o) J2105+1908 \\
 \includegraphics[width=0.3\textwidth]{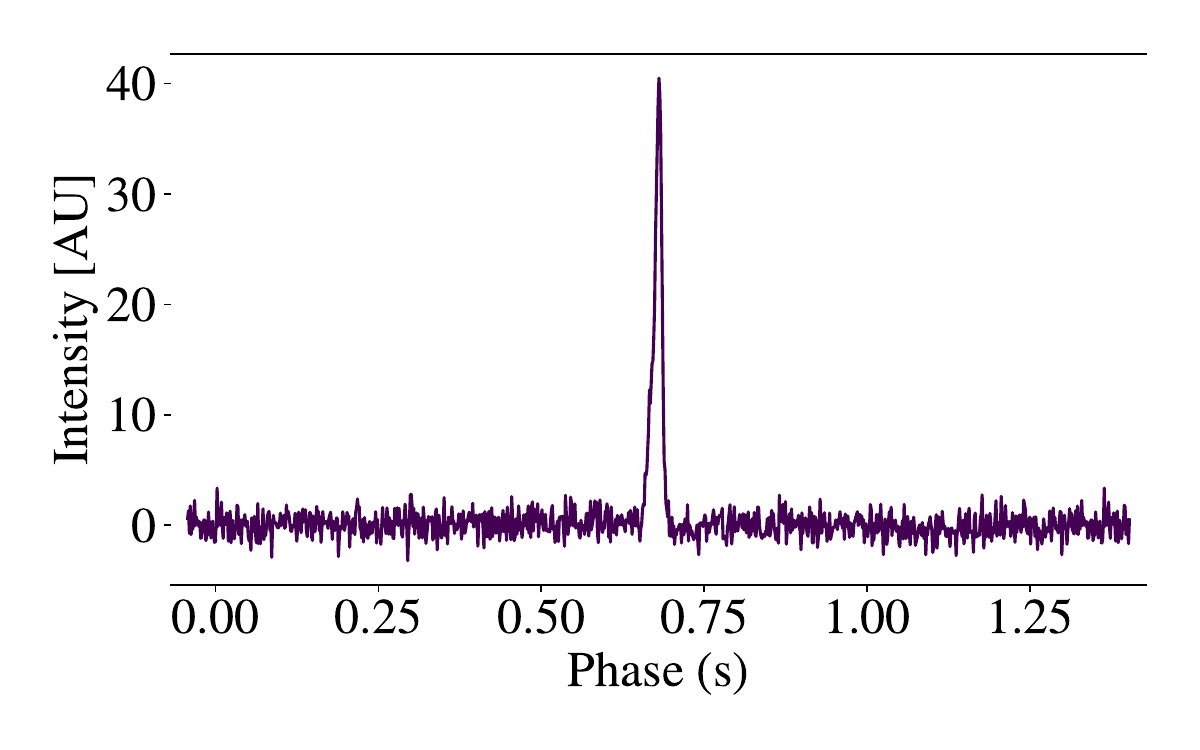} & 
 \includegraphics[width=0.3\textwidth]{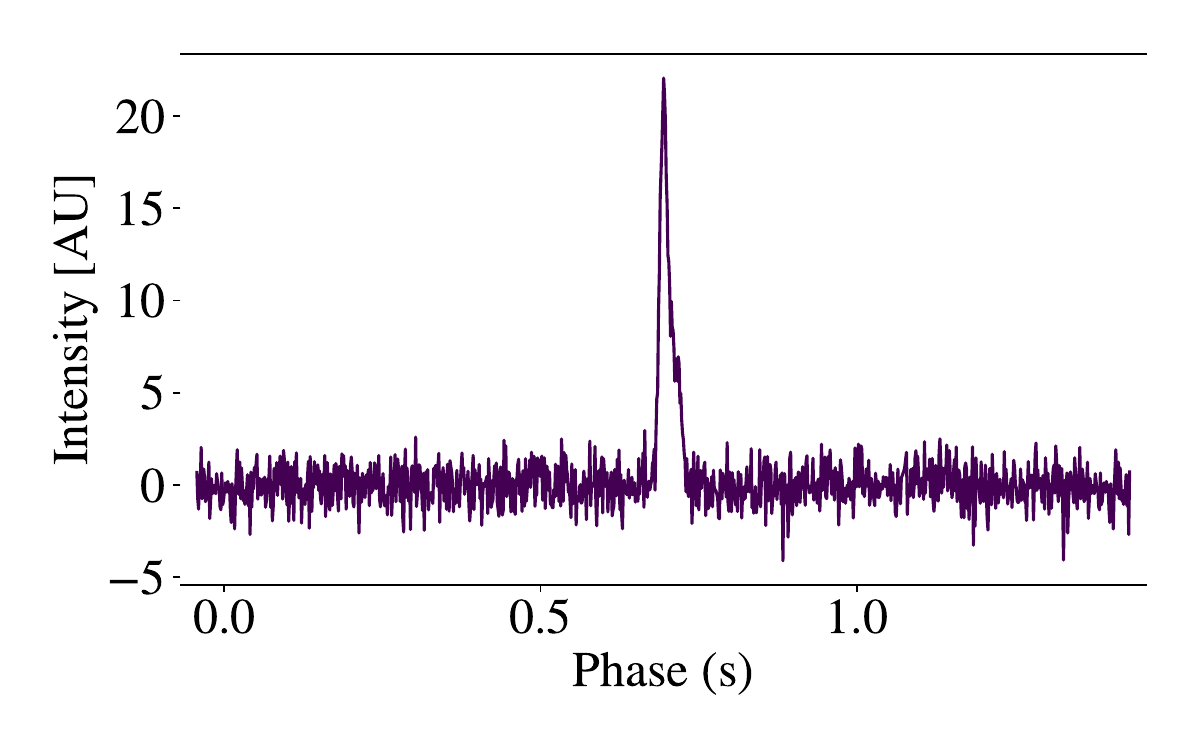} \\
 (p) J2202+2134 & (q) J2347+0300 \\
    \end{tabular}
    \caption[Folded profiles of sources timed with \ac{ilofar}.]{The Stokes $I$ frequency-averaged folded profiles of the sources detected and monitored as a part of this work. The x-axis labels cover the full folded pulse profile of each source in seconds, while the y-axis contains off-axis-normalised emission in arbitrary units.}
    \label{fig:timingprofiles}
\end{figure*}

Figure~\ref{fig:timingprofiles} contains the 1-dimensional, folded, dedispersed and time-scrunched profiles for the $\timedmonitored$ pulsars monitored as a part of this work.

\end{appendix}

\end{document}